\documentclass[traditabstract]{aa} 

\usepackage{natbib,amsmath,amssymb}

\newcommand{\ha}{\mbox{H$\alpha$}}
\newcommand{\hi}{\mbox{H{\sc i}}}
\newcommand{\hii}{\mbox{H$_2$}}

\newcommand{\msol}{\rm M$_\odot$}
\newcommand{\msolkpcs}{\rm M$_\odot$ pc$^{-2}$}

\newcommand{\gala}{M99}

\newcommand{\kms}{km~s$^{-1}$}

\newcommand{\atan}{\,{\rm atan \;} }
\newcommand{\vt}{$v_\theta$}
\newcommand{\vr}{$v_R$}
\newcommand{\vz}{$v_z$}
\newcommand{\vc}{$v_c$}
\newcommand{\los}{l.o.s}

\usepackage{graphicx,amssymb,amsmath,natbib,hyperref,multirow}
\usepackage[usenames,dvipsnames]{xcolor}
\bibpunct{(}{)}{;}{a}{}{,} 
\usepackage{rotating}
\usepackage{txfonts}

\definecolor{red}{rgb}{0.99,0.0,0.0}
\newcommand{\stars}[1]{\textbf{\textcolor{red}{STARS}}}
\definecolor{green}{rgb}{0.33,0.63,0.13}
\newcommand{\mole}[1]{\textbf{\textcolor{green}{MOLECULAR GAS}}}
\definecolor{blue}{rgb}{0.0,0.0,0.99}
\newcommand{\atom}[1]{\textbf{\textcolor{blue}{ATOMIC GAS}}}

\begin{document}

\title{Asymmetric mass  models of disk galaxies - I. Messier 99}
\titlerunning{Asymmetric mass  models of disk galaxies - I. Messier 99}

 \author{Laurent Chemin\inst{1,2} \and Jean-Marc Hur\'e\inst{1,2} \and Caroline Soubiran\inst{1,2} 
 \and Stefano Zibetti\inst{3} \and St\'ephane Charlot\inst{4} \and Daisuke Kawata\inst{5}}

\institute{Univ. Bordeaux, LAB, UMR 5804, F-33270, Floirac, France \email{astro.chemin@gmail.com}
\and CNRS, LAB, UMR 5804, F-33270, Floirac, France  
\and INAF-Osservatorio Astrofisico di Arcetri, Largo Enrico Fermi 5, I-50125 Firenze, Italy 
\and Institut d'Astrophysique de Paris, CNRS \& Universit\'e Pierre \& Marie Curie (UMR 7095), 98 bis Bd Arago, F-75014 Paris, France
\and  Mullard Space Science Laboratory, University College London. Dorking, Surrey, RH5 6NT, United Kingdom}

   \date{Received \today; accepted XX}

  \abstract{Mass models of galactic disks traditionally rely on axisymmetric density
  and rotation curves, paradoxically acting as if their most remarkable
  asymmetric features, such as lopsidedness or spiral arms, were not important.  
   In this article, we relax the axisymmetry approximation and introduce a methodology that  
   derives 3D gravitational  potentials of disk-like objects  and robustly 
   estimates the impacts of asymmetries on circular velocities in the disk midplane. 
     Mass distribution models  can then be  directly fitted to asymmetric line-of-sight velocity fields.  
     Applied to the grand-design spiral M99, the new strategy shows that circular velocities are
      highly nonuniform, particularly in the inner disk of the galaxy, as a natural response to 
     the perturbed gravitational potential of luminous matter. 
     A cuspy inner density profile of dark matter is found in M99, in the usual case where luminous and dark matter share the same
        center.  The impact of the velocity nonuniformity is to make the inner profile less steep, although the density remains cuspy. 
        On another hand, a model where the halo is core dominated and shifted by 2.2-2.5 kpc from the luminous mass center 
      is more appropriate to explain most of the kinematical lopsidedness evidenced in the velocity field of M99. 
           However, the  gravitational potential of luminous baryons is not asymmetric enough to explain the 
      kinematical lopsidedness of the innermost regions, irrespective  of the density shape of dark matter. 
     This discrepancy  points out the necessity of an additional dynamical process in these regions: possibly a lopsided distribution of dark matter.}
   

   \maketitle

\section{Introduction}
\label{sec:intro}
 
Rotation curves and surface density profiles    of  galactic disks are the observational pillars most 
models  of extragalactic dynamics are based on. Rotation curves   are needed to constrain the total mass distribution, the parameters 
of dark matter haloes, or the characteristics of modified Newtonian dynamics, while surface density profiles are helpful to constrain the structural parameters of disks and bulges, 
and generate the velocity contributions of luminous matter essential to  mass models.
As the density and rotation velocity profiles are axisymmetric by construction,   
 mass models implicitly assume  that  the   rotational velocity is only made of uniform circular motions. 
 Though  attractive for its simplicity, this approach remains a reductive exploitation of 
 velocity fields and multiwavelength  images of stellar and gaseous disks, which are  information rich. 
  In particular, it prevents one from measuring the rotational support through  perturbations (spiral arms, lopsidedness, etc.), 
which are obviously  the most striking features of galactic disks. 
In an era of conflict between    
observations  and expectations from Cold Dark Matter (CDM)  simulations,  
the cusp-core controversy \citep[see the review of][and references therein, but see \citet{gov10}]{deb10}, 
it appeared fundamental to  assess the impact of such perturbations  on the shape of   rotation curves, 
  and more generally on mass  models and density profiles of dark matter. 
     
 This is the reason why efforts have been made to determine the kinematical asymmetries   
 inferred by perturbations or to model the effects of dynamical perturbations. 
 In the former case,   \citet{fra94} and \citet{sch97} initiated  
  the derivation of high-order harmonics with first-order kinematical components from gaseous velocity fields.  
  They argued that kinematical Fourier coefficients are useful to constrain  
    deviations from axisymmetry and 
   the nature of dynamical perturbations.  Using that technique, \citet{gen05} concluded, for instance, that the  kinematical asymmetries 
  in  the \hi\ velocity field of a  dwarf disk presenting a core-dominated dark matter halo (DDO 47)   
  could likely originate from  a spiral  structure. However, 
  their amplitudes were not high enough to account 
  for the velocity difference expected between 
    the CDM cusp and the cored halo preferred by the rotation curve fittings. 
    In the second case, \citet{spe07} proposed to fit a bisymmetric model of bar-like/oval distortion 
     to the \ha\ velocity field of another low-mass spiral galaxy (NGC 2976) and argued 
     that   negligible high-order Fourier motions in velocity fields do not necessarily imply  
     that the bisymmetric perturbation is negligible, and that the rotation curve should be similar to the underlying circular motions only if 
    the departures from circularity remain small. These authors also showed that the inner slope of the rotation curve of NGC 2976 is likely 
     affected by the bar-like perturbation.  Numerical simulations of barred disks  
     arrived at a similar conclusion about the impact of the bar on the inner shape of rotation curves  \citep{val07,dic08}.  
     \citet{ran15}  performed numerical simulations to determine a  
      corrected rotation curve for another barred galaxy (NGC 3319), free from the perturbing motions induced by the bar.        
    While these simulations demonstrate it is possible to hide cuspier DM distributions into artificial cored distributions under the effect of stellar 
    bars, they perfectly illustrate the difficulty of performing mass models and constraining the shape 
    of dark matter density profiles from observations of barred galaxies.   Other numerical models based on closed-loop orbits showed 
    that shallow kinematics and core-like haloes could actually be explained by cuspy triaxial distributions of dark matter viewed with particular projection
     angles \citep{hay06}.
    Finally,  other studies  focused on extracting in \hi\ spectra the line-of-sight (\los)  velocity components  supposed to 
  trace  the axisymmetric rotational velocities better than the components based on more usual  intensity-weighted means \citep{oh08}. 
  Applied to two dark matter dominated disks, NGC 2366 and IC 2574, which are   prototypes of galaxies whose dark matter density conflicts 
  with the cosmological cusp, this method yielded steeper rotation curves in the inner disk regions. However, the velocity differences 
  with the intensity-weighted mean velocity curve were not sufficient to   
  reconcile the observation with the CDM cusp.
    
\begin{figure}[t]
\centering
\includegraphics[width=0.7\columnwidth]{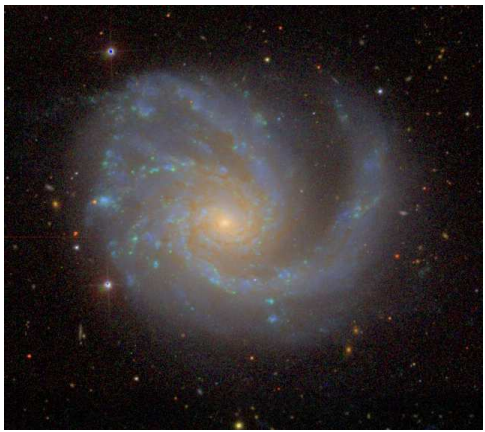}
\caption{Composite SDSS $gri$-image of the grand-design spiral galaxy M99. North is up; east is left. The image size is
$7\arcmin \times 6.3\arcmin$.}
 \label{fig:m99}
 \end{figure}

  In this context, in this article we propose a new approach to model the mass distribution of disk galaxies.
 Our   strategy goes beyond the   decomposition of rotation curves and fully exploits the bidimensional distribution 
 of luminous matter, thus the asymmetric nature 
  of  stellar and gaseous  disks. 
  Our approach determines the 3D gravitational potential of any disk-like mass component through hyperpotentials theorized 
   by \citet{hur13}. It then derives the corresponding circular velocity map in the disk midplane, 
   which allows us to determine where and to which extent the circular 
   motions should deviate from axisymmetry.
   A 2D mass distribution model can then be directly fitted to a \los\ velocity field, 
   by adding the 2D velocity contributions from luminous baryons  to that from  the missing matter. The impacts of the velocity asymmetries 
   on the mass models and structure of dark matter haloes can then been investigated by comparing with results 
   obtained with the axisymmetric mass models.   
   
    We apply that methodology to a prototype of unbarred, spiral galaxy Messier 99, whose general properties are presented 
    in  Section~\ref{sec:pres}. The  axisymmetric mass model of the  high-resolution rotation curve of M99 
    is detailed in  Section~\ref{sec:axiapp}. This Section also presents  a more elaborated axisymmetric mass model,   
    fitted directly to a high-resolution velocity field of M99. The  basis of the derivation of the 3D 
    gravitational potentials for luminous matter is described in  Section~\ref{sec:asymapp}, which also 
    presents  the inferred   gravitational potentials, accelerations, and circular velocities
     for the contributions from the stellar, atomic , and molecular gas disks of M99.
    The  meaning of these asymmetric outputs is discussed in Section~\ref{sec:nbodysims}. 
    We then perform asymmetric mass distribution models of M99 (Section~\ref{sec:fitasym}), 
 and compare them to  the   axisymmetry-based models of  Section~\ref{sec:axiapp}. 
 The conclusions for M99 and the prospects of our new strategy for galactic dynamics are finally given in Section~\ref{sec:conclusion}.

\section{The grand-design spiral galaxy Messier 99}
\label{sec:pres} 
\subsection{A brief presentation}

 We selected the galaxy Messier 99 (M99 hereafter) because it is a SAc 
  type disk harboring a very prominent spiral structure (Fig.~\ref{fig:m99}).
   Located in the Virgo Cluster \citep[adopted distance  of 17.1 Mpc,][]{fre94}, observing M99 is  a good opportunity to benefit from high-sensitivity and high-resolution multiwavelength 
observations of the stellar disk and  interstellar medium. 
The integrated \ha\ profile from data of \citet{che06} has a width at 20\% of the maximum \ha\ peak of 216 \kms. Combined with a small disk inclination \citep[20\degr,][]{mak14}, this implies a  massive
 galaxy with most of rotation velocities greater than 245 \kms.
 This makes it an ideal target to study the structure and kinematics 
of the disk in detail and to test our new mass modeling strategy. 
 
 The spiral structure is asymmetric. A one-arm mode dominates the  \hi\ disk
 \citep{pho93,chu09}, while the stellar distribution exhibits more than one single arm (Fig.~\ref{fig:m99}). 
 \citet{pho93} proposed a scenario where the \hi\ arm is triggered 
 by gas infalling and winding on the disk, resulting from a tidal encounter. 
 Based on numerical simulations, \citet{vol05}   mimic   the asymmetric  disk and perturbed \hi\ kinematics by a flyby of a massive companion, coupled with 
 ram pressure stripping 
from the Virgo intracluster medium (ICM). These authors argued that M99 is entering the Virgo cluster for the first time.  
Further numerical models of \citet{duc08} also explained   the origin of the large-scale \hi\ tail, which apparently connects M99 to VIRGOHI21, a $10^8$ \msol\ \hi\ cloud 
wandering in the Virgo ICM \citep{min07}, by a tidal interaction that was with another candidate companion than  in \citet{vol05}. 
In principle, mass distribution models should only be carried out with targets supposedly in dynamical equilibrium. 
The possibly infalling \hi\ mass of $\sim 10^8$ \msol\ only represents  $\sim$ 2\% of the total neutral gas 
mass of M99, and about 0.2\% of the total luminous mass  (Section~\ref{sec:axiapp}). This should be not enough to affect the large-scale dynamics of M99. 
Furthermore, ram pressure stripping predominantly affects the outskirts of the neutral atomic gas component, not the inner, densest regions of the gaseous disk of M99.

\subsection{High-resolution \ha\ kinematics}
\begin{figure*}[th!]
\includegraphics[width=0.33\textwidth]{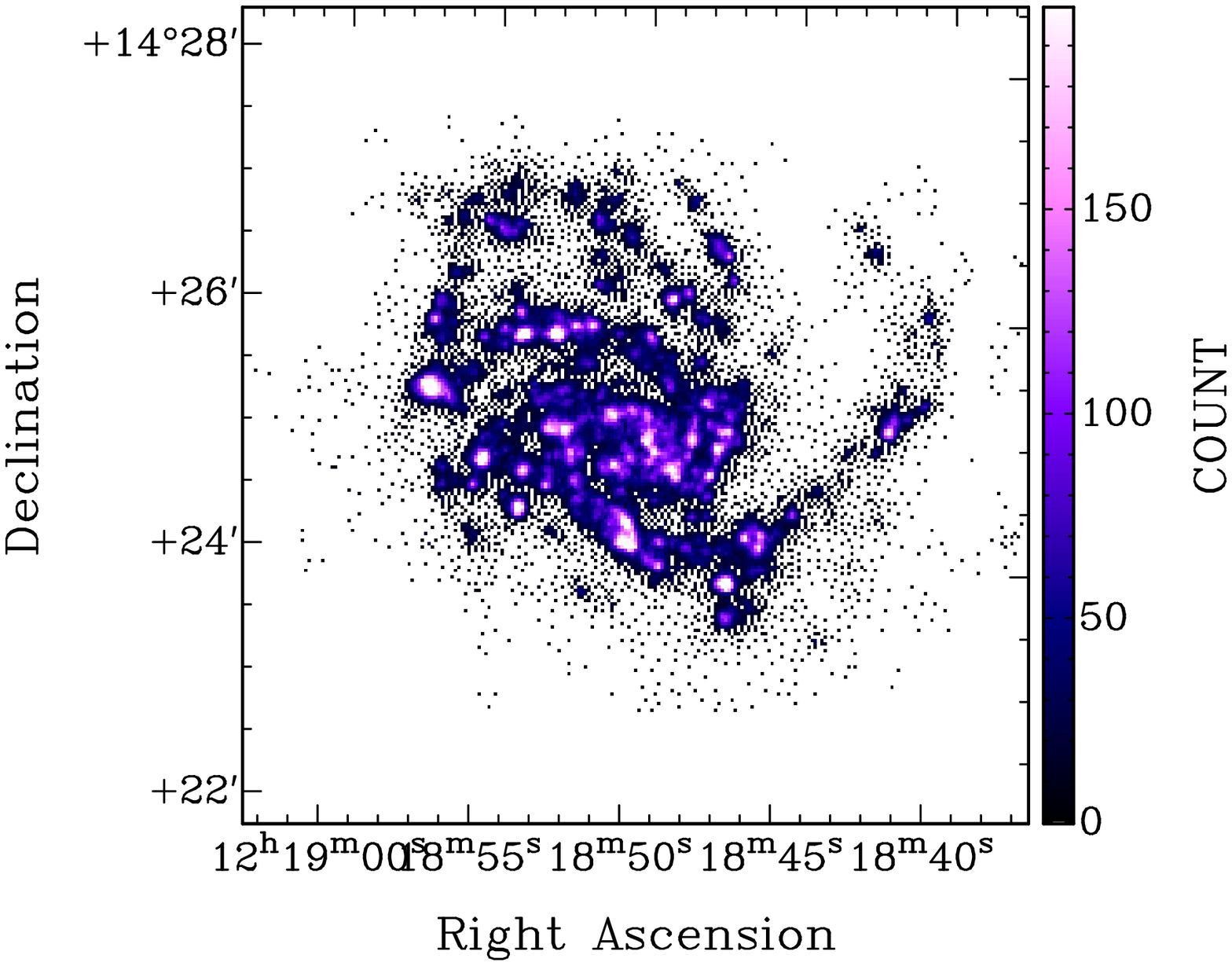}\includegraphics[width=0.33\textwidth]{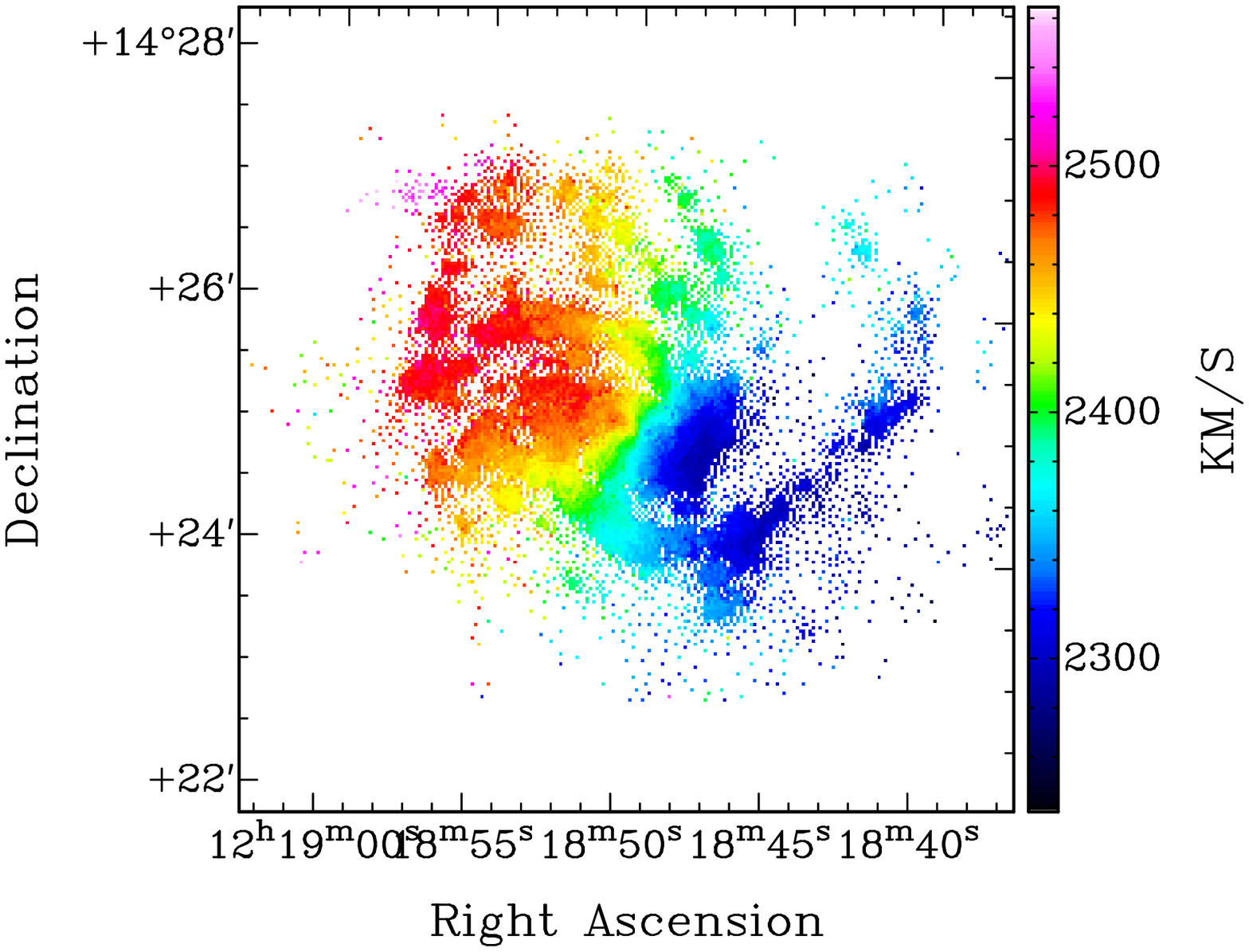}\includegraphics[width=0.33\textwidth]{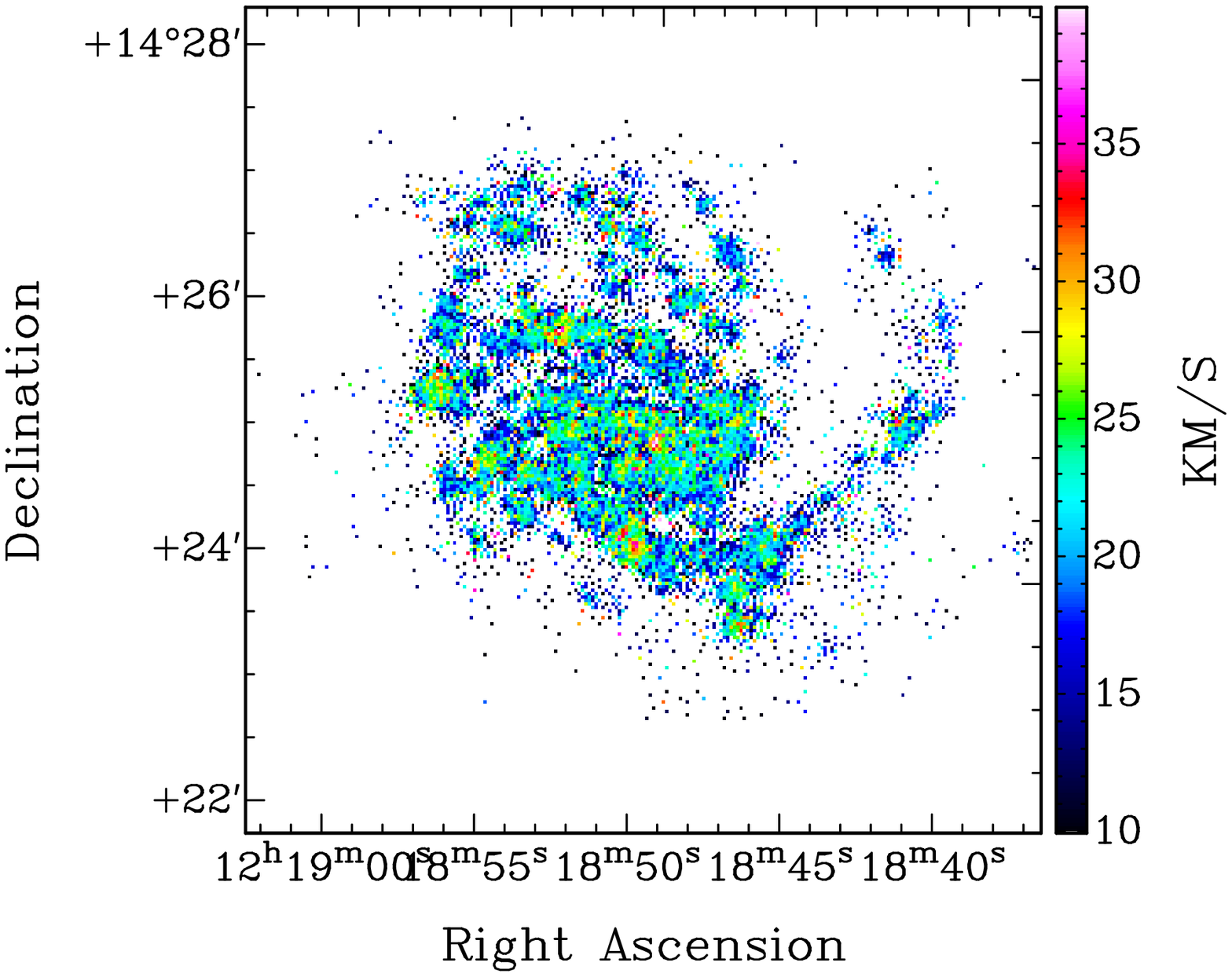}
\caption{\ha\ integrated emission, velocity field, and velocity dispersion map of M99 (from left to right, respectively). }
 \label{fig:havfobs}
 \end{figure*}
 
Optical long-slit and resolved observations  also revealed perturbed kinematics of the ionized gas disk of M99 \citep{pho93,kra01,che06}.  
In particular, \citet{kra01} argues for the presence  of a possible stellar bar that would impact the \ha\ kinematics in the inner 1.5 kpc. 
Asymmetric motions have also been clearly evidenced along the \ha\ spiral arms \citep{pho93,che06}, as well as a lopsided \ha\ velocity field \citep{che05a}.
The kinematical data we use  to model the mass distribution of M99 are from the 3D spectroscopy survey of Virgo cluster galaxies 
by \citet{che05b, che06}. In their catalog of  \ha\ velocity fields of bright Virgo spiral and irregular disks, 
the  Fabry-Perot observations of M99 have angular and spectral samplings of
1.6\arcsec\ (130 pc at the adopted distance) and $\sim 8.2$ \kms. 
The \ha\ velocity field of \citet{che06} resulted from an adaptive binning  and a spatial interpolation of the \ha\ datacube, following prescriptions given in \cite{dai06}.  
 We did not use these  interpolated data but an improved version of the binned datacube of \cite{che06}, where new bins 
 are now made of a unique pixel  located at the  barycenter of   initial bins. Every bin has thus an equal angular size,  in accordance with the 
kinematical analysis of many other \ha\ Fabry-Perot velocity fields \citep{epi08b,epi08a}.  Figure~\ref{fig:havfobs} shows the revisited integrated 
 \ha\ emission,  velocity, and dispersion velocity fields of M99.  We restricted the \ha\ kinematics to $R=11.5$ kpc, which corresponds to a distribution of
9002 velocity pixels. This allowed us to derive accurate velocity and velocity dispersion profiles.
Beyond the optical size of the stellar disk, the \hi\ kinematics is too scattered to infer useful kinematical information. 

\section{Axisymmetric mass modeling of M99}
\label{sec:axiapp}

A traditional axisymmetric mass model consists in decomposing a rotation curve 
into  contributions from luminous baryons (stellar and gaseous disks, bulge, etc.) and dark matter. 
The fitting procedure yields  fundamental scale parameters for the hidden mass component, 
and eventually a factor that  enables the scaling of  luminosities into surface densities for stars (the 
mass-to-light ratio, M/L).  
This section details the rotation curve decomposition, which we refer to   as the  1D axisymmetric case  hereafter, as   
the rotation velocity only depends on one angular scale, which is the galactocentric radius. 

This Section also envisages a second  strategy, where the axisymmetric modeling is  carried out directly from the velocity field. The motivation for this is, first, that 
circular motions are expected to dominate the kinematics and, since rotation curves stem from  velocity  fields, one may logically perform 
  mass models   with 2D resolved data instead of 1D curves. The second motivation is that the development of a pipeline   that makes fittings in 2D  
  is mandatory for the new asymmetric methodology discussed from Sections~\ref{sec:asymapp} to~\ref{sec:fitasym}. 
  The most natural way to understand the impact of the asymmetries on the mass models 
  is, thus, to  fit to the \ha\ velocity field of M99 2D velocity models  built under axisymmetry assumptions. 
  In short, one cannot directly compare the results from the decomposition of the rotation curve with those presented in Sect.~\ref{sec:fitasym} for
   the asymmetric fittings of the velocity field. 
  One needs an intermediate step between them, which we refer to as the  2D axisymmetric case hereafter.  
 Though the data to be fitted are not totally axisymmetric because of the evident signatures of, for example, spiral arms, 
 we nonetheless call this 2D case axisymmetric because the individual contributions from 
 dark matter, stars, and gas are  distributed axisymmetrically.  

 This Section is therefore organized as a traditional axisymmetric mass   model of a galactic disk. 
 We first derive the rotation curve of M99 and estimate the asymmetric drift contribution, and then we present   
 multiwavelength observations needed to infer the mass 
 surface density profiles and the velocity contributions of the stellar and interstellar matter. 
 The  results of the mass modeling  are then discussed  to investigate which dark matter halo shape is more appropriate. 
 In addition, we discuss the 2D axisymmetric modeling. 
 
\begin{figure*}
\includegraphics[width=0.33\textwidth]{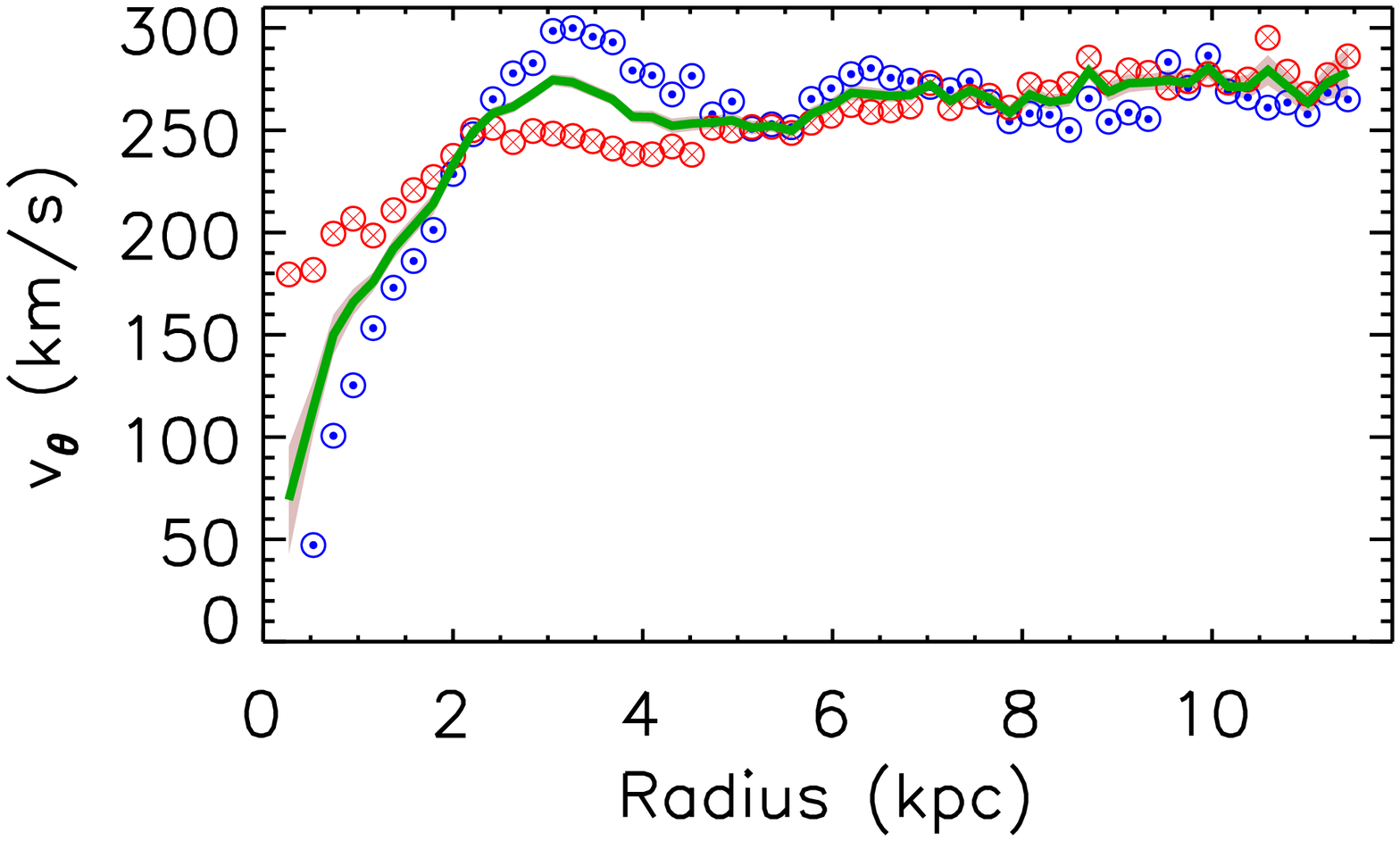}\includegraphics[width=0.33\textwidth]{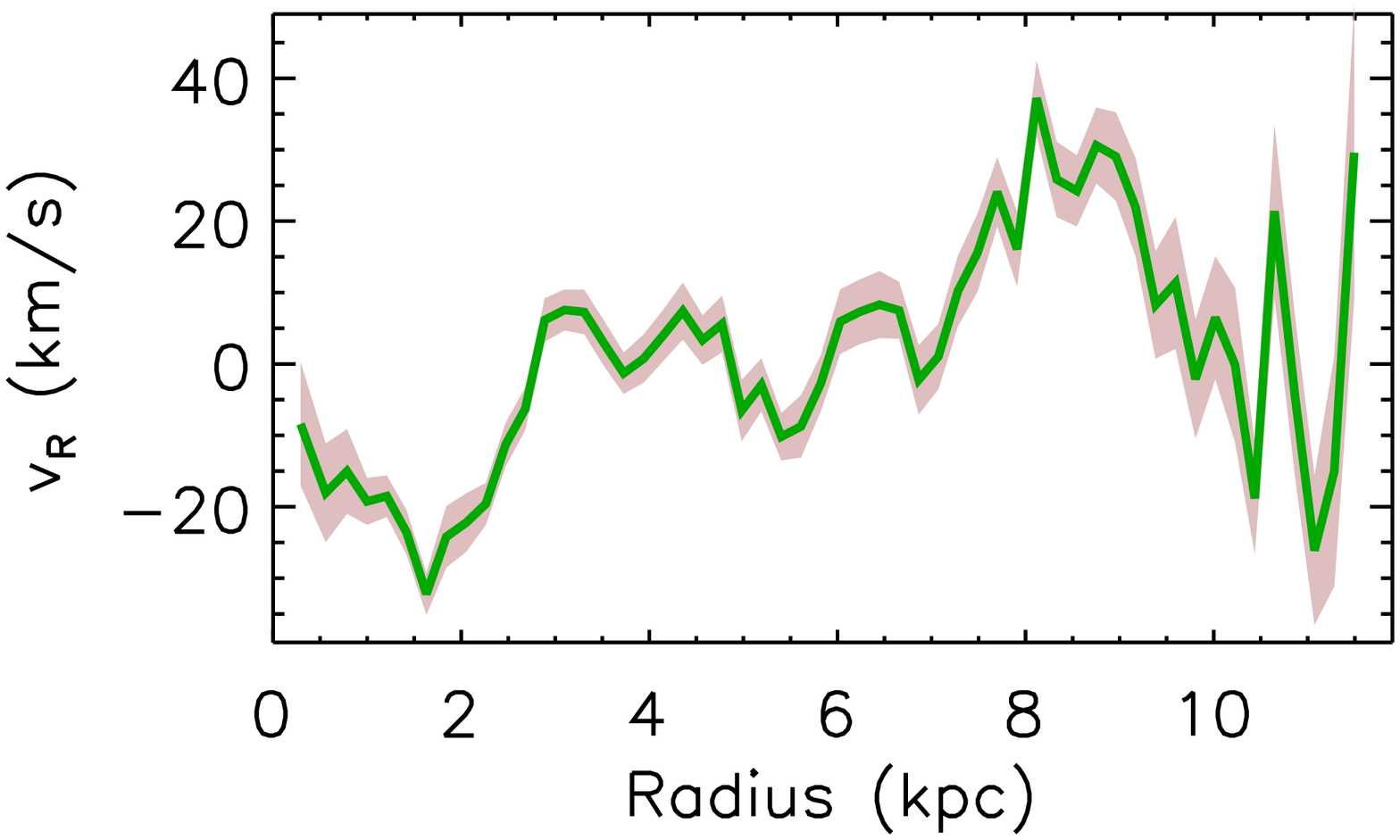}\includegraphics[width=0.33\textwidth]{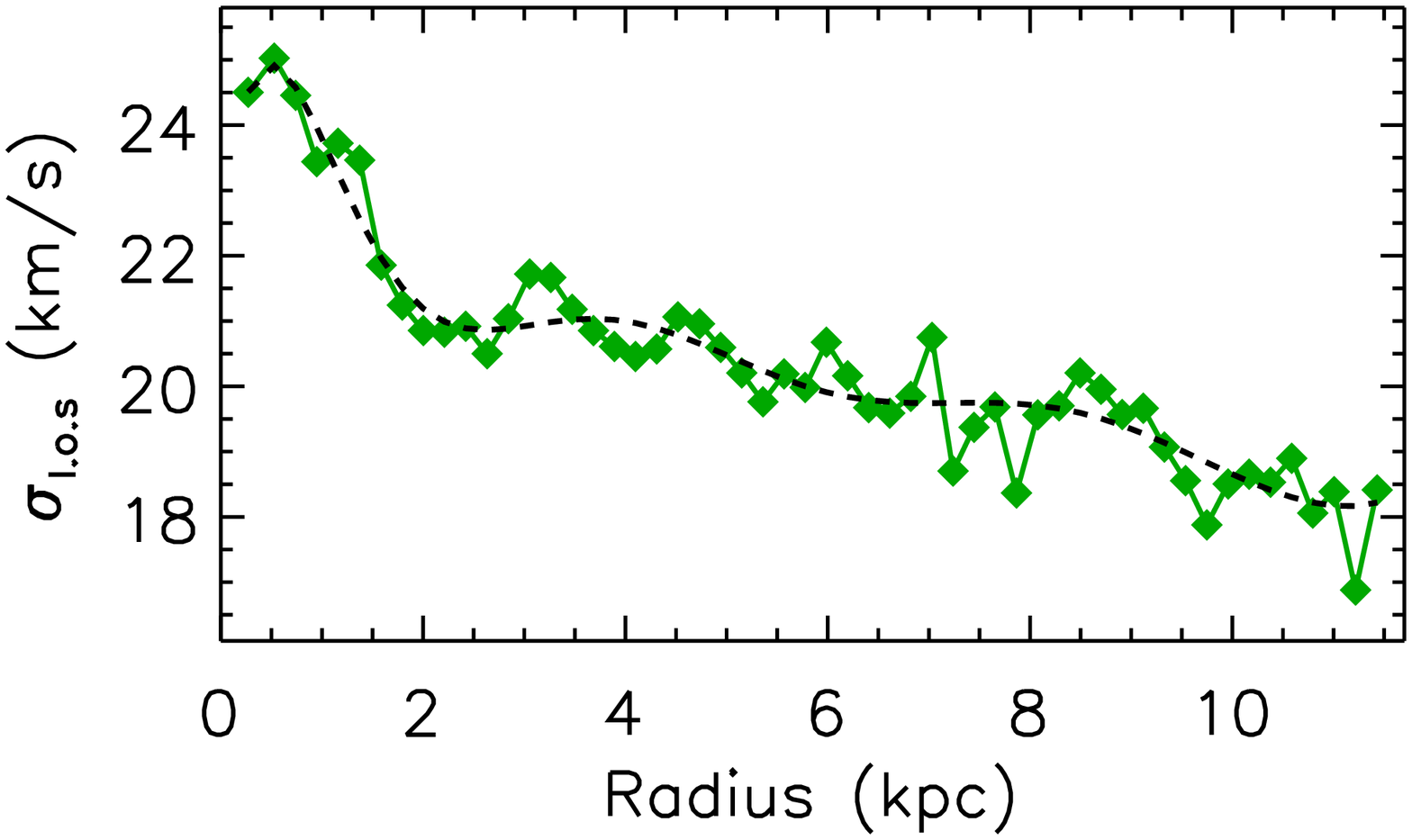}
\caption{\textit{Left and middle:} Profiles of tangential and radial  velocity  of M99. 
 Shaded area indicates the $1\sigma$ r.m.s. from the fittings. A green solid line represents the fitting for the whole disk, while 
 blue dotted  and red crossed circles are respectively the results for the approaching and receding sides fitted separately.  
 \textit{Right panel:} \ha\ velocity dispersion of M99. Symbols represent the observed dispersion; 
   the dashed line is a smooth model of the observation used to derive the asymmetric drift.}
 \label{fig:harc}
 \end{figure*}

\subsection{Tangential and radial  velocities in M99}
\label{sec:rcvr}
As the mass distribution modeling first needs  a rotation curve, 
the 3D velocity space (\vr,\vt,\vz) is deduced 
from fitting to the \ha\ velocity field of M99  the following model: 
\begin{equation}
  v_{\rm l.o.s} =   v_{\rm sys} +   (v_\theta \cos\theta + v_R\sin\theta) \sin i + v_z \cos i  
  \label{eq:vtanvradvf}
,\end{equation}

where $v_{\rm l.o.s}$ is the \los\ velocity, $v_{\rm sys}$ and $i$ are the systemic velocity and the inclination of the M99 disk, 
and $\theta$ is the azimuthal angle  in the deprojected orbit.
In the axisymmetric approach, these velocities are assumed uniform, i.e. only dependent on $R$. The azimuthal velocity is the rotation curve. 
The \vr\ and  \vz\ components are usually omitted in kinematical studies because they are generally assumed to be negligible. 
As seen below, it is  the case for \vr\ but since one of our goals is to study the  impact of the radial motions on the mass modeling, we 
decided to fit them. Deriving \vt\ with or without \vr\   does not impact the shape or amplitude of the 
tangential component.  
Then, a problem in leaving \vz\ free  in Eq.~\ref{eq:vtanvradvf}  is that it should exhibit artificial variations if  
 the galaxy has a kinematical lopsidedness. The reason for this  is that  the systemic velocity  must  be naturally impacted by a lopsidedness (see Appendix~\ref{sec:fouriervf}), 
 but since variations of $v_{\rm sys}$ are not allowed here, the $v_z \cos i$ term absorbs the kinematical signature of lopsidedness. 
 As both the gravitational potential of luminous matter and the \ha\ kinematics of M99 are lopsided (Section~\ref{subsec:potaccmap} and Appendix~\ref{sec:fouriervf}), 
 the best solution   to avoid such artificial  \vz\ variations   is 
 thus to assume $v_z=0$ hereafter. Other face-on, grand-design spirals of similar star formation activity to M99  
  are known to have negligible vertical motions \citep[e.g., NGC 628,][]{kam92}.
   
Since the \ha\ gas is well confined in the optical disk that is not warped, 
we have not allowed $v_{\rm sys}$, $i$, the position angle of the disk major axis ($\Gamma$)
and the coordinates of  the dynamical center to vary with radius. 
We chose  $\theta = 0\degr$ aligned with the semimajor axis of the receding side of the galaxy disk. 
We used an inclination of 20\degr, which is the one of the stellar disk, i.e. the photometric value. This value differs from the kinematical value of $31\degr \pm 6\degr$ derived in \citet{che06}. 
The photometric value is more appropriate for the mass modeling than  larger inclinations (see Section~\ref{sec:lumvelo}).
 Nonlinear Levenberg-Marquardt fittings were performed with uniformly weighted velocities.   
We found $v_{\rm sys}=(2398 \pm 0.5)$ \kms and   $\Gamma = (67\pm 0.5) \degr$. 
The kinematical center appears slightly offset from the photometric center (peak of the stellar density). However that difference 
is not significant owing to the measured uncertainties \citep[see also][]{che06}. 
For simplicity, we adopt the photometric center as the kinematical center. 
These results are in very good agreement with \citet{che06}. 
The \ha\ systemic velocity agrees   well with the centroid of the   integrated \ha\ profile from our dataset (2392 \kms) or the value found by \citet{chu09} from the integrated \hi\ profile of M99 (2395 \kms).
Then, we  derived \vr\ and \vt\  with all other parameters fixed at these adopted values.
We chose an adaptive angular sampling with a ring size of at least 2.5\arcsec\ (well larger than the seeing of the observations) and with 
 a minimum number of 30  pixels per ring to ensure good quality fits. With these rules, radial bins are fully uncorrelated.  

 Figure~\ref{fig:harc} shows the resulting profiles of \vt\ and \vr, whose values are reported in Tab.~\ref{tab:vtvrsigmalos} of Appendix~\ref{sec:vtvrsigmalos}. 
 The rotation curve is regular. It shows a peak at $R \sim 3$ kpc, 
 which roughly corresponds to the radius equivalent to 2.2 times the stellar disk scalelength (see Section~\ref{sec:surfdens}), and a
 flat part at larger radii ($v_\theta \sim 270$ \kms). This rotation
  curve is consistent with the \ha\ measurements presented in \citet{pho93}, \citet{kra01}, and  \citet{che06} after scaling at similar inclinations. We also 
verified the good consistency with  CO kinematics from \cite{ler09}.  
 The rotation  curves for the approaching and receding halves of the \ha\ disk have also been fitted separately. They are  asymmetric 
 in the inner 2 kpc, and around 3-3.5 kpc. This is one of the signatures of the kinematical  lopsidedness of M99. 
    The Very Large Array (VLA) 
   data did not allow us to perform a reliable tilted-ring model of the \hi\ velocity field  
   for $R>11.5$ kpc. The scatter of the \hi\ kinematics is indeed too large to consider an outer \hi\ rotation curve, and constrain 
   the parameters of a disk warping, if any warp exists in M99.  
    This has nonetheless no consequence on the results presented hereafter because \hi\ densities are small 
     and the gravitational impact of the atomic gas disk is   negligible in the overall mass budget.
 
  The rotation of M99 is clockwise, assuming trailing spiral arms. With this rule, positive \vr\ (negative, respectively) 
 corresponds to  motions radially oriented inward (outward). 
 As a consequence,  the fitted profile presents globally inward radial motions in M99, except   in the inner $R = 2.7$ kpc, between $R=4.9$ kpc and $R=5.9$ kpc 
 and  locally at $R = 3.7$ kpc and $R=6.9$ kpc, which are consistent with \vr\  directed outward. Beyond $R\sim10$ kpc, the scatter of 
 the radial component becomes  larger, though  this scatter is mainly consistent with a velocity directed outward.
 The formal error  from the fittings are small at these radii and remain negligible relative to the amplitude of \vr.
  
 \begin{figure*}[t!]
 \begin{center}
  \includegraphics[width=0.9\textwidth,trim=0 0 60 0,clip=true]{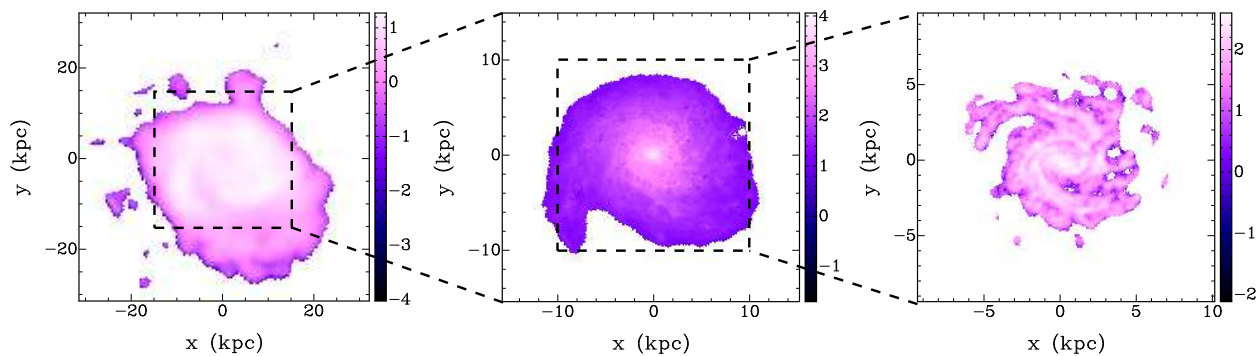}
   \caption{Surface density maps for the atomic gas, stellar and molecular gas disks
    of M99 (from left to right, respectively). Densities (in \msolkpcs) are shown in a decimal logarithmic scale. }
 \label{fig:density}
 \end{center}
 \end{figure*}

 It is expected that the observed tangential velocity of the kinematical tracer differs from its circular velocity because   
of asymmetric drift. Starting from   Eq. (4.227) of \citet{bin08}, and assuming that random motions drives the gas pressure, 
an isotropic dispersion ellipsoid, and that the product $v_R v_z$ is independent of $z$, 
the   circular velocities, $v_c$, are deduced from the tangential motions by
  \begin{equation}
   v^2_c = v^2_\theta  -   \sigma_{\rm l.o.s}^2  R \left(\frac{d \ln\rho}{dR}  + \frac{d \ln\sigma^2_{\rm l.o.s}}{dR}\right) 
   \label{eq:pressurecorrection}
   ,\end{equation} 
where $\sigma_{\rm l.o.s}$ is the observed \los\ dispersion and
 $\rho$ is the total (atomic+molecular) mass volume density that can be replaced by 
 the surface density $\Sigma$ if we also assume a constant disk thickness. 
The atomic and molecular gas densities are those presented in Section~\ref{sec:surfdens}. 
The \ha\ velocity dispersion profile, corrected from instrumental broadening, is shown in 
Fig.~\ref{fig:harc} and listed in    Appendix~\ref{sec:vtvrsigmalos}.  The mean dispersion is 20 \kms\ for a standard deviation of $\sim$2 \kms,  
 which is comparable to values found for many other star-forming galaxies of similar morphology and mass \citep[][]{epi10,kam15}. 
 The profile variations of the density and dispersion remain too  small   to imply a significant gas asymmetric drift.    
Equation~\ref{eq:pressurecorrection}    yields an almost constant pressure support, well represented by an average value 
$\langle v_c - v_\theta \rangle \sim 1$ \kms. Because of this minor correction relatively to \vt, asymmetric drift  is neglected hereafter. 
Such values are fully consistent with results found
for other   dwarf or massive disk galaxies 
 \citep[][]{deb02,gen07,swa09,dal10,wes11,mar13b}.

\subsection{Stellar and gaseous surface densities}
\label{sec:surfdens}
  
  The mass models need   velocity contribution from  luminous matter.  
  We have considered individual contributions from a molecular gas disk (mol), an
  atomic gas disk (atom), a stellar disk ($\rm \star,D$), and a  stellar bulge ($\rm \star,B$). The corresponding velocity components 
  are deduced  from mass surface densities.
   
 The molecular gas disk surface densities are from CO 1--0  mm observations 
 of \cite{rah11} from the Combined Array for Research in Millimeter-wave Astronomy (CARMA) array, at an angular resolution of 4.3\arcsec.
 The CO 0-th moment map  has been translated to H$_2$  surface densities using a conversion factor of $\rm 1.8 \times 10^{20}\ cm^{-2}\ (K\ km\ s^{-1})^{-1}$. 
 The \hii\ gas mass is $\sim 5\times 10^9$ \msol. 
   The atomic gas disk surface densities  come from the Very Large Array  
   \hi\ survey of Virgo Cluster galaxies by \cite{chu09}, originally from \cite{pho93}, at an angular resolution of  
  $\sim$ 25\arcsec.   Using the adopted distance, the \hi\ gas mass is $\sim 5\times 10^9$ \msol, which is thus 
  the same as the \hii\ mass.
  The total mass density maps for the atomic and molecular contributions are finally obtained by multiplying the
   \hi\ and \hii\ densities by the usual factor of 1.37 to take  the contribution of elements that are heavier than hydrogen into account. 
   
  The stellar mass surface densities are estimated using the method of \cite{zib09}, based on the pixel-by-pixel 
  comparison of optical and near-infrared (NIR) colors with a suite of stellar population synthesis models. 
  \citet{zib09} and \citet{zib09b} provide details of
   the image reduction and signal-to-noise enhancement via adaptive smoothing, which allows one to extract accurate (error $< 0.05$ mag) 
   surface brightness at each pixel in $g$, $i$, and $H$-bands, respectively, and, in turn, $g-i,$  $i-H$ colors. 
   We compute the same colors for a set of 150,000 composite stellar population synthesis (SPS) models 
   with variable star formation histories (exponentially 
   declining plus random bursts), metallicities, and two-component dust attenuations as in \citet{cha00}, following the prescriptions of  
   \citet{dac08}. The models are binned in the $g-i$, $i-H$ space and the median  M/L  in $H-$band is computed for each bin. 
  For any given pixel, the measured $g-i$ and $i-H$ colors select the bin in the model libraries and the corresponding M/L is 
  assigned to the pixel. By multiplying the $H$-band surface brightness with this M/L, we obtain the stellar mass surface density. 
With respect to the SPS library adopted in \citet{zib09}, the only difference is that the base of simple stellar populations (SSP) 
is not the so-called CB07 version of the \citet[][BC03]{bru03} models, but the 2012 updated version of the original 
BC03 SSPs\footnote{\url{bruzual.org/~gbruzual/bc03/Updated_version_2012}}. In fact, in recent years, many observations 
have shown that models with a very strong contribution by TP-AGB stars \citep[e.g.,][CB07]{mar05} fail to 
reproduce the optical-NIR spectral energy distribution of galaxies in the low- to intermediate-redshift Universe 
\citep[e.g.,][]{kri10, zib13, mel14}, while models with a  more moderate 
TP-AGB contribution (e.g. BC03) work  better. This motivates our decision to opt for BC03 SSPs. The M/L in 
$H-$band estimated with TP-AGB light models are typically $0.1-0.2$ up to $0.3$ dex higher than estimated with TP-AGB heavy
models \citep[see Fig. 3 of][]{zib09}. The 2012 update of BC03 introduces some improvements in the treatment of the stellar remnant, 
which results in larger M/L by roughly 10\% for old stellar populations. 
  The pixel scale of the stellar mass map is 1.6\arcsec. The stellar mass of M99 is $\sim 4.2\times 10^{10}$ \msol. 
  In total, it   yields a luminous mass of $5.2\times 10^{10}$ \msol.

\begin{figure*}[th]
\begin{center}
\includegraphics[width=0.45\textwidth]{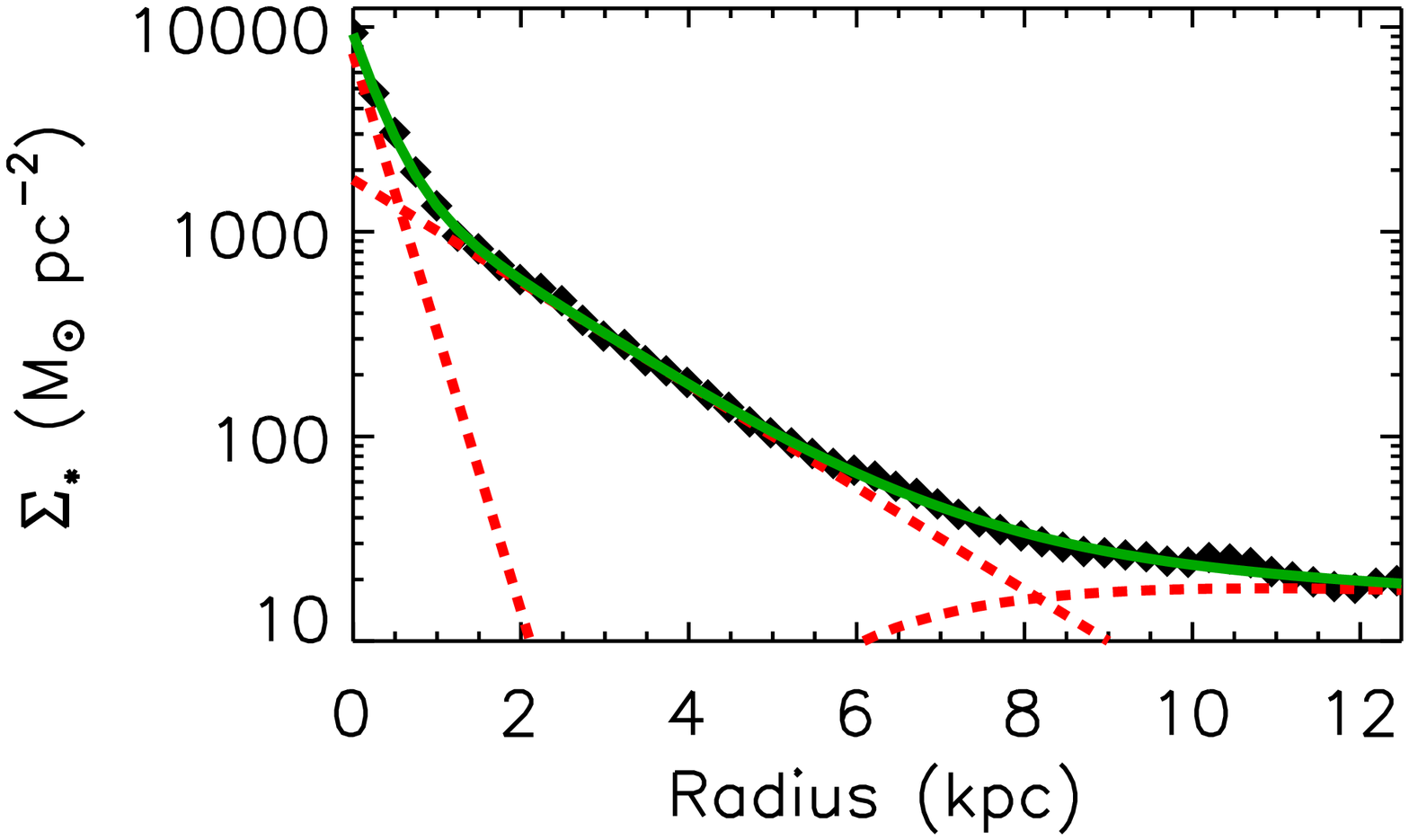}\includegraphics[width=0.45\textwidth]{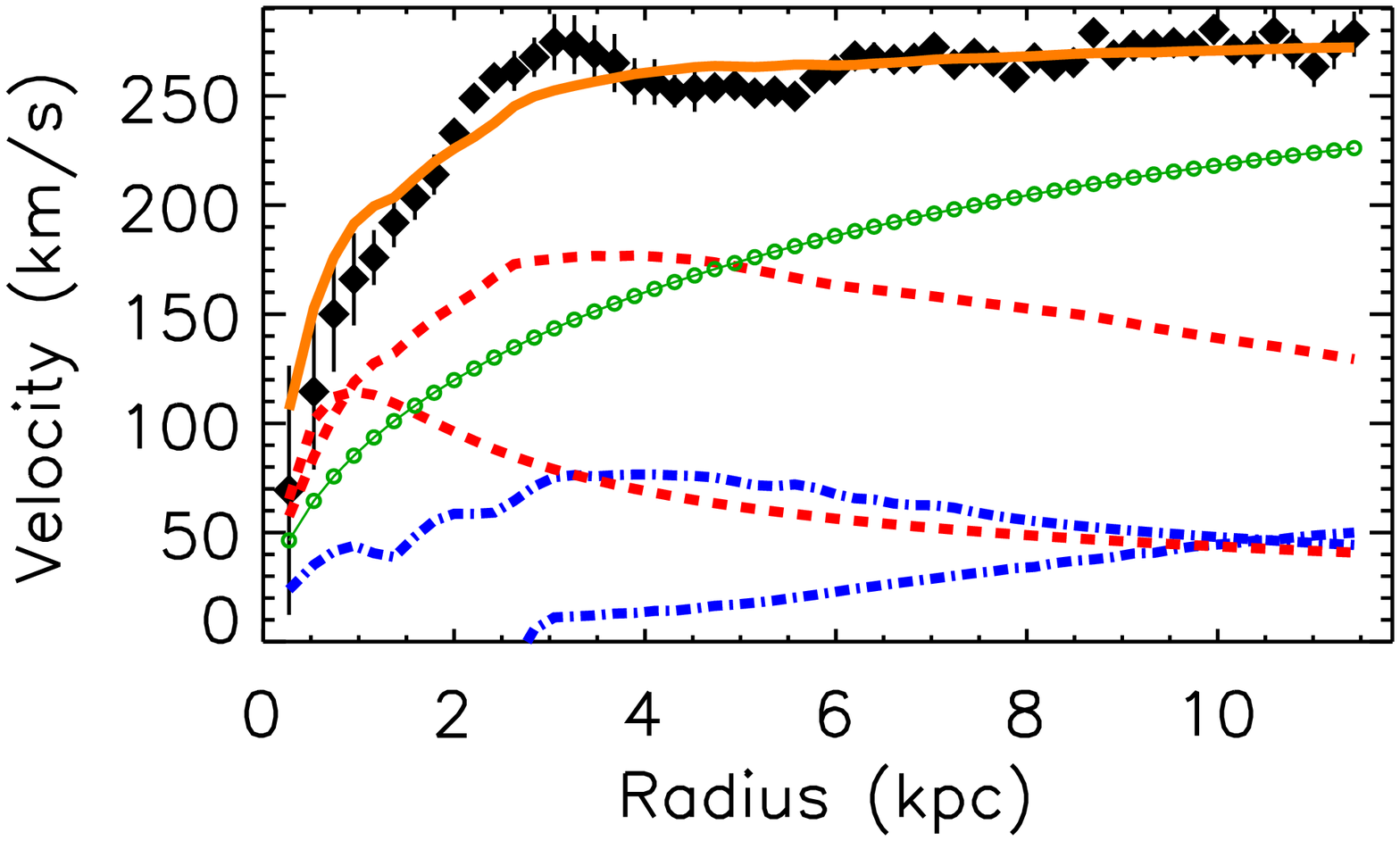}
\caption{\textit{Left}: Mass surface density profile of stars in M99. A  bulge-disk decomposition model (green solid line) to the observed profile (symbols) is seen, as well as  
bulge and disk components (red dashed lines). \textit{Right}: Axisymmetric mass distribution model of M99. 
The rotation curve is represented with filled symbols, and its model with an orange solid line. 
Contributions from the stellar bulge and disk are shown with red dashed lines, from the atomic and molecular gas disks with blue dash-dotted lines, and from 
the dark matter component with green open symbols. The dark matter model is the best-fit NFW halo whose parameters are given in Tab.~\ref{tab:paramdm}.}
 \label{fig:bulgediskdecompostion}
\end{center}
 \end{figure*}

   The   images have been deprojected with  constant kinematical parameters as function of radius, 
  using $i=20\degr$ and a position angle of 67\degr. 
    We set the $x>0$ axis of our deprojected frame to be coincident with to the semimajor  axis of the receding half of the disk.  
  We have also  assumed that all luminous mass contributions share the same dynamical center,  that is, fixed at the 
  position of the photometric center. 
  Figure~\ref{fig:density} shows the resulting surface density maps. The prominent spiral structure of M99 is 
observed at all angular scales with more obvious spiral patterns for the molecular gas and stellar components.  The outer 
stellar spiral arm ($x<0$) coincides well with the outer spiral arm of  the atomic gas disk.

\subsection{Luminous and dark matter velocity contributions}
\label{sec:lumvelo}

 The following    model is fitted to a kinematical observable
 \begin{equation}  
  v^2_{\rm \theta, mod}  =   v^2_{\rm DM}  + v^2_{\rm lum} 
  \label{eq:vmod1d}
,\end{equation}  
 where   $v_{\rm DM}$ the circular velocity contribution from the missing mass (dark matter, or DM), and   
 $v_{\rm lum}$ that from the total luminous mass, given by
 \begin{equation}
  v^2_{\rm lum}   = v^2_{\rm atom}  +v^2_{\rm mol}  +v^2_{\rm \star,D}  +v^2_{\rm \star,B}  .
  \label{eq:modelaxi}
 \end{equation}

Equations~\ref{eq:vmod1d} and ~\ref{eq:modelaxi} assume  tracers at centrifugal balance, which is probably not totally verified because of the disk asymmetries. 
In our 1D axisymmetric approach,  azimuthally averaged surface density profiles have been derived from the resolved density maps of Sect.~\ref{sec:surfdens}. 
The stellar mass surface density profile is shown in Fig.~\ref{fig:bulgediskdecompostion}. The   bulge component has a mass of $\sim 4.7\ 10^9$ \msol\ and
  follows an exponential law of central density and scalelength $(\Sigma_0,h) \sim$
 (7450 \msolkpcs, 0.3 kpc). The main stellar contribution comes from a disk with a scalelength of $h_\star=1.7$ kpc 
 and a central density of $\sim$ 1800 \msolkpcs. The  density profile  
 is better modeled with the addition of an outer truncated component having 
 a scalelength of $\sim 20$ kpc for a central density of $\sim$ 40 \msolkpcs. 

 The circular velocity  derives from the radial acceleration $g_R$, as $v_c^2 = -Rg_R$, assuming a pressureless mass component.  
 The circular velocity contribution $v_{\rm \star,B}$ of the bulge  has  been 
 derived from the bulge density assuming a  spherical bulge. 
The circular velocity contribution of the stellar disk $v_{\rm \star,D}$ has been deduced from a residual density profile  
 obtained by subtracting the bulge density to the total stellar density profile. The main disk and    outer  components are thus contained in this residual profile, and 
 we do not differentiate between both disk parts hereafter.  
 The  vertical density law of the  stellar disk cannot be measured directly for an almost face-on galaxy like M99.
We have  assumed  it follows a sech-squared  law \citep{van81} with a constant scaleheight of
  $\sim 0.35$ kpc.  This value corresponds to 20\% of the M99 disk scalelength,  following results found for edge-on disks  \citep[e.g.,][]{yoa06}. 
The vertical structure of gaseous disks is  less observationally constrained than for stellar disks. 
We considered that the gaseous disks are thin structures of negligible scaleheight.
 
Figure~\ref{fig:bulgediskdecompostion} shows the individual velocity profiles. One sees that the stellar disk   dominates 
the other luminous contributions, except in the inner kpc where it is comparable with the bulge contribution. 
We emphasize  that using an inclination of, for example, $\sim$ 30\degr\ as in \citet{che06} would   systematically make the stellar contribution
overestimate the rotation curve in the inner disk regions. 
Unless the stellar mass of M99 has been significantly overestimated 
or the asymmetric drift has been significantly underestimated, 
using a lower kinematical inclination than those given by \citet{pho93}, \citet{kra01}, or \citet{che06} 
is the best solution to reconcile the \ha\ kinematics with the   photometric inclination and the results from models of  stellar populations synthesis. 

\begin{figure*}[t!]
\hspace*{2cm}\atom\ \hspace*{4.3cm}\stars\ \hspace*{3.7cm}\mole\
\begin{center}
\includegraphics[height=12.5cm]{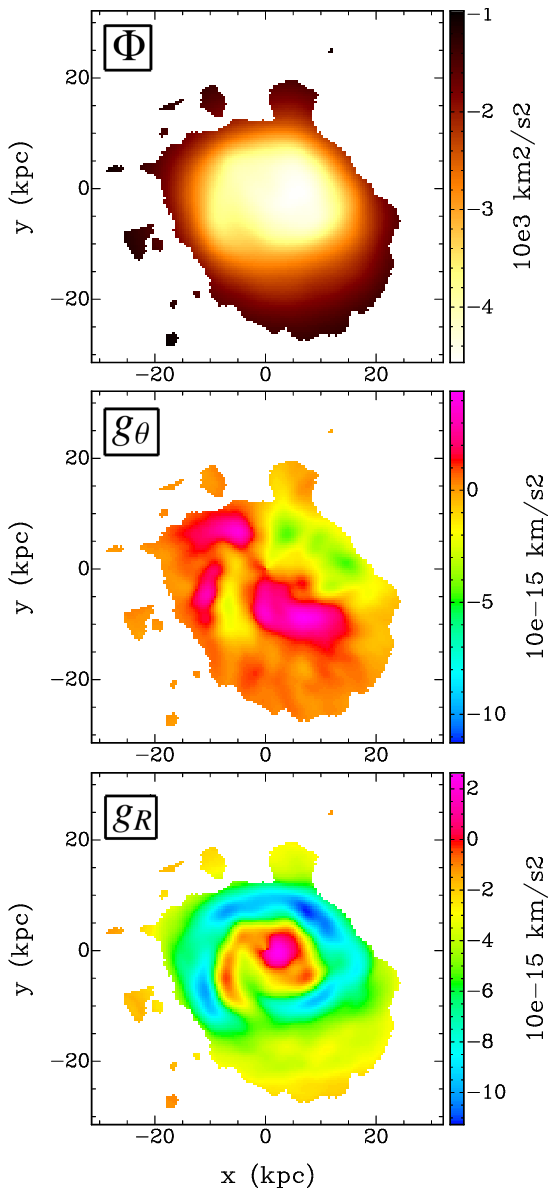}\includegraphics[height=12.5cm]{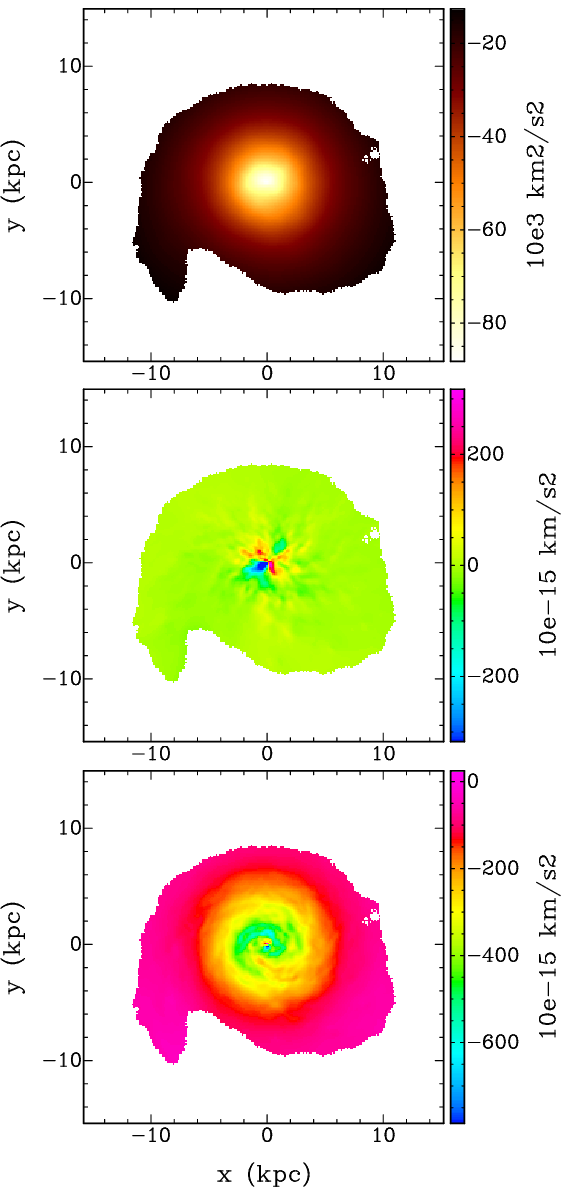}\includegraphics[height=12.5cm]{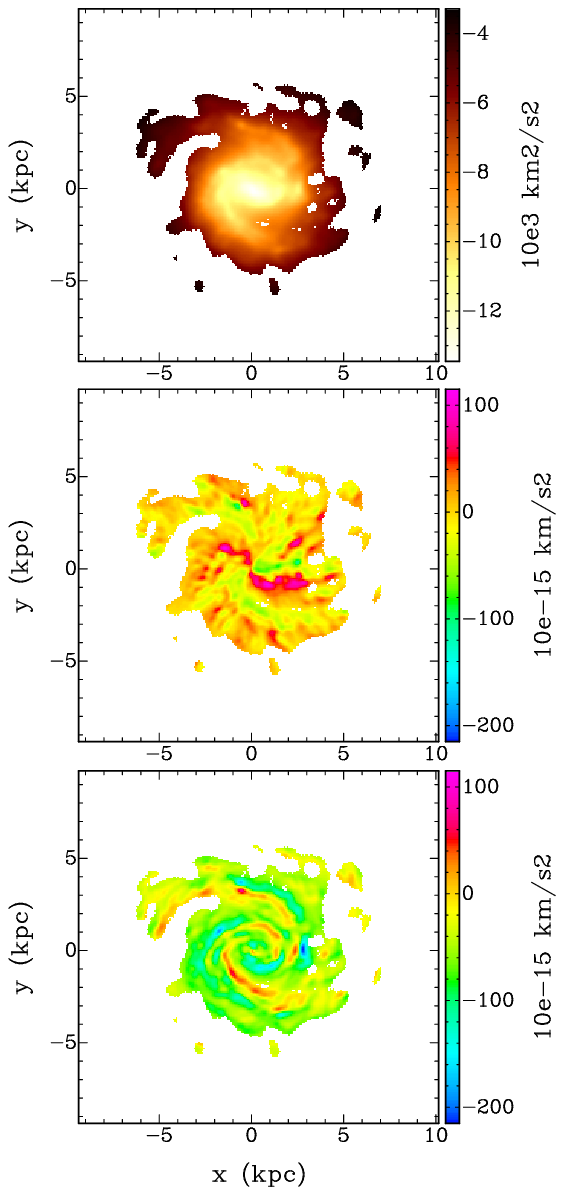}
\caption{Gravitational potential ($\Phi$),  tangential, and radial acceleration ($g_\theta$ and $g_R$, respectively) maps of   
 the disks of the atomic, molecular gas, and stellar components  of \gala. These maps are for the z=0 kpc midplane. 
 The axisymmetric contribution from the bulge potential is not included in the stellar disk component.}
 \label{fig:maps}
 \end{center}
 \end{figure*} 
 
The larger concentration of molecular gas in the inner disk causes 
$v_{\rm mol}$ to dominate over $v_{\rm atom}$; the latter velocity steadily increases toward the outer regions and peaks 
at a radius larger than the last point of the $\ha$ rotation curve. 
The radius of the peak of the stellar disk contribution $R= 2.2  h_\star = 3.7$ kpc coincides well with that of the inner peak of the rotation curve. 
This confirms our finding of a small scalelength for the stellar disk of M99. At this radius,  if the velocity contribution of the stellar disk   
  dominates the rotation curve  at a level of more than 75\%\ ($v_{\rm \star,D}/v_\theta > 75 \%$) for galaxies similar in morphological type to M99, 
  then  the stellar disk is said to be maximum \citep{sac97}.
We measure a  stellar disk velocity fraction of 67\%\ at $R=3.7$ kpc. Also, taking  the bulge contribution into account leads to a stellar fraction of 72\% and all luminous matter gives 
a velocity fraction of 78\%. The stellar disk of M99 is thus not  maximum, and the luminous baryons are barely maximum   with this definition. 
 This confirms that dark matter is a significant contribution to the total mass budget of M99, as often observed in  nearby spirals  \citep[e.g.,][]{ber11,wes11,mar13b}. 

 The velocity contribution of dark matter is assumed to be that of a spherical halo ($r=R$) whose center coincides with that of the disk of luminous matter 
 (hereafter centered-halo case). The DM halo models we fitted are the Einasto model (EIN hereafter), 
  the cuspy model inferred from cosmological simulations (the Navarro-Frenk-White model; NFW hereafter), and the core-dominated  model (pseudoisothermal sphere; PIS hereafter).

 The mass density profile of the Einasto model \citep{nav04} is defined as  
         \begin{equation}
         \rho_{\rm EIN}(r) =\rho_{-2} \exp\left(-2n\ \left(\left({r \over r_{-2}}\right)^{1/n}-1\right) \right) \ .
         \label{eq:rhoein}
         \end{equation}
 Here $r_{-2}$ is the characteristic scale radius at which the density profile has a 
 logarithmic  slope of $-2$, $\rho_{-2}$ is the scale density at that radius, and $n$ is a dimensionless index 
         that shapes the profile.   The circular velocity implied by the Einasto model is 
         \begin{equation}
         v^2_{\rm EIN}(r) =  4\pi G n r_{-2}^{3} \rho_{-2}\ {\rm e}^{2n}\ (2n)^{-3n} \gamma \left(3n,\xi\right) \ r^{-1}
        \label{eq:vein}
        ,\end{equation}
where   $\rm \gamma(3n, \xi)=\int_0^\xi{e^{-t} t^{3n-1}~dt}$  is the incomplete gamma function and $\xi=2n (r/r_{-2})^{1/n}$ \citep{car05,mam05,ret12}.  
This three-parameter model has  the flexibility to choose between a steep, intermediate, or   shallow density profile, depending on the value of the index.
At fixed characteristic density and radius,  models with small (large) indices correspond to shallow (steep, respectively) inner density profiles \citep{che11}.

 The density   of the NFW profile \citep{nav97} is
   \begin{equation}
   \rho_{\rm NFW}(r) = 4\,\rho_{-2} \,{r_{-2}\over r}\,\left({r_{-2}\over r+r_{-2}}\right)^2 \ ,
   \label{eq:rhonfw}
   \end{equation}
 
corresponding to a circular velocity profile 
         \begin{equation}
         v^2_{\rm NFW}(r) =  v_{200}^2  \left(\log(1\!+\!\eta)\!-\!\eta/(1\!+\!\eta)\right)\!/\!\left(x(\log(1\!+\!c)-c/(1\!+\!c))\right)
        \label{eq:vnfw}
        ,\end{equation}
        
where $x=r/r_{200}$. The parameter $r_{200}$ is the virial radius derived where the density equals 200 times the critical density  $3H_0^2/(8\pi G)$ for closure of the Universe; and $\eta=cx$, where $c$ is the 
concentration of the DM halo.  We fixed the Hubble constant at the value found by the Planck Collaboration $H_0=68$ \kms\ Mpc$^{-1}$ \citep{pla15}. 
The two   parameters are   the scale velocity $v_{200}$ and the halo concentration $c$. The NFW halo is said to be cuspy  because the density scales as $r^{-1}$ in the center, which is steeper compared with 
the density profile of the pseudoisothermal sphere
     \begin{equation}
     \rho_{\rm PIS}(r) = \rho_0 \,{r_c^2 \over r^2+r_c^2} \ ,
     \label{eq:rhoiso}
     \end{equation}
     which thus tends to the constant value $\rho_0$ in the core region of the halo. This model is therefore   referred as a core.
  The parameter  $r_c$ is  the characteristic core radius of the halo. 
 The density profile implies a circular velocity 
         \begin{equation}
         v^2_{\rm PIS}(r) =  4\pi G \rho_0 r_c^2 (1-r_c/r \atan(r/r_c))
        \label{eq:vpis}
        .\end{equation}
 
Each of $v_{\rm EIN}$, $v_{\rm NFW}$, or $v_{\rm PIS}$  replaces $v_{\rm DM}$ in Eq.~\ref{eq:vmod1d}.

 \subsection{Mass distribution modeling}
 \label{sec:massmod1}
   The rotation curve is the observable for the 1D modeling, while  the \ha\ velocity field of M99  has been fitted in the
 2D case.
 For that, a modeled \los\ velocity map is built following Eqs~\ref{eq:vtanvradvf} and~\ref{eq:vmod1d} with fixed kinematical parameters, 
 using axisymmetric velocity contributions from dark matter, molecules,   atoms, and  stars.  
   In practice, it is the bidimensional tangential velocity only that is modified in Eq~\ref{eq:vtanvradvf}, 
   as the parameters of dark matter vary while fitting the observation. 
   However, we have built two \los\ projections: with  and without fixed radial   motions. 
   The modeling without \vr\   assumes $v_R=0$, while that with fixed \vr\  uses the axisymmetric profile derived in Section~\ref{sec:rcvr}. 
    Modeling with noncircular motions is referred as the 2D+\vr\  case hereafter.
   Though not perfect, as \vr\ is not  dynamically motivated unlike \vt, the 2D+\vr\ case
    remains a simple method to estimate the global impact of   noncircular motions in the dynamical modeling 
    and the inner shape of the dark matter density.
 
   The SPS modeling of \cite{zib09} already provide the scaling of the  stellar mass. 
  Our models  only have  those of the DM component as free parameters.  
 Fittings were thus carried out with 52 d.o.fs (9000, respectively)  for the NFW and PIS models, and 51 d.o.fs (8999) for the EIN model  in the 1D  axisymmetric case (2D). 
 As in Sect.~\ref{sec:rcvr}., we used uniform weightings  for the 2D models. 
   Rotation curve decomposition often uses normal weightings that are the inverse of the squared  uncertainties on rotation
    velocities, $\Delta_{v_\theta}$.  We defined $\Delta^2_{v_\theta}$ as the quadratic sum of  the formal error from the rotation curve fitting 
    with a systematic error (half the velocity difference between the approaching and receding disk halves). 
    The error distribution is not Gaussian, which prevents us from using normal weightings. 
     We thus used the number of points per radial bin as the weighting function of  velocities.  
   Though this is not  homogeneous  to the 2D weighting function, it turned out
    to be the most appropriate way to account for a distribution of velocities in the rotation curve 
    comparable to the pixel distribution in the velocity field. 

   Appendix~\ref{sec:tablesresults} (Tab.~\ref{tab:paramdm}) reports the fitted  parameters of the different halo models. 
  The quoted parameter errors correspond to the formal $1\sigma$ error from the fittings.  
  Both 1D and 2D fittings are correct and yield parameters in good agreement within the errors. The 2D modeling 
  yields    more constrained parameters than for the 1D case. We also
  note the degeneracy for the halo parameters of the Einasto model. 
  This degeneracy is partially raised in the 2D case, as  the Einasto index   is about 2.5 times more constrained.
   We base the analysis upon the   Akaike Information Criterion \citep[AIC;][]{aka74} to compare the different halo models, following  \citet{che11}. 
    The  criterion is defined by ${\rm AIC}=2N+\chi^2$, where $N$ is the number of parameters to be fitted 
    by the modeling ($N=2$ for NFW and PIS, $N=3$ for EIN).  
   Since the AIC involves both the $\chi^2$   and   the number of parameters,
     an AIC test is then appropriate to compare models that are not nested and  do not have a similar numbers of parameters, such as the NFW, PIS, and EIN forms. 
        An AIC test cannot be invoked to rule out a specific model, but instead helps us 
     to decide which model is more likely than others.  \citet{che11} applied this criterion to analyze rotation curves of disks from
     the \hi\ Nearby Galaxy Survey \citep{wal08,deb08} and found that in most of configurations the Einasto model is an improvement with respect to  the NFW and PIS models.     
  As the smaller the AIC, the more likely one model with respect to another, we then compared two models by deriving the difference between their respective AIC 
  (Tab.~\ref{tab:diffaic1}). Reporting AIC differences explains why Tab.~\ref{tab:paramdm} does not list the $\chi^2$   of each fitting.
  
 The Einasto  and NFW cusps are found to be   the most likely models, compared with the pseudoisothermal sphere. 
 We estimate a DM density slope at the first point of the rotation curve
 $R=0.27$ kpc of $-0.88 \pm 0.37$ for the 1D case ($-0.87\pm 0.33$ for the 2D case)
 for the Einasto model, thus slightly shallower than for the NFW cusp ($\sim -1$). 
  The AIC difference between the NFW and Einasto cusps implies that the three-parameter model is more likely. However, it is difficult to judge the real pertinence of this result
   because of the degeneracy of the Einasto model parameters.  The best-fit 1D mass model with the NFW halo is shown in Fig~\ref{fig:bulgediskdecompostion}.
    Results for the 2D model that take  the radial  motions into account   follow the same trends, and an inner DM 
    slope of $-0.84 \pm 0.31$ is deduced for the EIN model. This slope is only marginally  shallower than for the model without \vr. 
    The noncircular radial motions have a  negligible impact on shaping the DM density profile of  M99, at least within this axisymmetry approach.  

    \begin{figure}[t]
\begin{center}
\includegraphics[width=0.9\columnwidth]{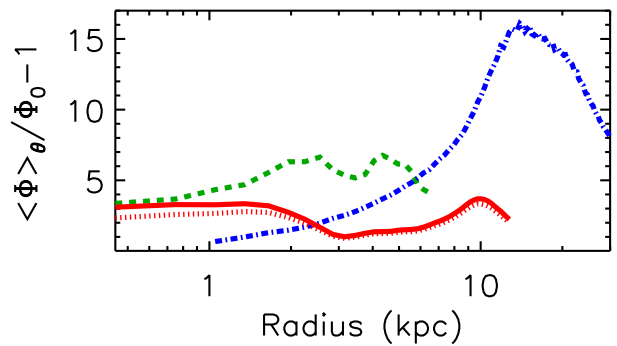}
\caption{Fraction (in \%) of the perturbed potential  to the unperturbed potential   for luminous matter in M99. The red solid (dotted) line is for the stellar disk (stellar 
bulge+disk) potential, green dashed line for the molecular gas disk, and  blue dash-dotted line for the atomic gas disk.}
 \label{fig:fracpotpert}
\end{center}
\end{figure}
\begin{figure*}[t]
\begin{center}
\includegraphics[height=4.45cm]{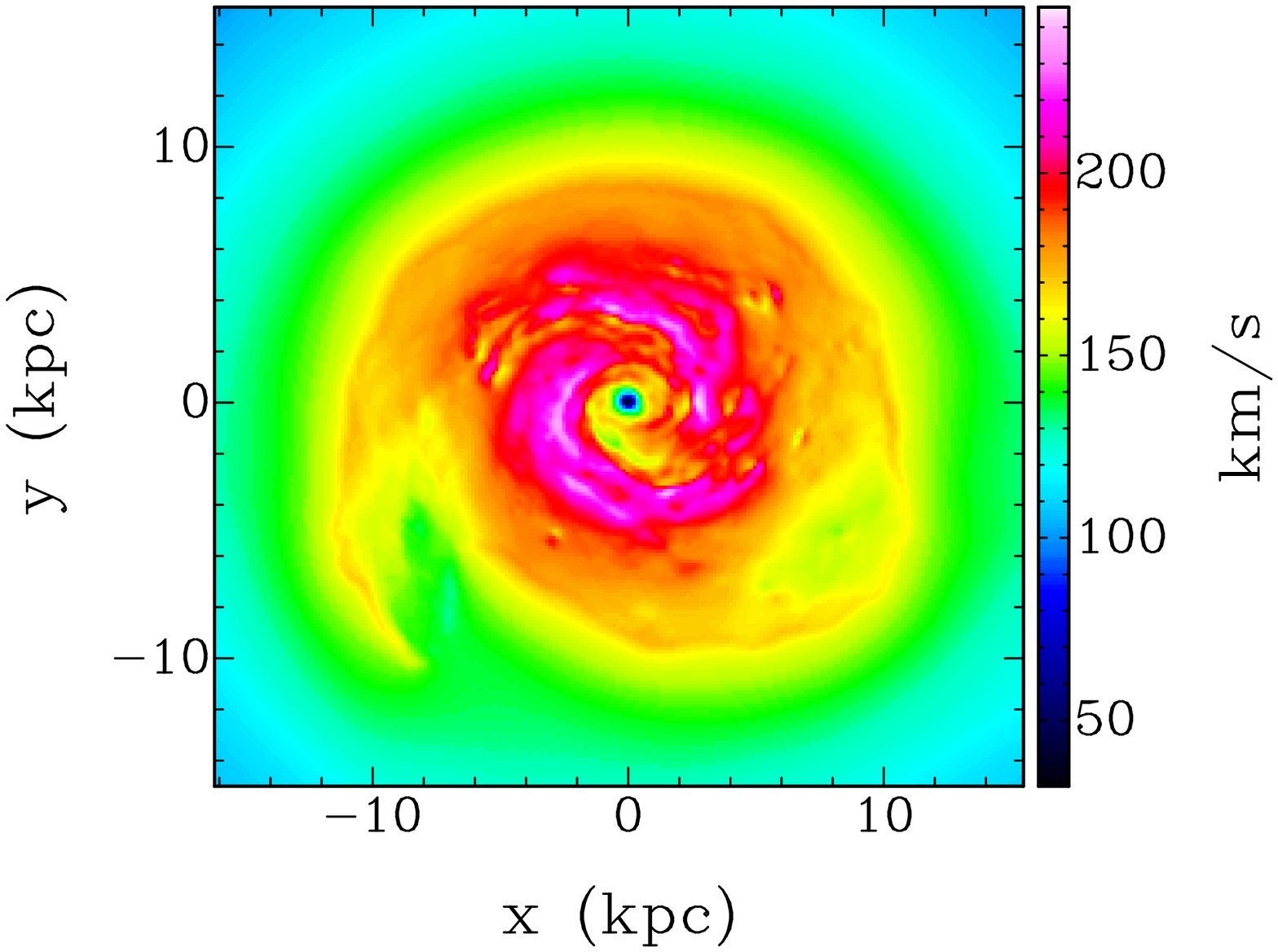}\includegraphics[height=4.45cm]{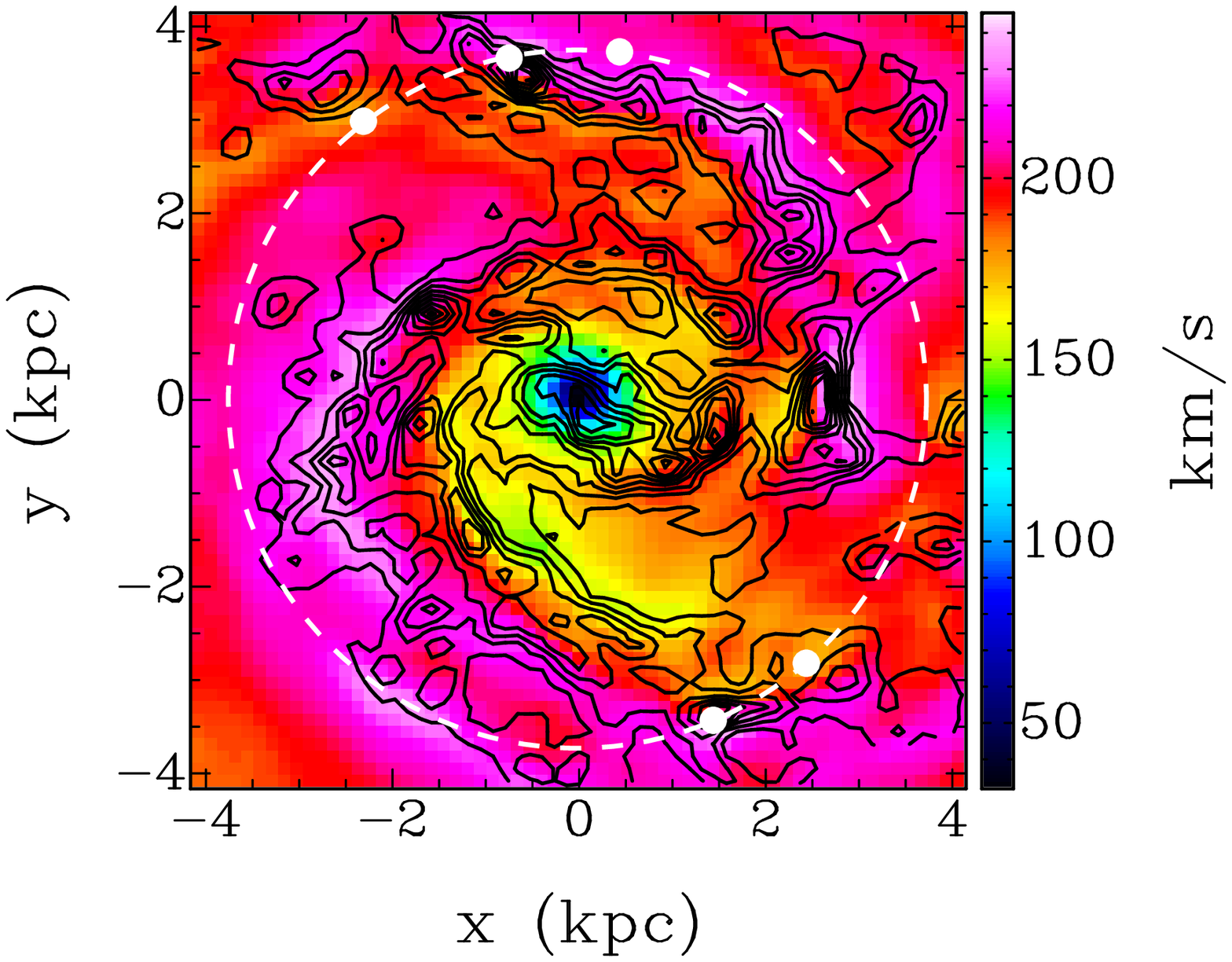}\includegraphics[height=4.45cm]{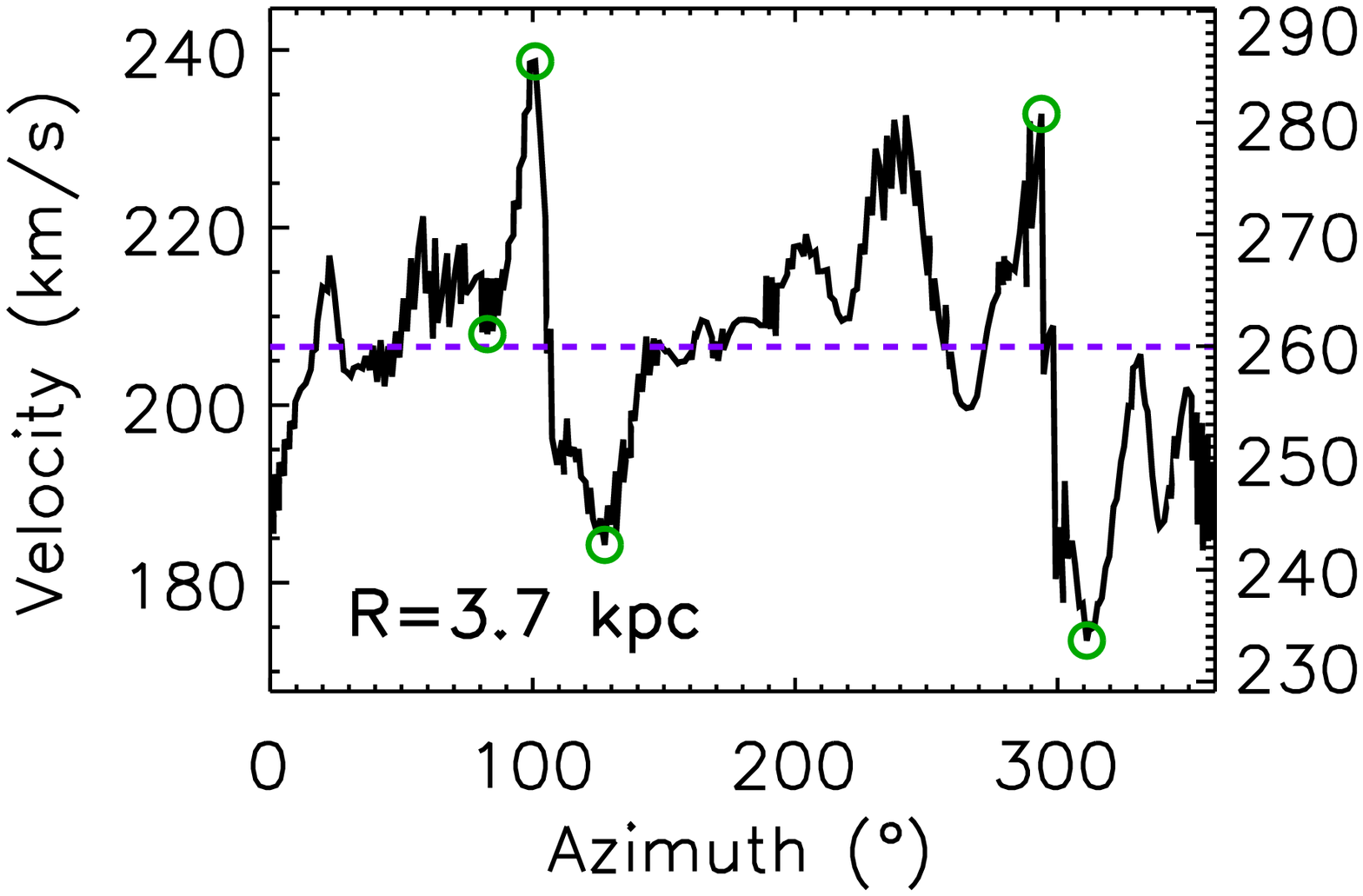}
\caption{\textit{Left}: Circular velocity field for the contribution of luminous matter only. 
\textit{Middle}: Zoom in the inner 4 kpc with contours representing the emission from   molecular gas. 
A dashed circle shows the location of $R= 2.2  h_\star = 3.7$ kpc, where $h_\star$ is the stellar disk scalelength. 
\textit{Right}: Variation of velocity with azimuth at $R= 3.7$ kpc.  Open circles represent the velocities of 
  four points inside,     down- and upstream  of the spiral arms, indicated with white filled circles in the middle panel. The 
  dashed line is the axisymmetric value. The velocity scale of the left axis is for  luminous matter only, while that of the right axis
   takes  an additional contribution from dark matter into account.}
 \label{fig:vtotlum}
 \end{center}
 \end{figure*} 
 
 \section{Dynamical asymmetries of M99 luminous matter}
\label{sec:asymapp}

The benefit from a 2D analysis should become  more interesting  if 
the velocity contribution  from luminous matter could stick to the asymmetric reality of the luminous matter. 
The objective of this section is thus to describe the methodology and products of the asymmetric approach (gravitational potentials, radial, and tangential forces), 
and in particular the  resolved circular velocity contribution  of luminous matter to be used by
 the 2D asymmetric modeling presented in Section~\ref{sec:fitasym}.
 
 \subsection{Methodology}

The major difference between the axisymmetric case and the asymmetric approach is that we need to derive the 3D, asymmetric gravitational potential of luminous baryons beforehand.
We thus computed the potential in Cartesian coordinates $(x,y,z)$, which enables us to derive both radial
and azimuthal forces at any desired $z$. This can be derived
independently for each stellar or gaseous contribution. 
The gravitationtal potential $\Phi$ of the mass distribution is in
principle deduced from the convolution of the volume mass density by the
Green function, namely 
\begin{equation}
 \Phi = -G \iiint\frac{d\rho'}{|\vec{r}-\vec{r}'|},
\label{eq:pot}
\end{equation}
where $G$ is the gravitational constant. However, the Green function
written by  $1/|\vec{r}-\vec{r}'|= [(x-x')^2+(y-y')^2+(z-z')^2]^{-1/2}$,
is well known to diverge at each point where $x=x'$, $y=y'$ and $z=z'$.
This function renders any direct estimate of $\Phi$ inaccurate and generally
encourages modelers to incorporate a softening length to bypass the
divergence. Here, we use the new formalism presented in \cite{hur13} who
showed that the Newtonian potential is exactly reproduced by using an
intermediate scalar function ${\cal H}$, namely
$\Phi = \partial_{xy}^2 {\cal H}$. In 3D, this hyperpotential is written as
\begin{equation}
{\cal H}(x,y,z)  = -G \iiint_{\Omega'}{\rho(x',y',z')}
\kappa^{xy}(X,Y,Z)dx'dy'dz'
,\end{equation}
with $X=x-x'$, $Y=y-y'$ and $Z=z-z'$. The $\kappa$ function is a
hyperkernel defined by
\begin{equation}
\kappa^{xy}(X,Y,Z) = -Z \atan \frac{XY}{Z |\vec{r}-\vec{r}'|}+Y \ln
\frac{X+|\vec{r}-\vec{r}'|}{\sqrt{Y^2+Z^2}}\ .
\label{eq:k}
\end{equation}
This approach is particularly simple and efficient for 2D or 3D
distributions since ${\cal H}$ is, in
contrast to $\Phi$, the convolution of the surface or volume density
with \textit{a regular, finite amplitude kernel}. 
The methodology thus does not make use of a softening length in the derivation of the potential.
In practice, this convolution is performed using the second-order
trapezoidal rule and
the mixed derivatives are estimated at the same order from centered
finite differences. Furthermore, the  volume density of the tracer are deduced from a surface density map, considering 
 that the  vertical density follows a sech-squared or exponential law of constant 
 scaleheight with radius. The precision of these schemes is sufficient for the present purpose. 

The volume density of gas and stars were derived using their respective surface densities (Section~\ref{sec:surfdens} and Fig.~\ref{fig:density}).
Similar to the axisymmetric case, we  considered a sech-squared law, using a scaleheight of 0.35 kpc for the vertical variation of the 
density of the stellar disk, and that the molecular and atomic gas disks have negligible scaleheights. 
Once the 3D gravitational potential of a tracer is derived, the azimuthal and radial components of the gravitational acceleration are obtained  
from $g_\theta = - \nabla_\theta \Phi$ and $g_R = - \nabla_R \Phi$, respectively. 
From the 3D products, we can extract the gravitational potential, radial, and tangential accelerations  
in the galaxy midplane ($z=0$ kpc). This is necessary to fit the observed kinematics of ionized gas that is assumed to lie in that  plane.

\subsection{Asymmetric gravitational potentials and accelerations}
\label{subsec:potaccmap}

Figure~\ref{fig:maps} shows the  midplane  potential   maps for the  stellar and gaseous disk components of M99. 
As we are primarily interested  in the asymmetric components, the axisymmetric potential of the bulge is not represented here 
(see Fig.~\ref{fig:bulgepot} for its radial profile). 
We only show pixels for which the density is strictly positive  for each component, as  these are the most important regions for the analysis.
Beyond the observable extent of each disk, the potential well decreases smoothly with radius and has no interesting features that deserve to be shown.
 
 The absolute strength of the potential is  observed to increase from the atomic gas, to the molecular gas, and finally 
  to the stellar component. As expected, none of the 
  potentials show pure axial symmetry, as they exhibit lopsided 
  and spiral-like features. 
To quantify the nonaxisymmetric perturbations, we have modeled the gravitational potentials via a series of  harmonics, as detailed in Appendix~\ref{sec:fourierpot}.
We find that the stellar and gaseous potentials are indeed lopsided ($m=1$ mode), and exhibit spiral structures ($m=2$ modes)  and other less prominent $m=3$ perturbations. 
We also find that the stellar  potential is not barred and that the lopsidedness dominates the amplitude of the stellar 
perturbations in the inner disk region, while it totally dominates those in the disk of atomic gas. 
  We derived the azimuthally  averaged ratios 
   of total perturbed against unperturbed potentials to summarize the importance of the perturbations. These ratios are written as      
  $\langle \sum_{m \ne 0}  \Phi_{m} (R)\cos(m(\theta-\theta_m(R)) / \Phi_0 (R) \rangle_\theta =  \langle 
  \Phi(R,\theta)\rangle_\theta/\Phi_0(R)-1$, following Eq.~\ref{eq:fourierpot} and are shown in Fig.~\ref{fig:fracpotpert}. They show  
  that the degree of perturbation significantly increases with decreasing mass density.  
  The stellar potential is preferentially less disturbed in regions where perturbations of the molecular gas potential 
  are stronger. On overall average,  
  we find that the gravitational  potential is  disturbed at the level of $\sim 10\%$ for the atomic gas component, 
   $\sim 5\%$ for   molecules, and $\sim 2\%$ for stars. The addition of the  bulge potential   to the axisymmetric part of the potential of the stellar disk only marginally 
   modifies the ratio of perturbations for  the stellar component. 
  
 Figure~\ref{fig:maps} also  shows the tangential and radial  accelerations in the midplane. 
  The most important results   are  that   $g_\theta$ is far from  negligible and then that  
 $g_R$  and $g_\theta$ are strongly  asymmetric. 
  The  morphology of the $g_R$ maps is intrinsically linked to  spiral structures in the density and potential maps, 
 while structures in the $g_\theta$ maps differ markedly from the other 
 density, potential, and radial acceleration fields.  
 The sign of $g_R$ is almost exclusively negative   except 
   on the trailing sides of the   gaseous arms, principally   at $R=1-2.5$ kpc and $R= 4.5-5$ kpc.   
   Interestingly, these radii  coincide or are very close to regions where  the ionized gas 
  has its radial motion outwardly directed (Section~\ref{sec:rcvr} and Fig.~\ref{fig:harc}).  
  The (absolute) radial force increases radially in the trailing sides of the (density) spirals and reaches local maxima. 
 Once the peak of density has been met, it decreases on the leading sides to reach  local minima. 
  As for the azimuthal acceleration, its structure is governed by striking alternations of positive and negative patterns. 
Basically, $g_\theta$ 
varies rapidly through the spiral arms, admitting a minimum where the density is maximum. 
 It also presents a large  gradient in the center of the stellar disk. 
We estimate that the mean tangential acceleration is about 70, 300, and 3950 times smaller than 
the mean radial acceleration for the atomic gas, molecular gas, and stellar disk (respectively). 
The tangential force is thus completely negligible over the entire disk on average, 
but this is only due to the alternating patterns. Indeed, this is not the case anymore on small scales 
since   $g_\theta$ and $g_R$ have comparable amplitudes.

  \begin{figure}[t]
\begin{center}
\includegraphics[width=0.9\columnwidth]{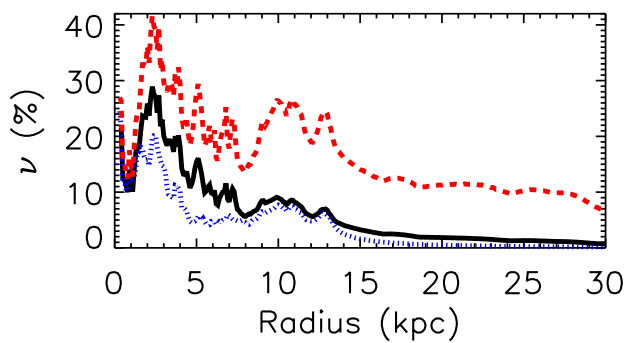}
\caption{Radial profile of the nonuniformity factor $\nu$, defined as the maximum velocity variation  
with azimuth relatively to the axisymmetric velocity. The solid line is for  total (luminous+dark) matter, blue dotted line for   stellar and dark matter (no gas), 
and  red dashed  line  for the total luminous matter (no dark matter).}
 \label{fig:nu}
\end{center} 
 \end{figure}
 
  \subsection{Nonuniform circular velocity}
  \label{sec:nonuniform}
  
On a global scale only, circular motions can almost be considered   as purely axisymmetric.
However, the implication  that both $g_\theta \neq 0$ and $g_R$ are asymmetric is  
 that the circular velocity  must be nonuniform.  To quantify the velocity nonuniformity, we  derived the
 circular velocity field $v_{\rm lum}$ for the total luminous matter following Eq.~\ref{eq:modelaxi}.
The major difference from the axisymmetric approach is that each circular velocity contribution   now depends  on $(x,y)$.  
That velocity field is built on a $(x,y)$-grid similar to  the stellar component. 
Velocity contributions of the gaseous disks have thus been interpolated at the nodes of the stellar grid. 
Figure~\ref{fig:vtotlum} shows the $v_{\rm lum}$ field for the whole disk as well as in a region focusing in the inner $R=4$ kpc.  

The velocity field   presents many spiral  patterns, which are mostly caused by  the perturbed stellar and molecular contributions. 
The map  admits extrema that depend on the location with respect to the spiral arm.
 Basically, the velocity sharply increases on the trailing sides of the spiral arms 
 where the density of stars and gas increases, then peaks at  higher densities, to sharply decrease   on the leading sides of the arms.
 Streaming motions  observed along spiral arms of galaxies, usually identified by wiggles in contours of \los\ velocities,  naturally find  
 their origin in the nonuniformity of circular motions.   Examples of modeled \los\ velocity field with apparent streaming motions 
 are presented in Section~\ref{sec:fitasym}.
    An example of highly  nonuniform circular velocities  is shown in the right panel of Fig.~\ref{fig:vtotlum} with 
  the azimuth-velocity diagram at $R= 2.2  h_\star = 3.7$ kpc.  
  At this radius, the stellar contribution   is maximum and the  axisymmetric circular velocity of total luminous matter 
  is $\langle v_{\rm lum} \rangle \sim 207$ \kms\ 
  (or 259 \kms\ when a contribution from, e.g., the best-fit NFW  halo of the 2D axisymmetric case of Section~\ref{sec:massmod1} is included). 
  The  overall variation of velocity, 65 \kms\ (52 \kms\
with DM), is  very significant;  the standard deviation, which is $\sim 13$ \kms\ (10 \kms), is significant as well.
    The  sharp gradients in the trailing sides of the spiral arms for azimuths 101\degr\ to 127\degr, and
   294\degr\ to 311\degr\ are of 54 and 59 \kms, respectively  (44 and 47 \kms\ with dark matter included). 
   Such nonuniformity is remarkable  considering the very  proximity of the  points 
   (e.g., $\theta=294\degr$ and 311\degr\ are separated by 1.2 kpc only).

  We  define a nonuniformity factor, $\nu$, as the maximum variation of circular velocity  relatively to the axisymmetric circular velocity (Fig.~\ref{fig:nu}). 
 The nonuniformity factor is important in the inner disk regions. As rule of thumb, it exceeds 10\% for $R \lesssim 2.2 h_\star$. 
 with a maximum of $\sim 30\%$ ($\sim 40\%$ without DM) at $R=2.5$ kpc. 
    This factor is less strong  in the outer spiral arm at $R \sim 10$ kpc, up to a level of $\sim 9\%$  (25\%\ without DM). 
   For $R>12$ kpc,  bumps can be identified as caused by the  $m=1,2,3$  perturbations in the gravitational potential of the atomic gas.  
   Here, however, $\nu$ smoothly decreases because of the dominant axisymmetric contribution of dark matter.
    A close inspection of the $v_{\rm atom}$, $v_{\rm mol}$, and $v_{\rm \star,D}$ maps  reveals the 
    important contribution of gas in the nonuniformity, thus in generating velocity wiggles in the M99 spiral arms. 
    This effect is shown with the dotted line that corresponds to the $\nu$ factor when the total gaseous component is omitted.  
   We estimate  that    gas  is responsible for about half of  $\nu$  and  the scatter of circular velocity at  $R=2.2  h_\star$,  
    although the stellar disk  velocity contribution is larger   by 100 \kms\ than that of total gas at this radius. 
    In the $R=9-10$ kpc spiral arms,  atomic gas contributes by up to $\sim$ 20\% to $\nu$. Self-gravity of gas 
    is therefore not negligible in M99,  in particular, in the densest disk regions.
    The impact of gas could have been even more important with a higher angular resolution for the atomic gas component. 
    Indeed,  the low angular resolution of the current \hi\ observations has smeared out  the surface density of the atomic gas, which  
    has then likely prevented us from deriving higher velocity contrasts through arm-interarm regions, 
    such as those evidenced within the stellar and molecular components.

    \begin{figure}[t]
\begin{center}
\includegraphics[width=0.9\columnwidth]{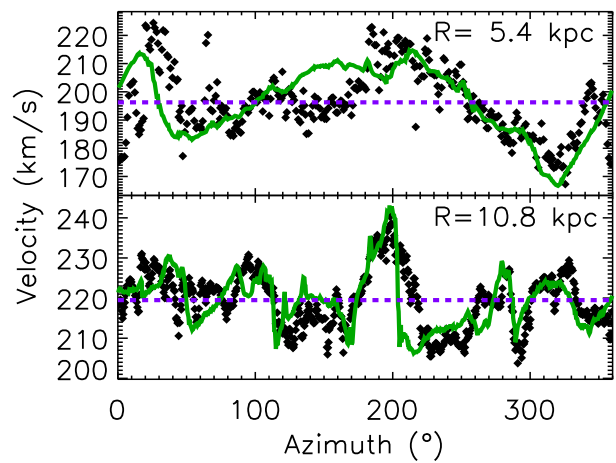}
\caption{Azimuthal velocity profiles at $R=5.4$ and 10.8 kpc in the simulation of the MW-like galaxy of \citet{kaw14}. Filled symbols are the tangential velocity of the gas component of the simulated galaxy. The green solid line
is the  nonuniform circular velocity predicted from the asymmetric methodology, the violet dashed line is the uniform circular motion.}
 \label{fig:compsimu}
\end{center} 
 \end{figure}

    \begin{figure*}[t]
\begin{center}
\includegraphics[width=0.33\textwidth]{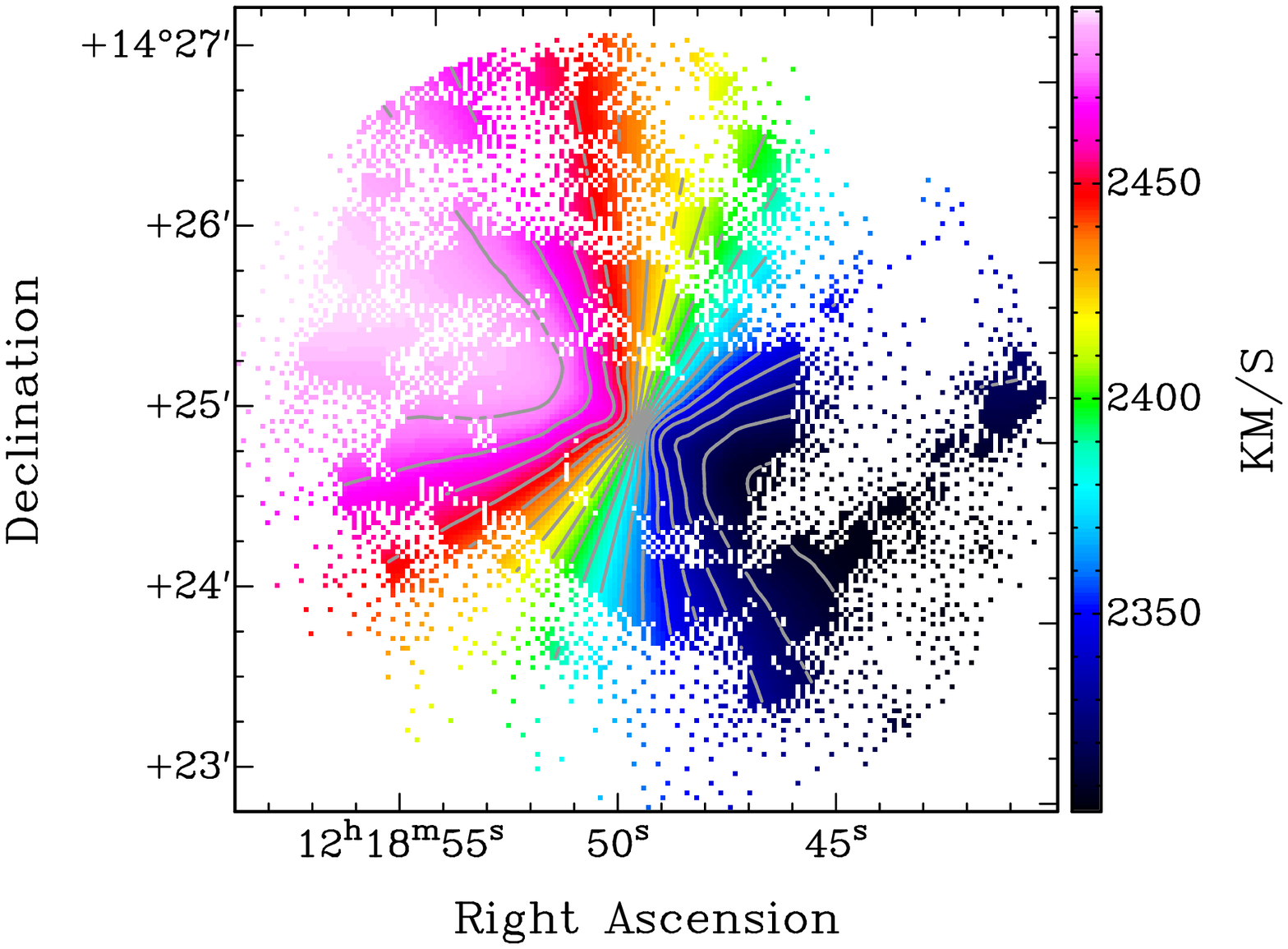}\includegraphics[width=0.33\textwidth]{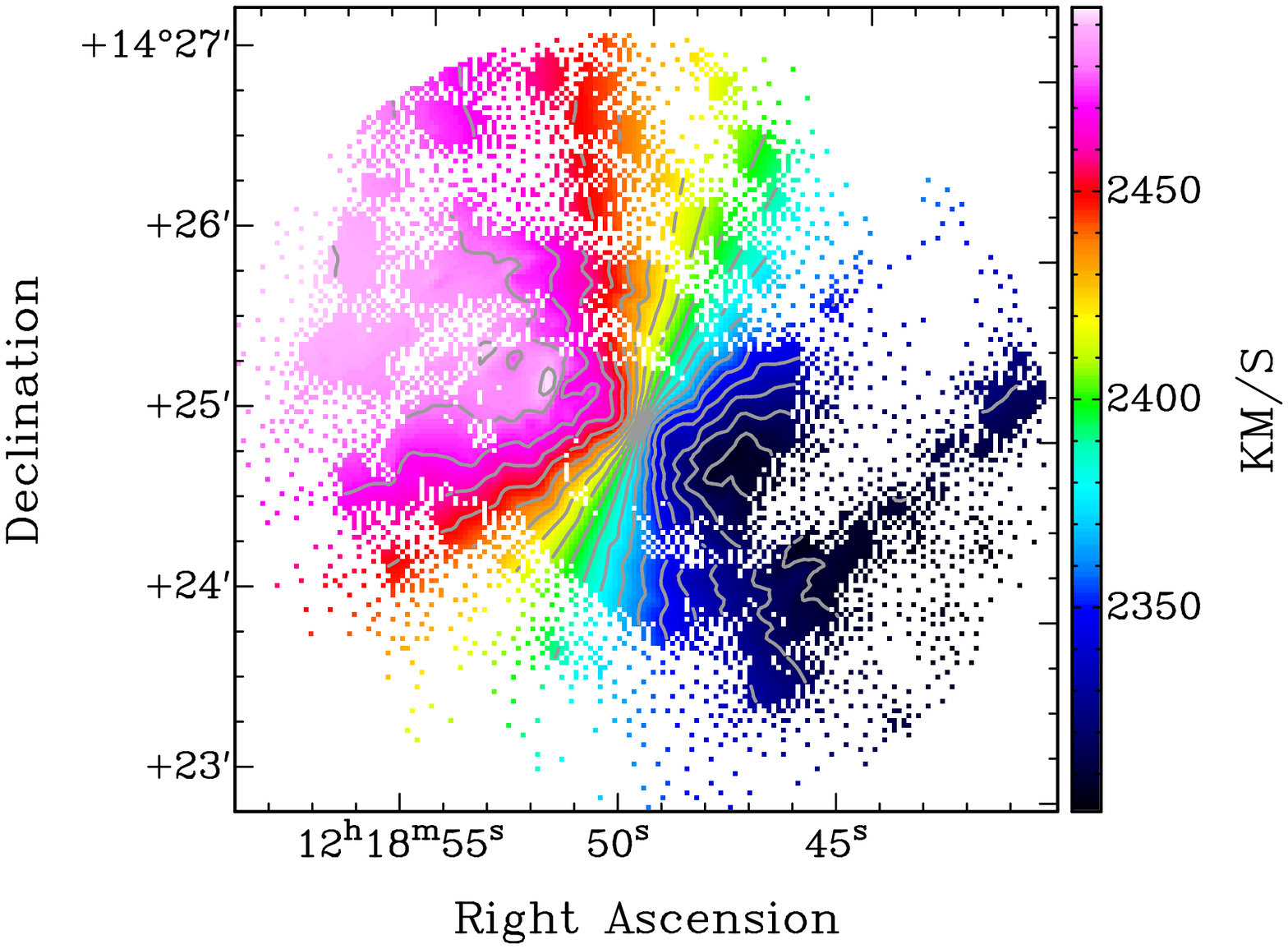}\includegraphics[width=0.33\textwidth]{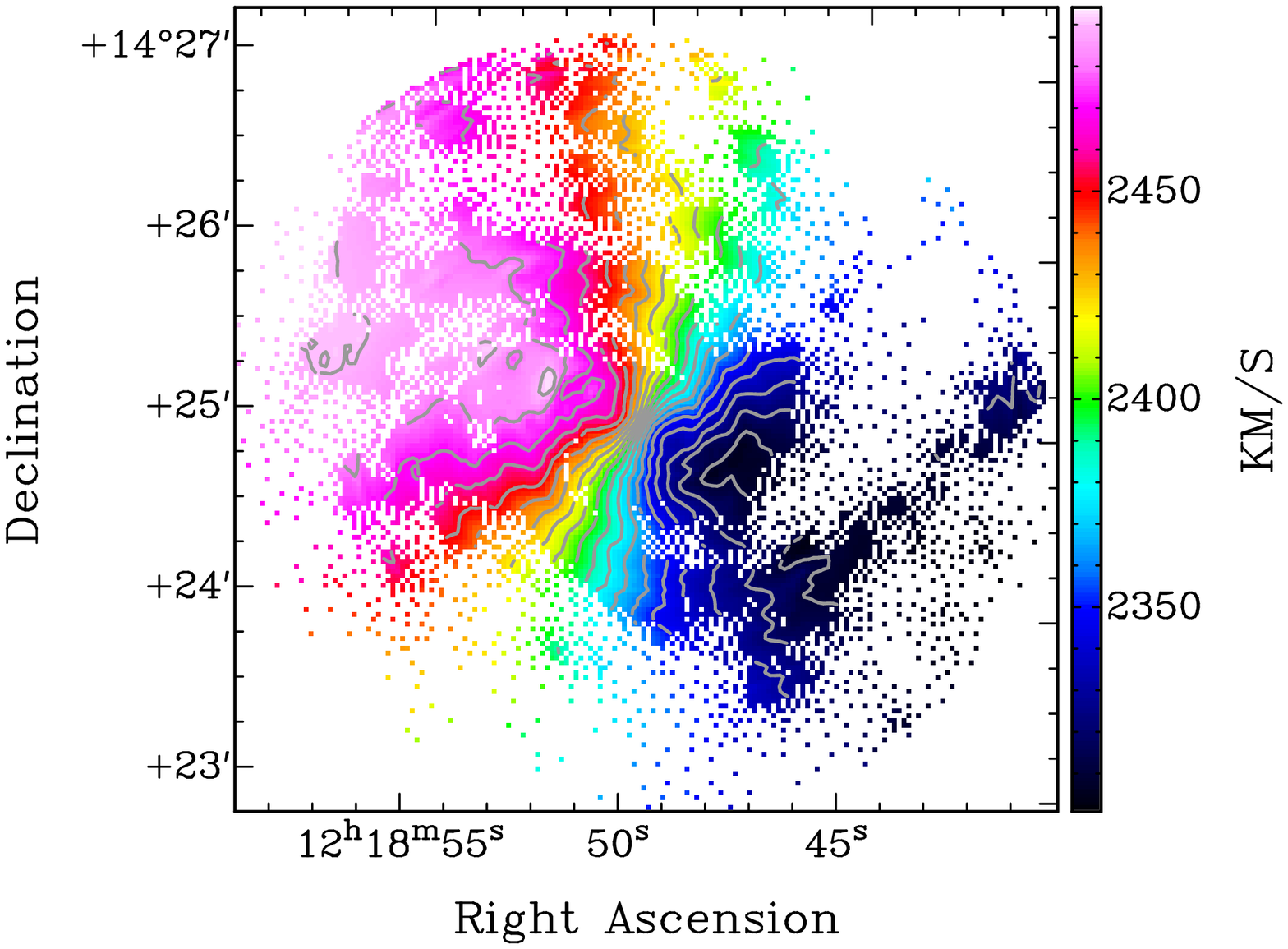}
\includegraphics[width=0.33\textwidth]{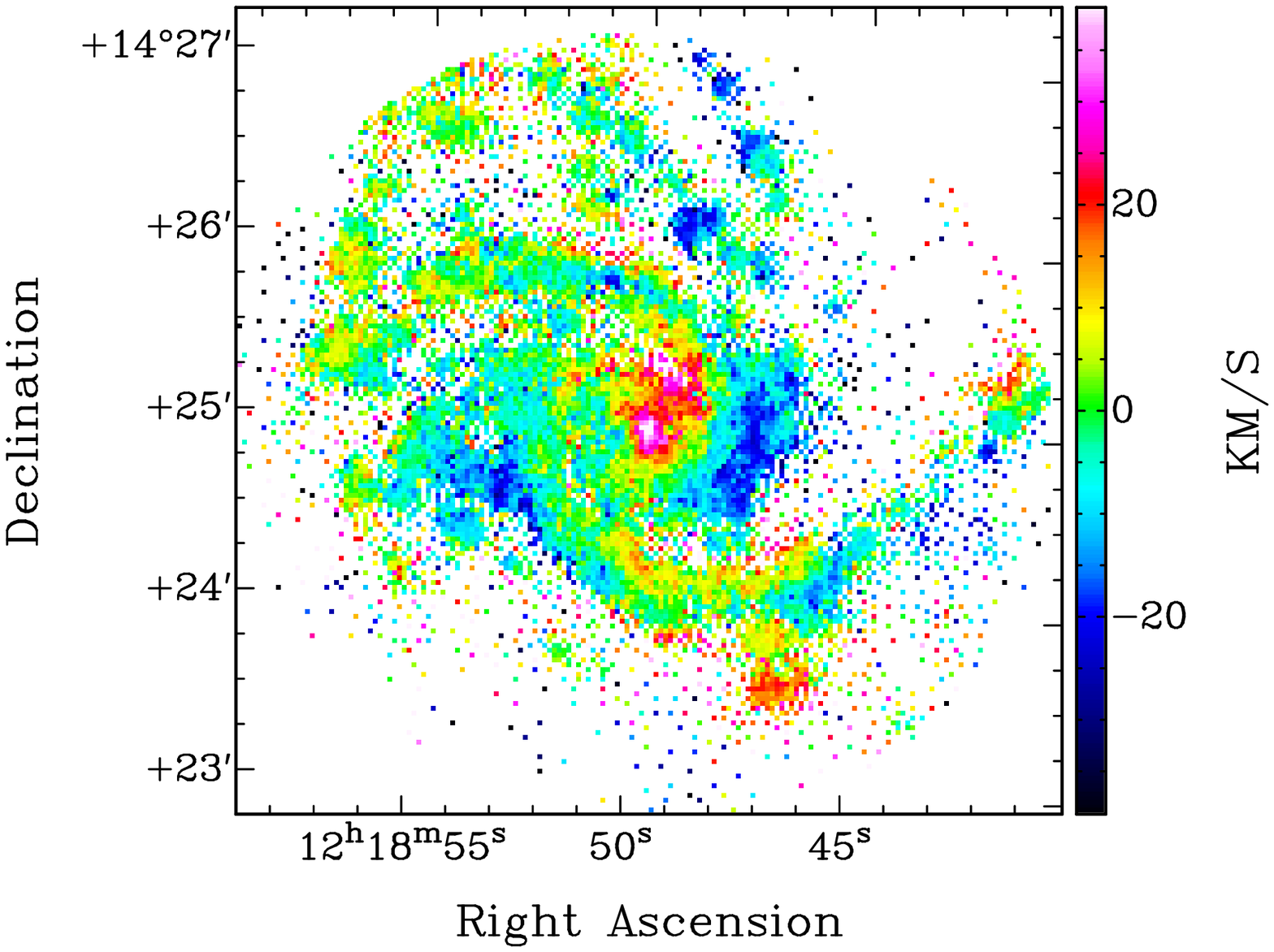}\includegraphics[width=0.33\textwidth]{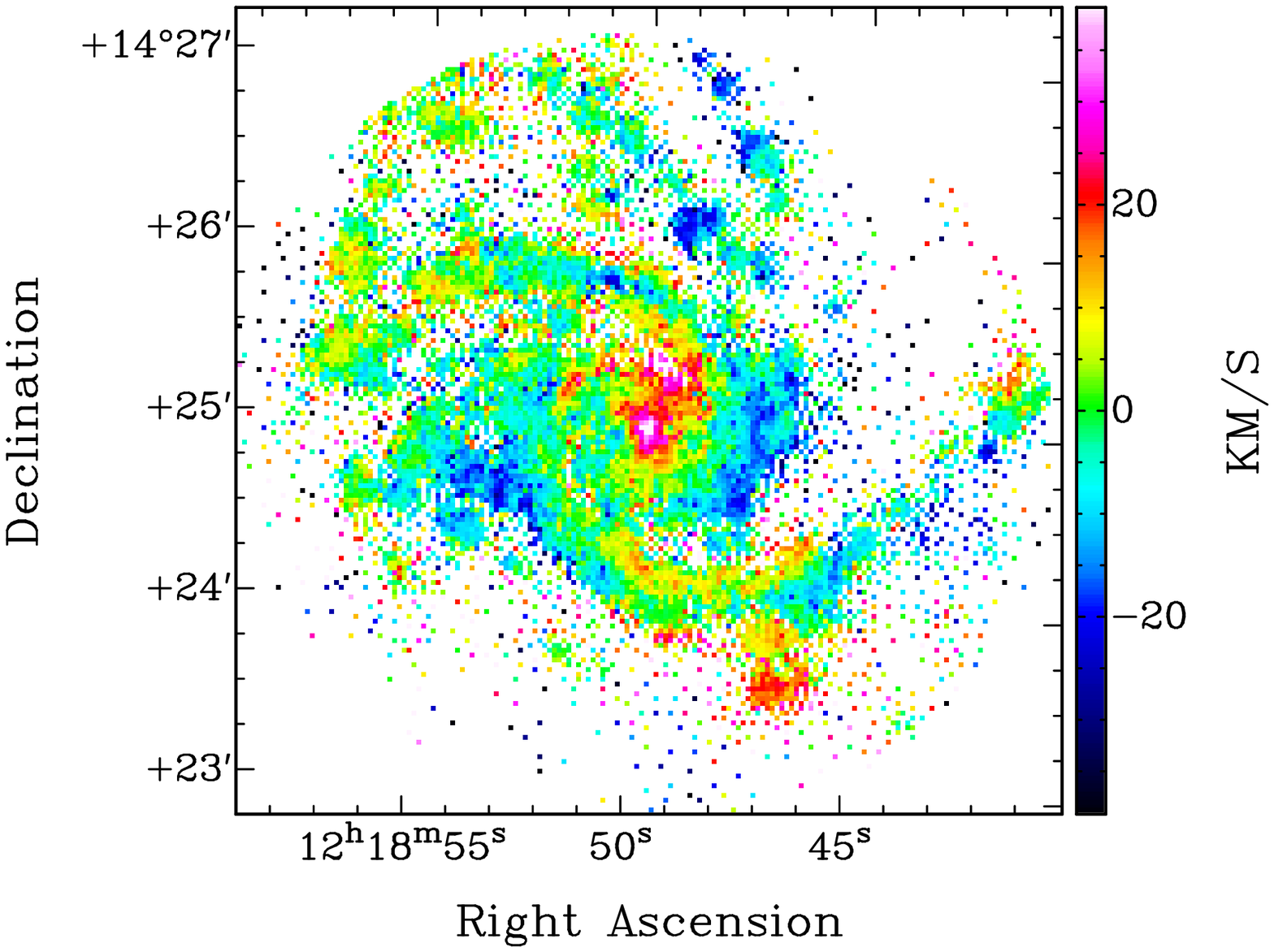}\includegraphics[width=0.33\textwidth]{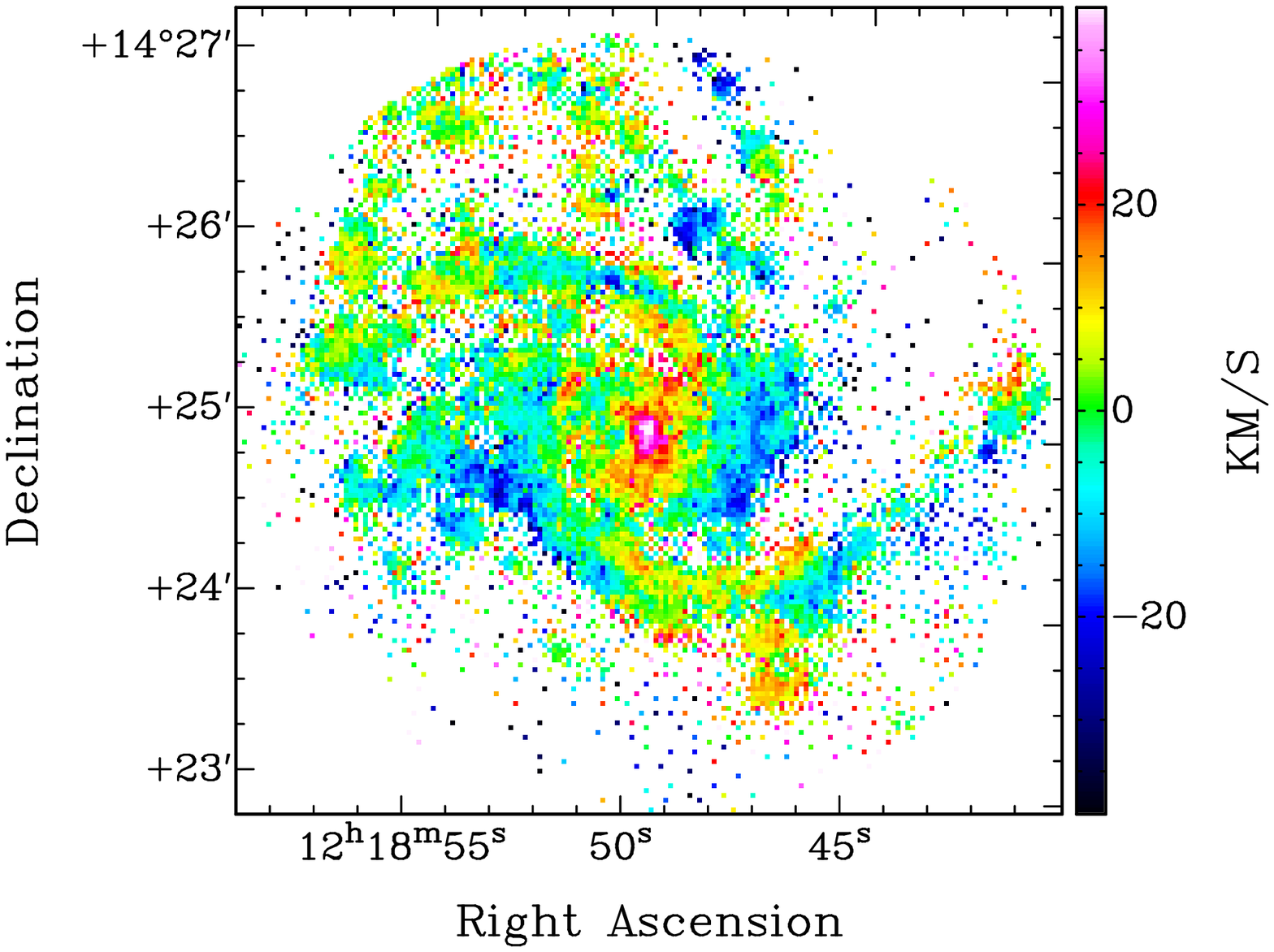}
\caption{\textit{Top:} Modeled velocity field of M99 in the 2D axisymmetric, asymmetric,  and asymmetric+\vr\ cases for the Einasto halo (from left to right, respectively). 
\textit{Bottom:} Corresponding residual maps (observed minus modeled velocities). Velocity contours are from 2300 to 2500 \kms\ in step of 10 \kms.}
 \label{fig:vfein2dasym}
 \end{center}
 \end{figure*}
  
\section{Comparison with previous works and numerical simulations}
\label{sec:nbodysims}

 It is important to clarify that  nonuniform  is not equivalent to  noncircular, although both phenomena have connections 
 as they are  caused by perturbed potentials.  
 In most kinematical studies of disk galaxies, noncircular motions are often only associated with asymmetries. 
  This is  however only part of the reality since  asymmetry/nonuniformity    applies  both to  noncircular and circular motions.  
 Our   method  can only estimate the degree of nonuniformity of circular motions for M99, however. It 
 should also be possible to  estimate this degree  for the noncircular (radial) component  with more advanced dynamical modeling, 
 but this is beyond the scope of the article. 
 
 Also,  the literature  usually refers  to circular as the axisymmetric value of the rotational motion only, while  
 departures from that mean value are often  called noncircular.  This circularity-axisymmetry association turns out to be
  inadequate since  departures from the axisymmetric velocities, such as those identified in Section~\ref{sec:nonuniform}, 
  directly stem from the radial force as a natural response to perturbed potentials. The  designation nonuniform for such departures is thus 
  more appropriate.    The interesting point for circular motions in M99 is that nonuniformity is the rule rather than exception, whereas axisymmetry is rarely expected to  occur.  This is the reason why we   make the difference throughout the article, on the one hand, between  the axisymmetric circular and the nonuniform circular components 
  and, on
the other hand, between the noncircular and the nonuniform circular components. 

The knowledge of nonuniform velocities and asymmetric radial and tangential forces in disk galaxies is not new.  First, in numerical simulations,  
the study of asymmetric motions has been shown to be powerful to understand the response of gaseous or stellar particles to barred, spiral, or lopsided potentials,  
\citep[e.g., among many articles,][]{com81,ath92,wad94,sel02,mac04,bou05,qui11,gra12,min12,ren13}. 
For instance, simulations are helpful to study the radial migration  of stellar particles through spiral arms and the 
dynamical effects of the spiral structure on radial and rotational velocities to predict  signatures to be detected by  kinematical samples of  
Galactic disk stars \citep{kaw14}. Simulations can also depict the strong influence of the dynamics in a stellar bar on   velocities across and along  
the bar axes in order to analyze the survival and merging of gaseous clouds and their implication on star formation \citep{ren15}.
Simulations can also show how the velocity nonuniformity  prevents us from measuring the correct shape of the Milky Way rotation curve from \hi\ measurements \citep{che15}. 
Second, the nonuniformity of the tangential and radial velocity is inherent to 
the theory of potential perturbations \citep{fra94,rix95,jog97,sch97,jog00,jog02,bin08}. Applications of the theory to observations has allowed the study of 
 the elongation of galactic disk potentials \citep{sch97,sim05} or to constrain   the amplitudes of   elliptical streamings  and of pure radial inflows along spiral arms
   \citep{won04},   which was made  possible by a decomposition of   \los\ velocity fields  into Fourier harmonics 
 (see Appendix~\ref{sec:fouriervf} for the case of M99). \cite{jog02} predicted the signatures of galaxy lopsidedness on rotation curves.  
 \cite{spe07} also illustrated the effects of nonuniformity  
 for a galaxy perturbed by an inner bar-like/oval distortion, but on \los\ motions and from a   different harmonic model 
  than in \citet{sch97} and \citet{won04}.

   Then, derivations of stellar potentials from photometry have already been carried out. 
   \cite{zha07} and \cite{but09}  determined the  stellar potential for $\sim 150$ 
   barred galaxies from near-infrared images to  
   measure phase shifts between the  density  and potential and to constrain  corotation radii. 
   The comparison with our work remains  limited as 
   these authors restricted the analysis to 1D calculations and assumed constant mass-to-light ratios over the disks. 
   \citet{kra01}  used NIR imagery with constant M/L as well to generate
   the gravitational potential of the stellar disk of M99. These authors ran
    hydrodynamical simulations  to follow the gas response to the asymmetric potential of stars and to 
    compare with \ha\ kinematics. Our analysis differs from that of Kranz et al. 
    since they proposed that the central spheroid is a stellar bar extending to about one disk scalelength;  they estimated this scalelength at 
    36\arcsec or $\sim 3$ kpc at our adopted distance, thus about twice the value we derived. 
    Our harmonic analysis of the stellar potential is not consistent with a bar because 
    the phase angle of the central $m=2$ perturbation varies at these radii (Fig.~\ref{fig:fourierpotphase}). 
    A comparison of dynamical asymmetries is impossible, however, as these authors have not shown  radial profiles 
    or a resolved map of the asymmetric stellar potential.

  An apparent difference between theoretical and numerical modeling with our approach is that  it is the tangential component that
 varies with azimuth in such modeling,  rather than the circular velocity, 
 while we stated that azimuthal departures from axisymmetry are those of \vc.  There are however striking similarities that are worth  reporting. 
Indeed, the azimuthally periodic patterns  evidenced  in the  \vc\ map of M99  
 are very reminiscent of those predicted for \vt\ by the theory or simulations for other galaxies.  
 This implies that the   circular velocity  can
 be a good approximation for the tangential component, $v_{\theta}=R\dot{\theta}$. 
 To verify this statement, we applied our methodology to  a 
   numerical simulation of a disk of similar mass and morphology as the Milky Way.  
   Details of the simulation are described in \citet{kaw14}, and we have used the same snapshot  as their Fig. 1. The simulation 
   followed the evolution of the gas and stellar disks with an N-body/SPH code \citep[\texttt{GCD+};][]{kaw03,bar12,kaw13,kaw14b}. 
 Figure~\ref{fig:compsimu} shows a strong correlation between our predicted \vc\ and the  simulated \vt\ of gaseous particles  (seen here 
 at, e.g., $R=5.4$ and 10.8 kpc).
 Both velocities strongly vary as a response to the gravitational impacts of the spiral arms. 
  Differences between \vc\ and \vt\ nonetheless exist, as seen, e.g.,   by a slightly smaller variation of \vc\ for $R=5.4$ kpc at $\theta=200-320\degr$, 
  or by \vt\ lagging (or exceeding)  \vc\ at, e.g.,  $\theta=120-170\degr$ for  $R=5.4$ kpc, or   $\theta=200-230\degr$ for $R=10.8$ kpc.  
   These offsets are caused by particles that are  losing (or gaining, respectively) angular momentum \citep[see also Fig. 4 of][]{kaw14},  and 
  that thus have  nonmarginal radial motions. The circular motion is therefore altered by noncircular motions at these locations. 
   Other notable differences are  \vt\ dips or peaks  occurring on very small angular scales, which are not present in \vc.
   They are disturbances of gas kinematics due to star formation and feedback processes in the simulation.
    Despite those   irregularities, the main result  is that \vt\ and \vc\ are very comparable 
  so that the 2D mass distribution models should   benefit from using nonuniform circular velocities.

\section{Asymmetric mass modeling of M99}
\label{sec:fitasym}

In our 2D asymmetric approach, the variations of \vt\ are   
therefore governed by those of \vc\ ($v_\theta \simeq v_c$). 
 The model velocity field is   supposed to be that of a tracer of negligible asymmetric drift, whose assumption should hold for gas 
in M99 (Section~\ref{sec:rcvr}). Furthermore, the asymmetric dynamical modeling does 
not try to  reproduce possible departures from circularity that  are locally associated with star-forming regions and feedback, such as expanding gas shells or gas accretion from galactic fountains. This Section presents the results of the asymmetric mass models using the 
inputs and observables decribed in the previous Sections. 
 
\subsection{Impact of nonuniform circular velocities}

We performed least-squares fittings of the  velocity field model  projected along the line of sight  
 to the observed \ha\ velocity field of M99, following Eq.~\ref{eq:vtanvradvf} and with uniform weightings. As for the axisymmetric case,  
we fitted models with and without radial  motions.  The modeling with \vr\ is hybrid between axisymmetry and 
 asymmetry as \vr\ is axisymmetric, unlike \vt.  
  In this section, the spherical dark matter halo is centered on the coordinates of the 
 center of mass of the luminous gas and stellar disks. Section~\ref{sec:shifted}  presents models with an alternative position of the dynamical center of dark matter.  
 Results of the nonlinear minimizations are listed in Tab.~\ref{tab:paramdm} of Appendix~\ref{sec:tablesresults} and the  differences 
 of Akaike Information Criteria are given in Tab.~\ref{tab:diffaic1}. 

 It is  found that the Einasto model is  more likely than the NFW cusp, which in turn is more likely than the  PIS model, irrespective of the 
 contribution from \vr. This result  confirms the trend observed for 2D axisymmetric modeling. 
 The  degeneracy of Einasto parameters is   insignificant, in contrast with  the axisymmetric results.
 Dark matter tends to   be slightly less concentrated than in the axisymmetric case. 
Moreover, the derived density slope  for the Einasto halo is $-0.72 \pm
0.27$ at $R=0.27$ kpc ($-0.70 \pm 0.26$ with \vr). 
The  impact of nonuniform circular motions is therefore to yield a less cuspy halo than in the axisymmetric case.  This trend  goes in the same direction 
 as the effect of noncircular motions, but at a larger extent. It seems likely that dealing with nonuniform, 
 noncircular velocities would amplify this trend as well.
 
  In Fig.~\ref{fig:vfein2dasym}, we show the velocity field of the best-fit model for the Einasto halo along with a corresponding map of residuals.  For comparison, the 2D axisymmetric model of the Einasto halo is also shown. 
 The velocity contours highlight the significant differences between both strategies. Not surprisingly, the asymmetric case  exhibits velocity 
 wiggles in the spiral arms that are  not observed in the axisymmetric case. The combination of 
 an asymmetric distribution of tangential velocities for luminous matter with a less cuspy dark matter halo 
 has improved the modeling of the \los\ kinematics.   
 We estimate that for the pixels where the residuals are lower (mainly on the leading sides of the spiral arms),   
 the median drop in residuals is about 20\% the amplitude of axisymmetry-based residuals. Other pixels  have seen their residual rising (mainly on the trailing sides), but at a smaller rate (10\%).
 Residual velocities are consequently less scattered than in the axisymmetric case. We measure a standard deviation of residuals 
 that is smaller  by up to 2 \kms\ inside $2 h_\star$ than in the axisymmetric case, i.e. in regions where the velocity contribution of luminous matter 
 is the most nonuniform (Fig.~\ref{fig:nu}).  The improvement is also reflected by positive $\rm AIC_{Axi.}-AIC_{Asym.}$  differences 
 (Tab.~\ref{tab:diffaic1}), and is effective for any halo shapes.
   
As for the beneficial consideration of   \vr\ in the modeling, it is also  verified by the  AIC tests 
irrespective of the halo shape and of the uniformity of circular motions.  Modeling has mostly been improved 
  in regions where \vr\  are larger ($R<1.5 h_\star$).  
 This trend was expected because   \vr\ results from a prior fitting of the \ha\ velocity field of M99. 
 The remarkable effect of \vr\ on the residuals is a motivation to mix both nonuniform circular and noncircular velocities in future dynamical works.

\subsection{A shifted dark matter halo in M99?}
\label{sec:shifted}
An important result  from the Fourier analysis of the potentials was to evidence the dominance of $m=1$ perturbing modes in inner regions of the stellar disk, through  
the entire \hi\ disk, and at some specific locations in the molecular disk. 
A detailed description of the effects of the potential perturbations on  velocities in the epicycle theory  has been given in \citet{fra94}, \citet{jog97}, \citet{sch97}, or 
\citet{jog00}.  \citet{sch97}  proposed that galaxy  velocity fields can be decomposed into harmonics to evidence signatures of dynamical perturbations (see Eq.~\ref{eq:fouriervf}).
They showed that a $m-$order perturbation of the gravitational potential generates kinematical Fourier coefficients of order $k=m-1$ and 
$k=m+1$ in velocity fields. Signatures of $m=2$ modes (a bar, spirals, an elongated dark matter halo, a bisymmetric warp) are identified in   
$k=1$ and $k=3$ kinematical harmonics. 
 The occurence of galaxy lopsidedness in terms of morphology and kinematics for stellar and gaseous components 
  \citep[e.g.,][]{rix95,zar97,sch99,bou05,ang06,ang07}  is    motivation  
  to explain the origin and variation of $k=0$ and $k=2$   coefficients in galaxies
    with a lopsided potential ($m=1$  perturbation). Additionally, \citet{sch97} studied the effects of lopsided  
    potentials in \los\ kinematics using toy models in which dark matter haloes were not centered on the gravity center of luminous matter.  

\begin{figure}[t]
\begin{center}
\includegraphics[width=0.9\columnwidth]{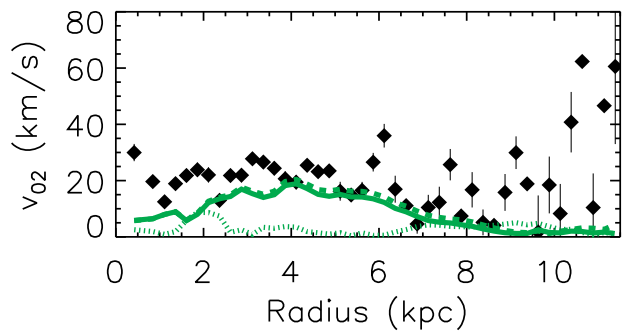}
\caption{Amplitude $v_{02}$ of the combined $k=0$ and $k=2$ harmonics of the \ha\ velocity field of M99 (filled symbols). The dotted lines indicate 
the result for the model velocity field of the centered-Einasto halo with \vr, while   the dashed and solid lines indicate models of the shifted-Einasto halo with and without \vr, respectively.}
 \label{fig:ampv02}
\end{center}
\end{figure}

The velocity field of M99  shows obvious signatures of a lopsided total potential.
The difference between the rotation curves of the approaching and receding sides of the M99 disk 
is  one such signature  \citep[for a detailed description of the impact of lopsidedness on rotation curves and velocity fields, see also][]{swa99,jog02,jog09,van11}. 
Appendix~\ref{sec:fouriervf}   details the analysis of the harmonic decompostion of the \ha\ velocity field of M99. 
  Figure~\ref{fig:ampv02} plots the amplitude  of the combined $k=0$ and $k=2$  coefficients, $v_{02}  = (v_0^2+v_2^2)^{0.5}$.
Basically, an asymmetry of $\sim 20$ \kms\ is detected up to $R \sim 7$ kpc, confirming the presence of a kinematical lopsidedness in M99. 
\cite{sch97} showed that such motions could be caused by systematic errors on the position of the mass center of the galaxy or on systematic velocity.
We reject the possibility that the systemic velocity is erroneous, as it agrees with 
the literature and other tracers. An error of systemic velocity would also shift the distribution of \los\ residuals, whose feature is not observed
in  our modeled residual maps.  As for an error on the position of the 
mass center, it is possible if the total mass center is not that of the disk of luminous matter (i.e. M99 is lopsided). 
By construction, our circular velocities contain hints of the $m=1$  perturbation  of the luminous potentials. 
Interestingly, the harmonic decomposition of these modeled velocity fields reveals a failure in our attempt to reproduce the observed $v_{02}$ amplitude. 
The modeled amplitude indeed never reaches 10 \kms\ (Fig.~\ref{fig:ampv02}, dotted line). 
In other words, the   lopsidedness of luminous matter is not strong enough to account for the overall kinematical lopsidedness.
 
\begin{figure*}[t]
\begin{center}
\includegraphics[width=0.31\textwidth]{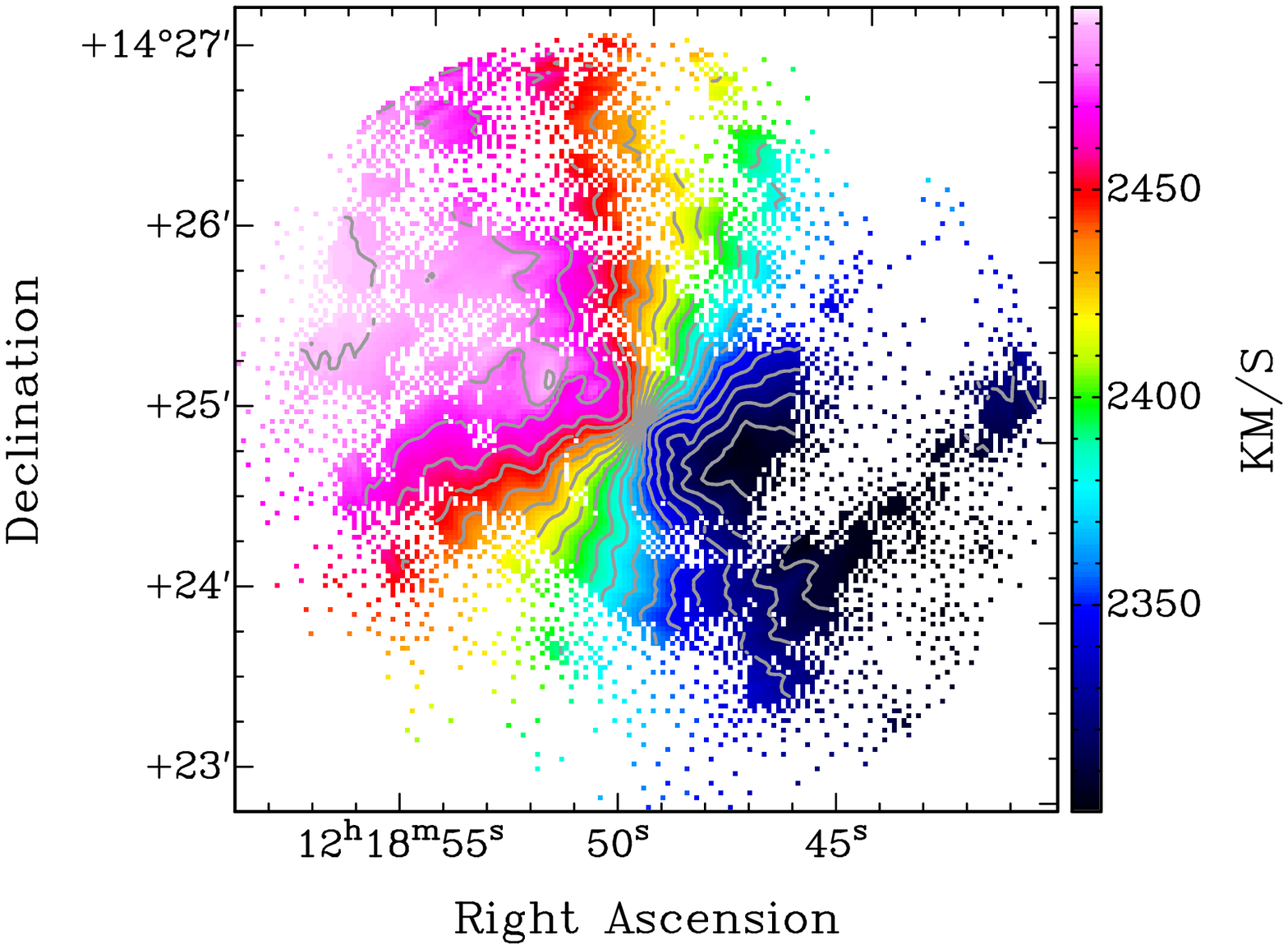}\includegraphics[width=0.31\textwidth]{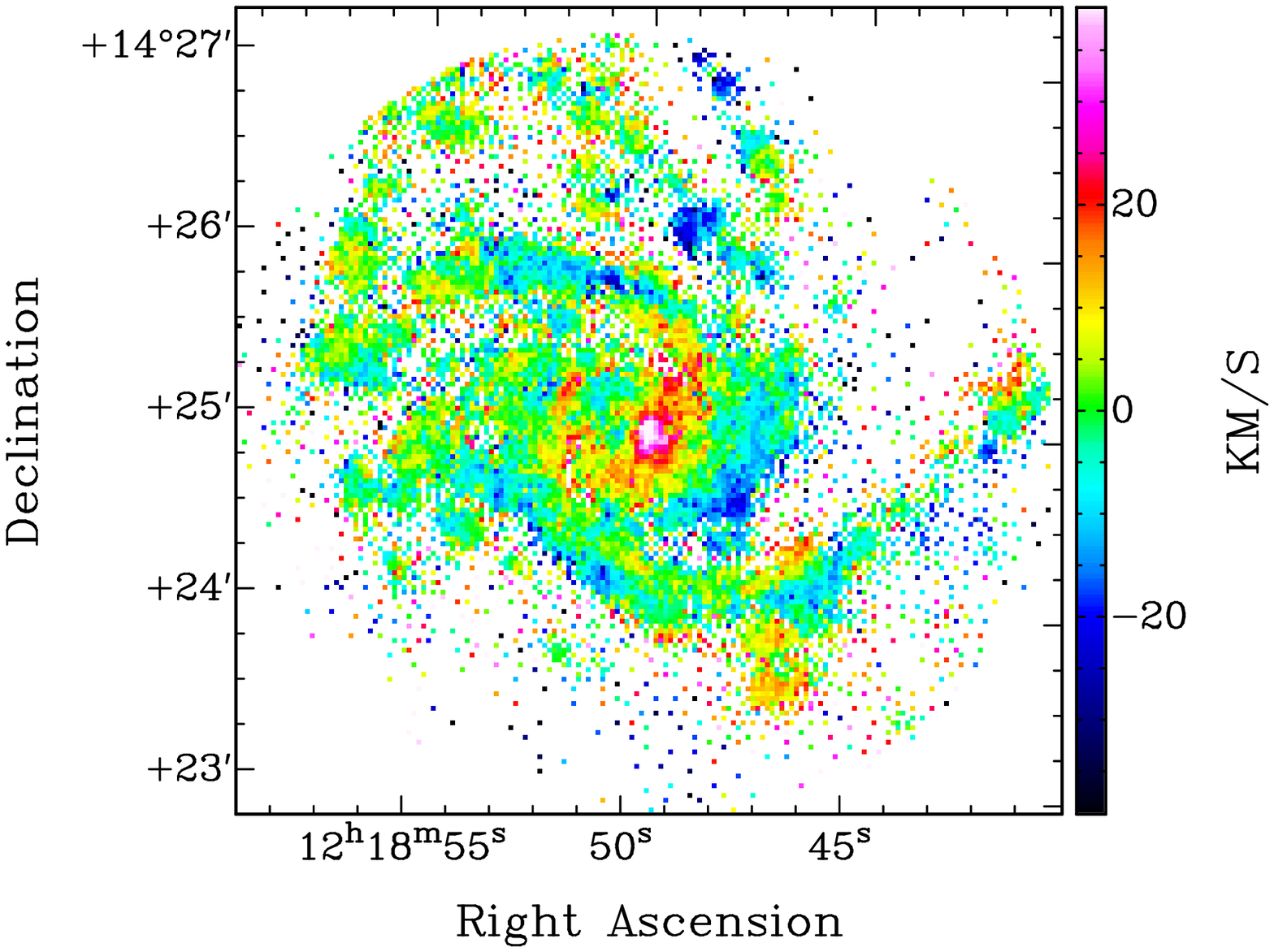}\includegraphics[width=0.38\textwidth]{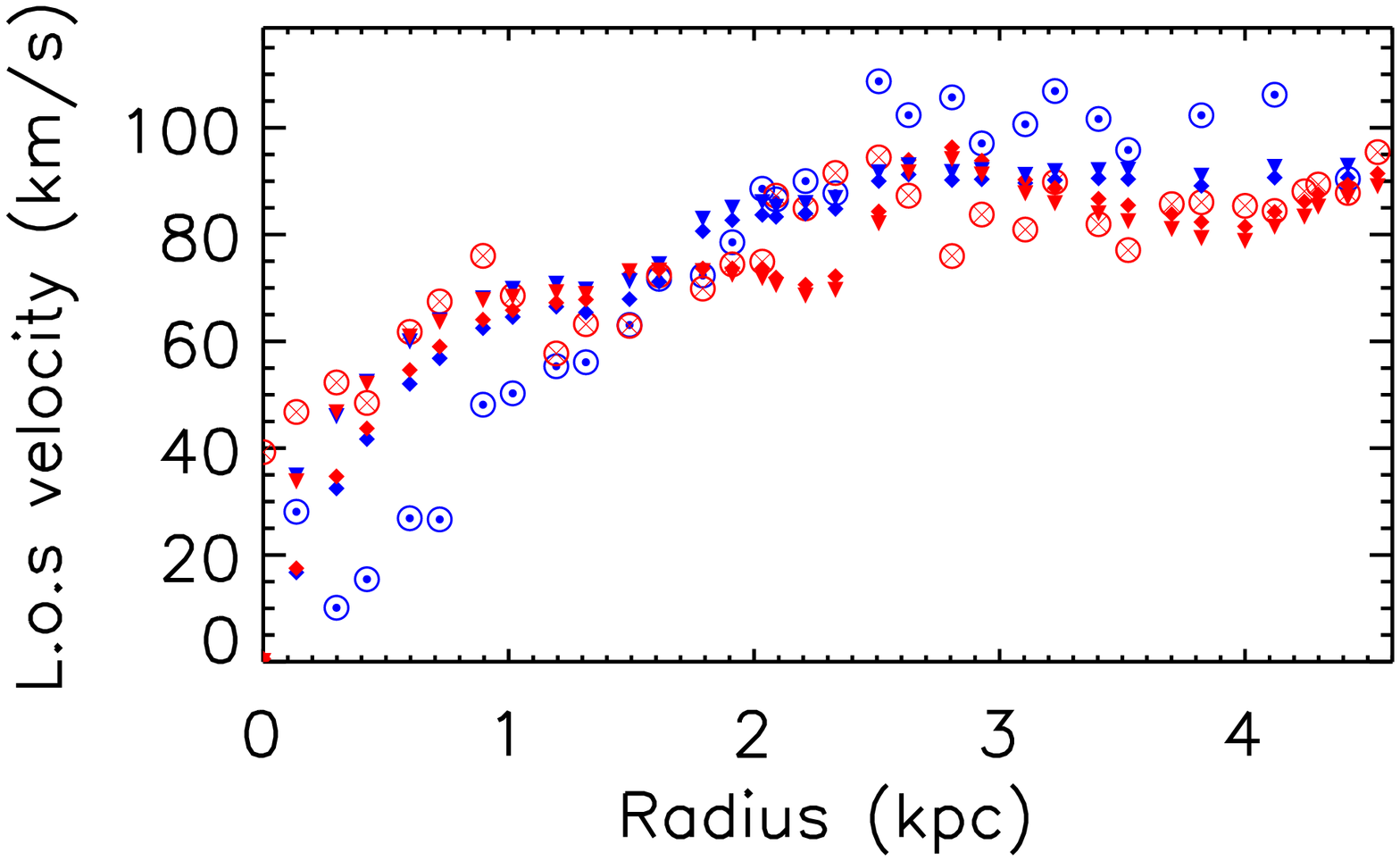}
\caption{\textit{Left:} Modeled velocity field of M99 from the 2D asymmetric case for the shifted Einasto halo with \vr\ motions. 
Contours are the same as in  Fig.~\ref{fig:vfein2dasym}. \textit{Middle:} Corresponding residual velocity field. \textit{Right:} Comparison of major axis \los\ velocities of the observation with the centered-halo and shifted-halo cases.  
Blue and red symbols indicate the approaching and receding sides, respectively. Open dotted and crossed symbols indicate the observed velocity field,  
 filled  diamonds, and triangles for the asymmetric centered- and shifted-Einasto models, respectively, with the contribution from \vr.}
 \label{fig:vfein2dasymshift}
 \end{center}
 \end{figure*}

In this section, we thus explore the possibility that the gravitational potential of total matter is additionally lopsided under the influence of
a spherical dark matter halo whose center is shifted with respect to the center of luminous baryons. 
For this purpose, we added two  parameters  to the modeling, which correspond to offsets $\delta_x$ and $\delta_y$ of   the dark matter center 
 to the coordinates $x-y$ of the center of luminous matter, respectively. The resolved velocity map of dark matter is 
estimated in a new grid of coordinates centered on $(\delta_x,\delta_y)$, and added quadratically to that of luminous baryons, 
where the latter remains the same as in previous sections, i.e. centered on $x=y=0$. For simplicity, we assume that the gravity center of dark matter 
remains in the disk midplane. Such least-squares fittings are referred to as
shifted-halo models in  Tabs.~\ref{tab:paramdm} and~\ref{tab:diffaic1} (Appendix~\ref{sec:tablesresults}). 
As a result, we find similarities with the centered-halo modeling: The Einasto model remains the model that is the most appropriate, 
the asymmetric case provides better results than the axisymmetric case, and the modeling with \vr\ provide better fittings than those without these motions.
However, there are numerous differences  between these models and the centered-halo configuration. First, 
the AIC tests   show that  all models with shifted dark matter haloes are
more likely than with centered dark matter haloes, and the formal errors on the
halo parameters are significantly smaller.
 Second, the center of the dark halo is found no more coincident with 
the luminous center. A mean total shift of $2.5\pm 0.2$ kpc ($2.2\pm0.3$ kpc) is implied by the asymmetric modeling  
 with  noncircular motions (without, respectively). The consideration of noncircular velocities thus tends to increase the DM shift. Compared to the axisymmetric modeling, the   nonuniformity of circular motions 
 also tends to increase the offset, but only at a level of 0.1 kpc. 
 A shift of $2.2-2.5$ kpc is significant as it corresponds to more than  15
 times the angular sampling of the   stellar density and Fabry-Perot interferometry maps. 
Third, the resulting   haloes  have become core dominated. 
The Einasto density profile is now considerably shallower in its center
(slopes at $R=0.27$ kpc of $-0.01$ and $-0.02$ ($\pm 0.01$) with and
without radial motions, respectively), and the pseudoisothermal sphere has become more likely than the NFW cusp. 
 Moreover, the concentration of dark matter is significantly smaller than in the centered-halo case. 
 Figure~\ref{fig:vfein2dasymshift} shows the modeled velocity   and residual velocity maps for the shifted Einasto halo with the \vr\ component.
 We estimate that on average, the beneficial impact of the shifted-halo occurs beyond $R \sim 3$ kpc. Inside that radius, the scatter of 
 residuals is roughly equivalent, and sometimes larger than in the centered-halo case, depending on the considered halo shape and the contribution from \vr. 
  The shifted-halo model correctly reproduces most of the $v_{02}$ amplitude (solid and dashed lines in Fig.~\ref{fig:ampv02}), implying that the assumption 
  that the total  potential is lopsided is valid.
  
  However, a lag with respect to the observed $v_{02}$ still exists in
   the innermost regions. In addition, as  in the centered-halo configuration, differences from the observation are not only evident in
  the center and along the minor axis of the residual maps,  but they are also observed along the major axis of the velocity fields. 
  Figure~\ref{fig:vfein2dasymshift} (right panel) indeed shows that none of our asymmetric models  
  succeeds in correctly reproducing  the   difference between the approaching and receding disk halves at small radii, whether
  the dark matter component is aligned with the luminous baryons or not. 
  More importantly, only the receding motions are fairly well reproduced and  the approaching motions for the centered or shifted halo models 
  remain quite similar to the receding motions.  Though promising, the shifted-halo  solution also turns out to be incomplete 
    to  explain the inner lopsidedness of  M99.   
   It appears here that an additional dynamical process causing more asymmetric motions  than our predictions 
   is required at small radius. What could explain this discrepancy? If it comes from the incorrect assumption that circularity dominates, then one should  question the relevance 
   of performing mass models of galaxies harboring lopsided kinematics and structure.  At the same time,   
     nonuniform \vr\ motions that would dominate the nonuniform \vt\ at these radii  cannot explain the discrepancy  because  \vr\ does not project along the major axis, 
   whereas the inner lopsided effect is  obvious along this axis. 
   We then reject the possibility of perturbed motions by a central bar because the stellar potential   rules out its presence, 
   and the kinematical signature of a bar should be bisymmetric.  
      A possible origin could be that vertical motions are finally not as negligible as initially thought. 
     It was shown that the \los\ width of the \ha\ profiles is slightly larger toward the center. Larger dispersions may be partly 
     explained by the observed larger nonuniformity of circular motions. It may be also due to a larger scatter of vertical motions,
     if the ellispoid of velocity dispersion is isotropic.  The impact of fixing $v_z=0$ instead of truly varying vertical motions with radius 
     is to generate artificial variations of $v_{\rm sys}$ or,  more precisely, of the $k=0$ Fourier coefficient (Appendix~\ref{sec:fouriervf}). 
     The reason for this bias is that  the \vz\ projection along the \los\  does not depend on $\theta$ but simply adds linearly 
     to the systemic velocity (Eq.~\ref{eq:vtanvradvf}).  Though $v_z \ne 0$  could explain some of the observed variation of the $k=0$ term  
     in the velocity field (Fig.~\ref{fig:fouriervfobs}), a \vz-induced mechanism  cannot impact the $k=2$ coefficients, however.  
     
    We are left with the hypothesis that our modeling could be improved by the action of circular motions with larger nonuniformity than predicted, particularly 
    for the approaching side of the galaxy because it is the most discrepant disk region with respect to the expectations.
    We reject the possibility that the   SPS   models yielded incorrect stellar masses and velocities for this disk half because they have been performed 
    following a homogeneous approach all through the field of view.  Instead, we think that a mechanism where the velocity contribution from dark matter is nonuniform  
    is more worthwhile to consider.
   It does not seem unrealistic that the gravitational potential of dark matter is lopsided as well, owing to the   perturbed  
   nature of both the stellar and gaseous components in M99.  In this case, our scenario  of shifted DM component with respect to the luminous disk 
   is anecdotal, as it would be only a signature of a more complex lopsided  distribution of  dark matter. 
   Further modeling of the velocity field of M99 is needed to constrain this possibly more asymmetric density and velocity distribution of dark matter.
  
  Off-centered density peak and lopsided distribution of dark matter haloes have been evidenced in cosmological hydrodynamics simulations  \citep{kuh13,sch15}. 
  The typical distance between the simulated dark matter density peak to the stellar density peak, or to the position of the minimum of the total gravitational potential,   
  is $\propto 100$ pc.  Interestingly, the presence of a   peak offset for the simulated halo
  of \citet{kuh13} is tightly linked to the core-dominated nature of its density profile. 
  The significantly larger shift found for M99   appears to conflict with 
  the collisional simulations, but the concomitance of the 
  inner density flatness with the halo shift in M99  is an observational support to the simulations.
    A dynamical mechanism that motivates a disturbed DM halo, and maybe a misalignment with luminous matter, could
  be a tidal event \citep[][]{jog09,van11}. 
  A tidal interaction between M99 and a massive companion has  actually occurred in the past Gyr, maybe 750 Myrs ago \citep{duc08}. 
  The interaction with the gravitational field of the  Virgo cluster is likely very active too, as M99 has  entered the cluster \citep{vol05}. If triggered by such tidal processes, 
  the DM core shift seems to be a long-lived dynamical event in M99.  
  However, a tidal scenario only seems problematic to explain the amplitude of the $v_{02}$ asymmetry 
   in the innermost regions. Such a scenario is  expected to enhance the strength of  lopsidedness  with radius \citep[e.g.,][]{jog09}, 
   while a roughly constant kinematical asymmetry is measured within $R \sim 7$ kpc. A mechanism other than a tidal event thus remains to be proposed for the origin of
   the innermost lopsidedness of the total mass distribution of M99, including that of the dark matter component.

\section{Conclusions}
\label{sec:conclusion}

 We have presented a new methodology to model the mass distribution of   disk galaxies directly from bidimensional   observables,   
   resolved stellar population synthesis models, high-sensitivity gas density maps, 
   and high-resolution velocity fields. The methodology  makes use of 
 hyperpotentials to derive the 3D gravitational potentials of luminous matter 
 and  the relevant azimuthal and radial  forces without the intervention of a softening length. The 3D strategy is advantageous to estimate the circular motions   
 through nonaxisymmetric features like spiral arms since the 2D distribution of circular velocity in disk midplanes naturally stems from  asymmetric potentials. 

 Applied to multiwavelength observations of the late-type spiral galaxy Messier 99, the method led us to the following results:
 \begin{itemize}
 
  \item The gravitational potential of the disks of stars, atomic, and molecular gas is perturbed by dominant
  $m=1$ and $m=2$ modes corresponding to  lopsidedness and grand-design spiral arms. The importance of the perturbations  decreases 
     with the mass surface density. Amplitudes of perturbations in the stellar disk thus represent no more than 4\% of that 
     of the axisymmetric mode, while they can be as high as 16\% for the atomic gas component.
  
  \item On a global scale, the radial force strongly dominates the tangential force, implying that M99 may be perceived as  totally dominated by uniform circular rotation.
  
  \item On local scales however, the radial forces are comparable to the tangential forces, implying that the circular motions are highly nonuniform.
  Nonuniformity of circular velocities turns out to be the rule, while  uniformity, i.e. axisymmetry, is the exception. The inner disk regions 
  are those where nonuniformity is larger.   The strongest variations occur through spiral arms. Gas self-gravity is not negligible
  in the densest parts of  spiral arms as it can account for, on average, up to  50\% of the nonuniformity.
  
  \item It makes it possible to fit 2D, asymmetric mass distribution models to the velocity field of the galaxy. 
  The modeled velocity fields harbors wiggles along the spiral arms, as a direct consequence of nonuniform circular motions. 
  The number of degrees of freedom has become considerably larger than for the rotation curve decomposition, yielding more constrained fittings and
  parameters for the  dark matter halo.   Compared to an axisymmetric viewpoint, the use of asymmetric velocities improves the mass modeling, as the scatter and amplitude of residual velocities 
  have decreased. Dealing with asymmetric circular velocities in M99   makes the density profile of  the dark matter component 
    less cuspy than in the axisymmetric case. This effect is more prominent for a dark matter model whose inner density slope is allowed to vary, such as the Einasto halo, 
  than for models with constant inner slope, such as the NFW cusp or the  pseudoisothermal sphere.
  
  \item The 2D strategy also makes it possible to take into account  (uniform) noncircular motions \vr\ in the mass modeling. Noncircular  
  motions  also make the DM density profile less cuspy in M99, but to a   lower degree than for the nonuniform  circular motions.

  \item On the one hand, the 2D asymmetric fittings prefer  cuspy Einasto or NFW haloes to the pseudoisothermal sphere in a traditional case where
  the dark matter halo is centered on the luminous mass component. On the other hand, a more likely model introducing a  dark matter halo   
  shifted from the luminous center requires a core dark matter halo.  
  That shifted-halo modeling  succeeds in reproducing a large part of the observed kinematical lopsidedness, contrary to the centered-halo 
  modeling.  However,   none of these models can reproduce the
  lopsidedness in the innermost regions of M99. All these results likely reflect the need for an asymmetric (lopsided) 
  dark matter distribution in M99. This 
  would not be surprising owing to the lopsided nature of disks of stars and gas. 
  Tidal effects from a companion and/or the Virgo cluster tidal field
    could partly explain   the lopsidedness. However, another process to be identified is needed to explain the innermost 
    lopsidedness of the total mass distribution and of the dark matter halo of M99. 
 
  \end{itemize}
  
  The new strategy is very promising for galactic dynamics.  We envisage  testing the reproducibility of such results on a larger sample of galaxies of various morphologies and masses. 
  Future papers from this series will focus on other massive late-type spirals, as well as lower surface density, dark-matter dominated disks, which are at the origin 
  of the  cusp-core controversy.  Velocity fields of these types of galaxies are rich with information that are worth investigating. 
   Furthermore, as resolved SPS models as in \citet{zib09}, \citet{cor14}, \citet{mei14}, and \citet{rah15} are in their early stages, 
  derivations of many more  stellar density maps is strongly encouraged to be able to quantify  the asymmetries of the gravitational potential of stellar 
  disks directly. This is more straightforward for the gas component since atomic and molecular surface density maps are routinely acquired with mm and cm arrays. 
  Finally, we hope that this research will have a broader impact because the concepts developed here
   can be easily transferable to the Milky Way and other types of disks, from protostellar disks  to high-redshift galaxies.

\begin{acknowledgements}
 Laurent Chemin acknowledges the financial support from CNES. 
 We are grateful to an anonymous referee whose suggestions led  
 to clarify the discussion, content, and structure of the manuscript.
 This research has made use of the NASA/IPAC Extragalactic Database (NED), which is operated by the Jet Propulsion Laboratory, California Institute of Technology,
  under contract with the National Aeronautics and Space Administration (\url{http://ned.ipac.caltech.edu}). 
  We acknowledge the usage of the HyperLeda database (\url{leda.univ-lyon1.fr}). The SDSS image is from \url{cosmo.nyu.edu/hogg/rc3}, prepared by David W. Hogg, Michael R. Blanton, and the Sloan Digital Sky Survey Collaboration.
\end{acknowledgements}

\bibliographystyle{aa}
\bibliography{cheminm99}

\begin{appendix}
 
 \section{Harmonics of the gravitational potentials of luminous matter}
 \label{sec:fourierpot}
 This Appendix presents the axisymmetric   potential of the bulge, and the analysis of the harmonic decomposition of the individual gravitational potentials 
 for the asymmetric stellar and gaseous disks of M99. 
 
\begin{figure*}
\hspace*{3.5cm}\atom\ \hspace*{3.4cm}\stars\ \hspace*{3.5cm}\mole\

\begin{center}
\includegraphics[width=0.3\textwidth]{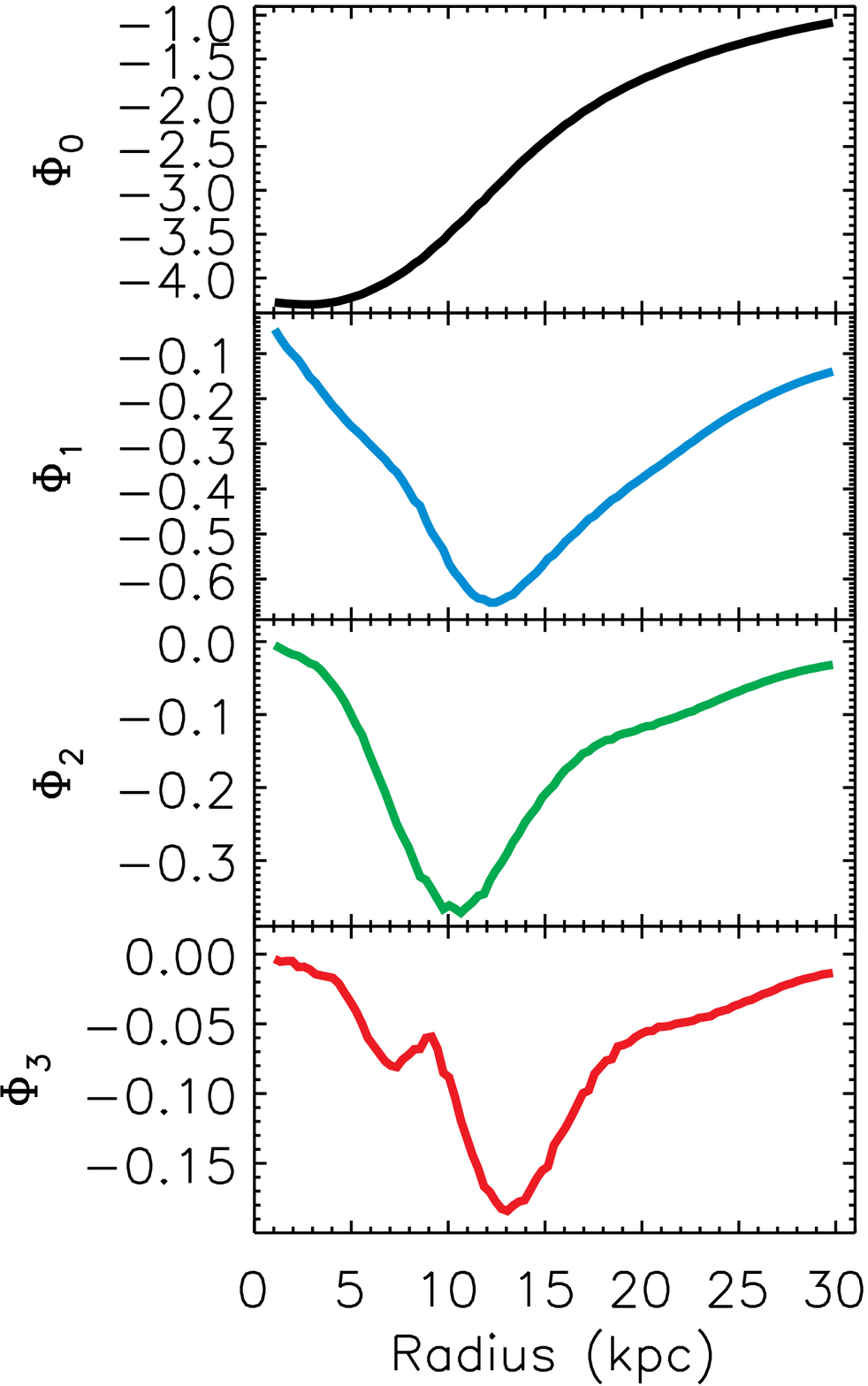}\includegraphics[width=0.3\textwidth]{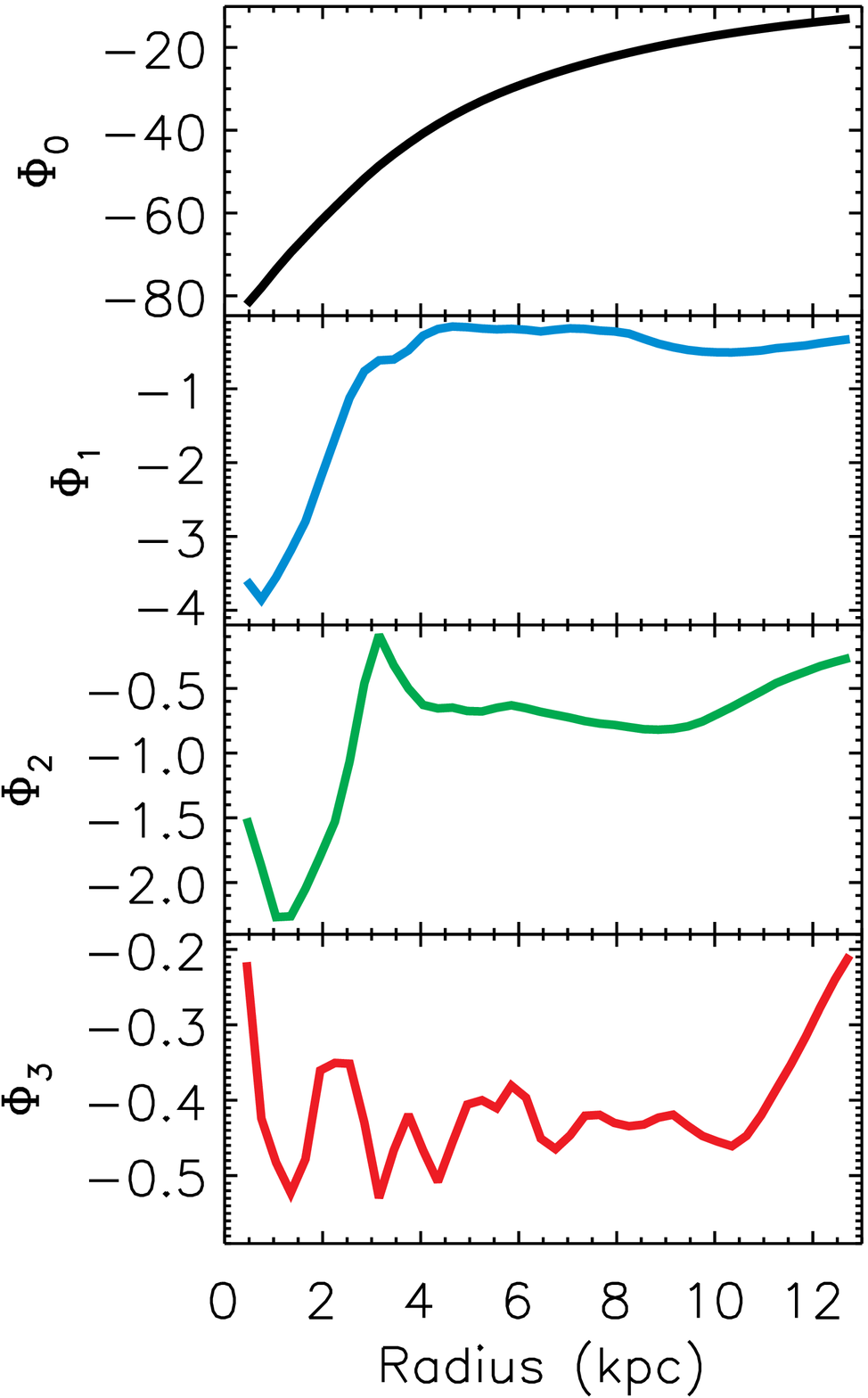}\includegraphics[width=0.3\textwidth]{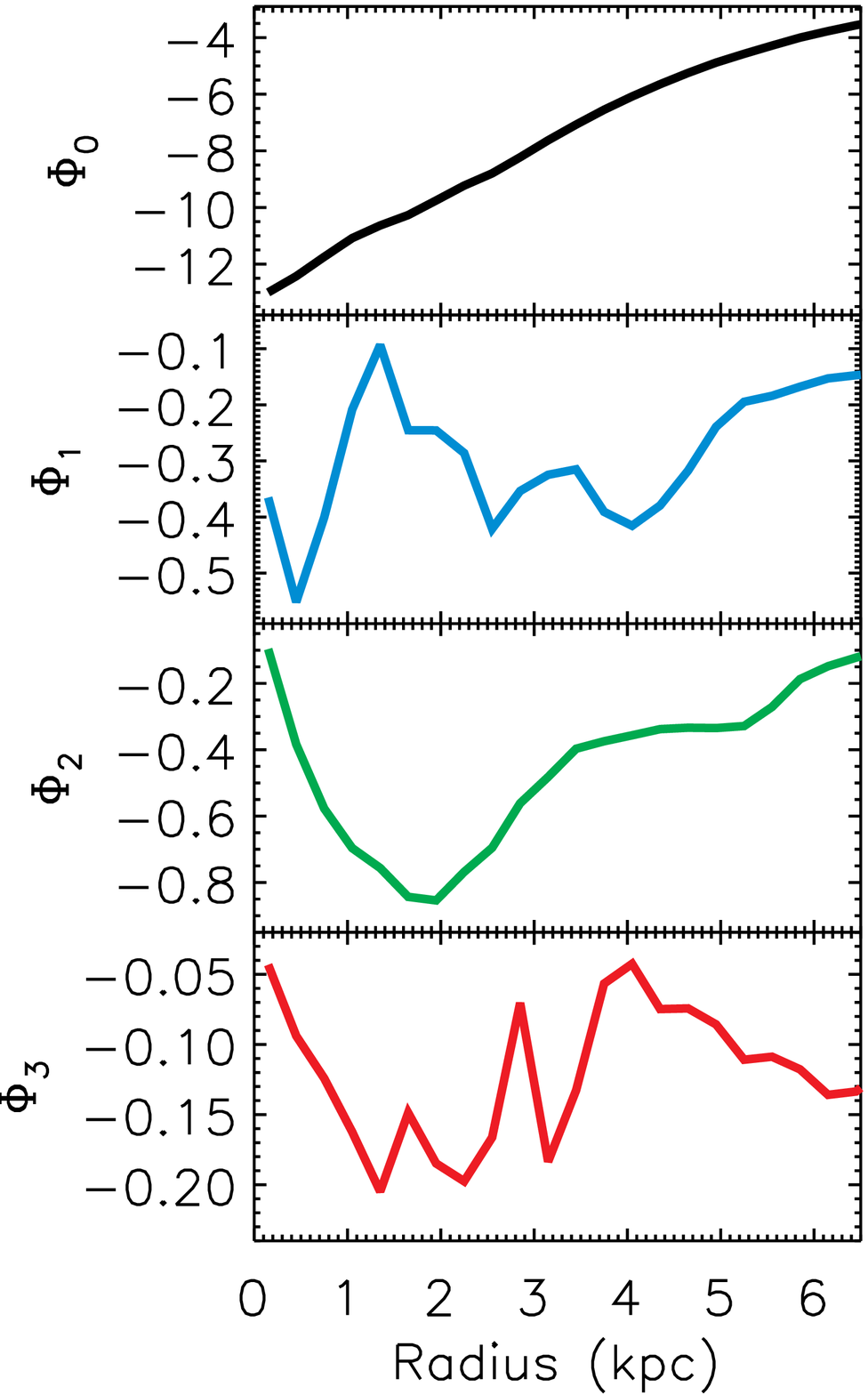}
\caption{Amplitude $\Phi_m$ of the harmonics of gravitational potential for the stellar and gaseous disks of \gala, in units of $10^3$ km$^2$ s$^{-2}$, with $m=0,1,2,3$. 
The left-hand column is for the atomic gas disk, the middle one for the stellar disk and the right-hand column for the molecular gas disk. For the stellar component, 
the axisymmetric potential of the bulge has been removed.}
 \label{fig:fourierpotamp}
\end{center} 
\end{figure*}

\begin{figure*}
\begin{center}
\includegraphics[width=0.3\textwidth]{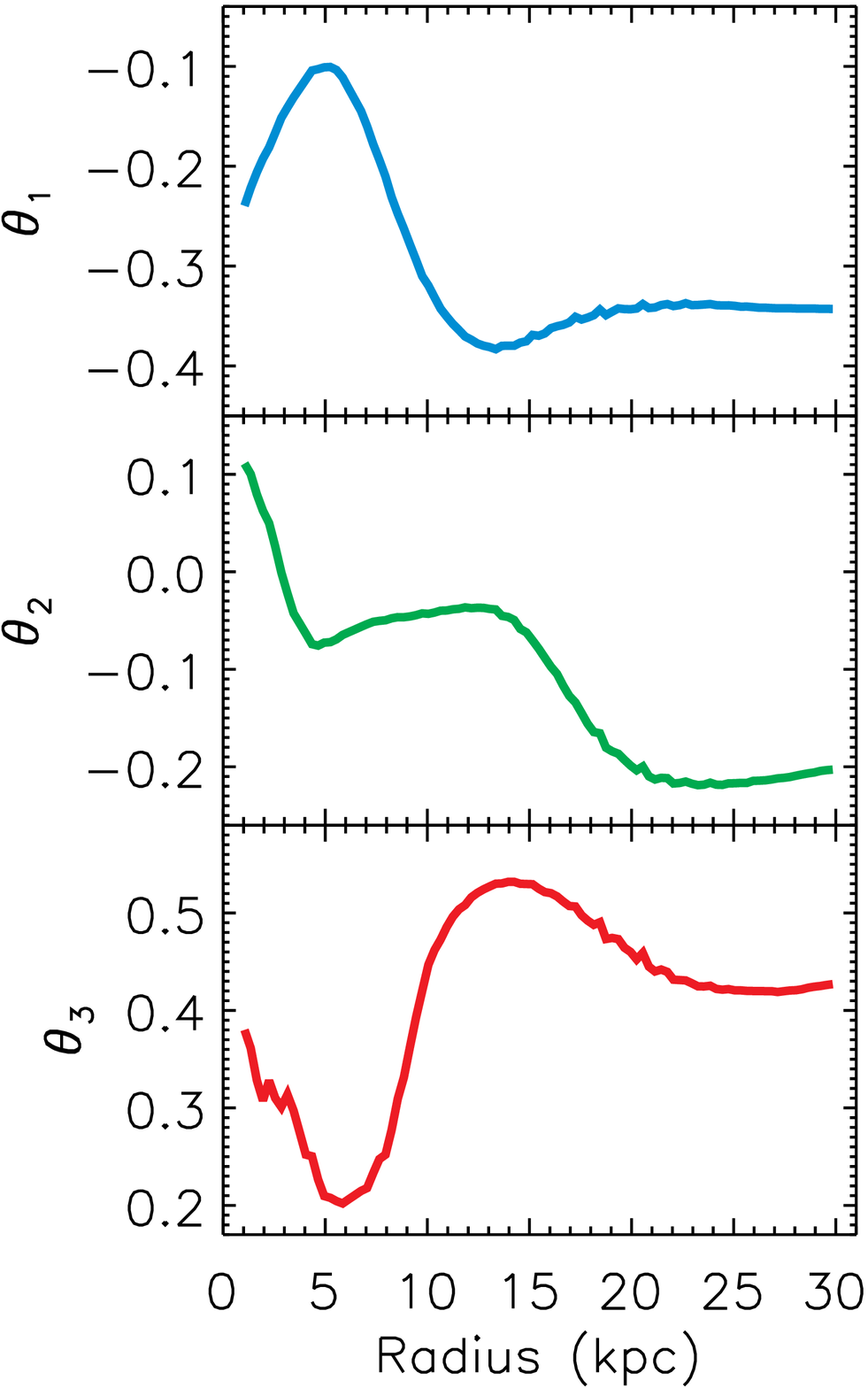}\includegraphics[width=0.3\textwidth]{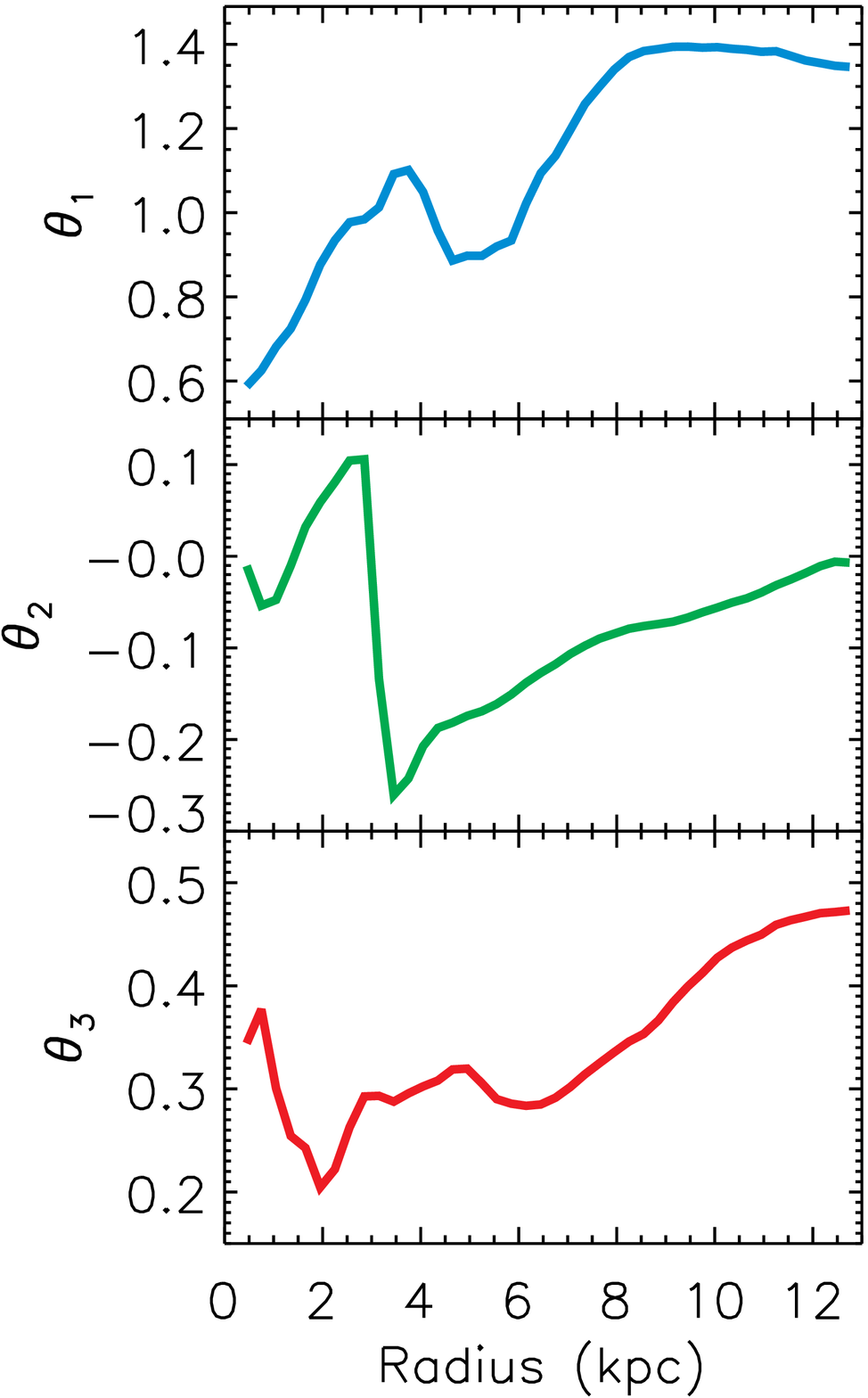}\includegraphics[width=0.3\textwidth]{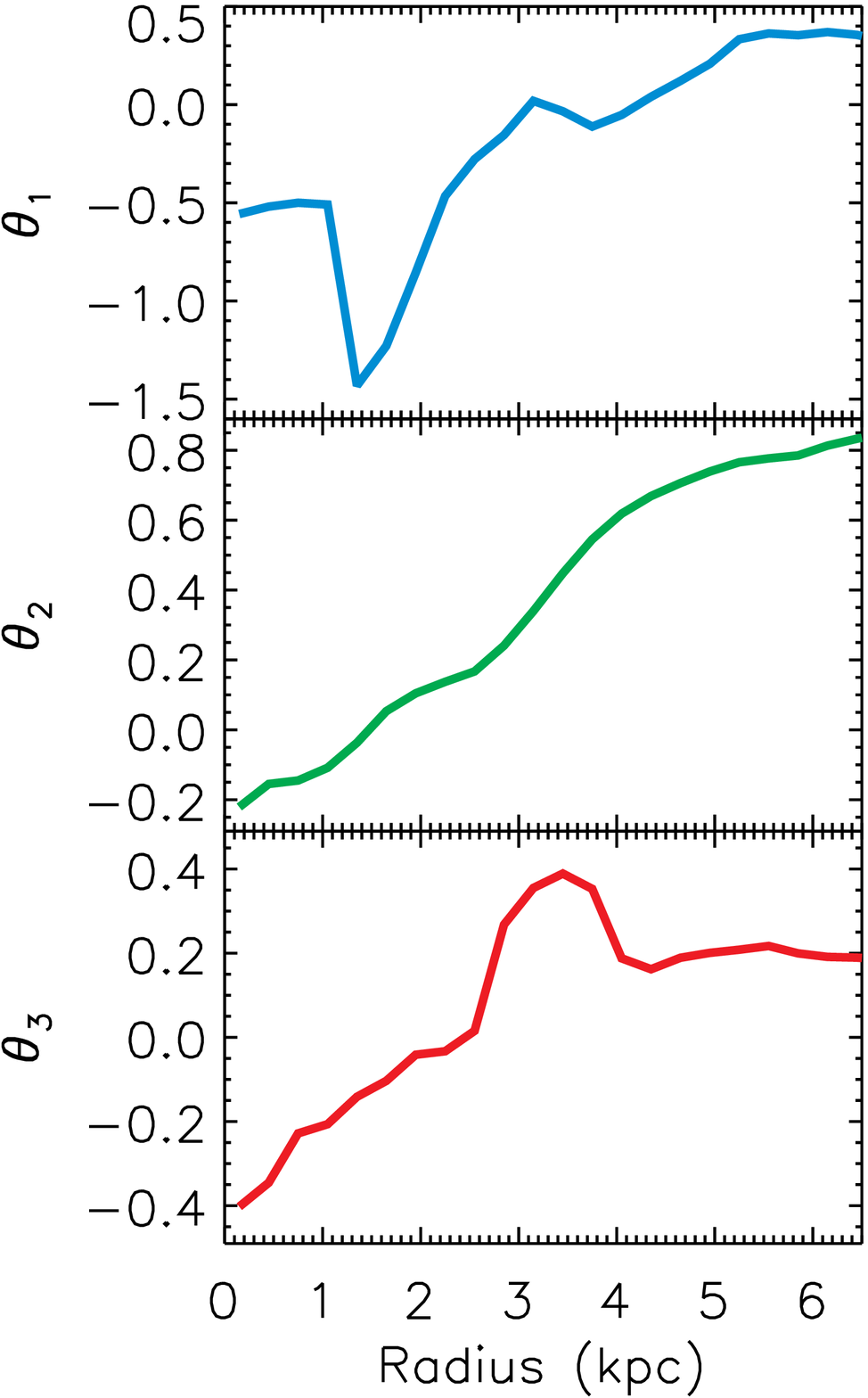}
\caption{Same as Fig.~\ref{fig:fourierpotamp}, but for the phase angle $\theta_m$, in units of $\pi$ rad, with $m=1,2,3$.}
 \label{fig:fourierpotphase}
\end{center} 
\end{figure*}

   For each   mass component but the bulge, the gravitational potential in the $z=0$ kpc midplane, $\Phi (R,\theta,z=0)$ (abridged in $\Phi (R,\theta)$),
  is given by
 \begin{equation}
  \Phi (R,\theta) =  \Phi_0 (R) + \sum_{\rm m=1}  \Phi_{m} (R)\cos(m(\theta-\theta_m(R)) \,
  \label{eq:fourierpot}
  ,\end{equation}

 where  $\Phi_0$ is the amplitude of the axisymmetric component, and 
 $\Phi_m(R)$ and $\theta_m(R)$ are the amplitudes and phases of the $m$-th harmonics.
Orders up to $m=3$ are enough for both stellar and gaseous disks.
The  results for $\Phi_m(R)$ and $\theta_m(R)$ are presented in Figs.~\ref{fig:fourierpotamp} and~\ref{fig:fourierpotphase}. 

 For the stellar disk, the $m=1$ and $m=2$ perturbations are stronger in the center. 
 The $m=1$ amplitude dominates the $m=2$ value inside $R=4$ kpc, reaching 5\%  of the unperturbed $m=0$ mode at $R \sim 0.7$ kpc. 
 The   $m=2$ perturbation is slightly stronger than $m=1$ at $R>4$ kpc. It reaches $\sim 3$\%  of the unperturbed   mode at $R=1$ kpc 
 and $\sim 4\%$ at $R=9.5$ kpc. 
 As these two modes lie at similar radii, it   implies that the central $m=2$ spiral   is strongly lopsided or, alternatively, 
 that a prominent $m=1$ spiral coexists  with the  $m=2$ spiral structure, but at different phase angle.
The  $m=3$ amplitude   is more scattered than the others, never exceeding 3\% that of the unperturbed  mode. 

 For the disk of molecular gas, it is the $m=2$ spiral pattern that   dominates the perturbations in majority with an amplitude reaching $\sim$9\% of the molecular $m=0$ coefficient at 
 $R=2$ kpc. We point out the smooth variation of the phase angle of the $m=2$ pattern as a function of radius. The $m=1$ amplitude is slightly  
 stronger than the $m=2$ mode at $R=0.5$  and $4$ kpc. The   $m=2$ spiral structure is thus lopsided inside $R=0.5$ kpc, or accompanied by a dephased single molecular arm.
 Here again, the  $m=3$ amplitude   is more scattered and never exceeds 3\%  of the $m=0$ amplitude within $R = 6$ kpc.

 For the atomic gas disk, the $m=2$ perturbation admits a minimum amplitude at $R \sim 10$ kpc, for a relatively constant phase angle 
  within $R=5-15$ kpc.  That $m=2$ mode coincides well with the spiral structure of the stellar distribution. 
  The maximum amplitudes of the $m=2$   mode reach $\sim 10\%$   that of the $m=0$  mode. 
  The amplitude of the $m=1$ mode is a minimum   at $R\sim12$ kpc and reaches more than 20\% that of the  $m=0$  amplitude beyond that radius.  
  Since   the $m=1$ spiral is an extension of one of the two spiral arms  (Fig.~\ref{fig:density}), 
   it explains the offset of about 2 kpc between the dips of the $m=1$ and $m=2$ potentials. Furthermore, since
   the $m=1$ mode dominates all other perturbations, it is likely that the outer one-arm  perturbation  
   is at the origin of the $m=2$ spiral structure of M99, and not the opposite.
    As for the $m=3$ mode, it is weaker and shows two major patterns at $7$ and $13$ kpc,  which are dephased by $\sim \pi/3$ from each other.
   Whether this mode is an actual organized $m=3$ spiral structure is a matter of debate. More simply, it likely  reflects the complexity of the  
   distribution and gravitational potential of the atomic gas beyond the stellar disk.
     For $R>20$ kpc,  the $m=2$ and $m=3$ modes also present features of very weak amplitude,  but which are evident in the  phase angle profiles.  
   This outer $m=2$ feature could be a genuine spiral structure. It could also be the signature of a \hi\ warp of the atomic gas disk, 
   even if \hi\ densities have been deprojected under a  constant inclination angle  (Section~\ref{sec:surfdens}). Gas accretion   on the \hi\ disk outskirts 
   after a tidal interaction with a companion, 
   as suggested by numerical simulations of \citet{vol05} and \citet{duc08}, is likely to generate a disk warping. 
    If it exists, this \hi\ warp cannot be bisymmetric because of the more prominent $m=1$ pattern. 
   This explains the difficulty we faced when fitting an axisymmetric tilted-ring model of the \hi\ velocity field at these radii.        

 \begin{figure}[t]
\begin{center}
\includegraphics[width=0.8\columnwidth]{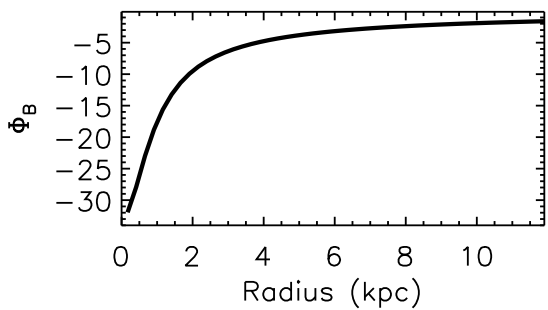}
\caption{Gravitational potential of the spherical stellar bulge of M99, in units of $\rm 10^3\ km^2\ s^{-2}$.}
 \label{fig:bulgepot}
\end{center}
\end{figure}

\begin{figure}[t]
\begin{center}
\includegraphics[width=0.8\columnwidth]{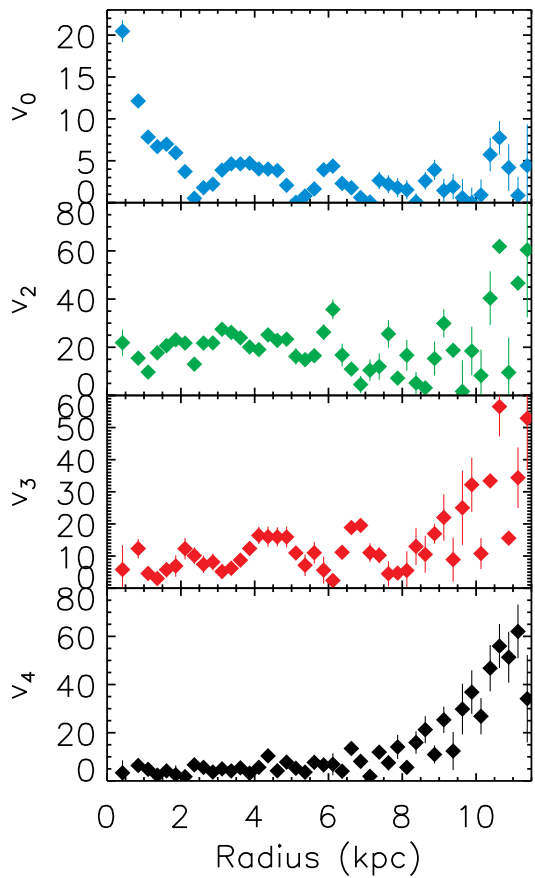}
\caption{Amplitude $v_k$ of the kinematical Fourier mode $k$ for the observed velocity field of M99  
 in units of \kms.}
 \label{fig:fouriervfobs}
\end{center} 
\end{figure}

 \section{Harmonics of the  \ha\ velocity field}
 \label{sec:fouriervf} 
 This Appendix  presents the results of the Fourier analysis of the \ha\ velocity field of M99.
 At a given radius, the  impact of kinematical asymmetries induced by perturbations, such as spiral arms, bars, lopsidedness, and warp,  makes 
 the velocity  fluctuate around the axisymmetric  component \citep{fra94,sch97,bin08}. The variations of \vt\ and  \vr\ are linked to the axisymmetric 
 \vc\ and the amplitudes,  phases, and pattern speeds of the gravitational potentials of the dynamical perturbations. 
   \citet{sch97} showed it is possible to   estimate the kinematical variations from a harmonic decomposition of velocity fields.  
   
 We  thus expand the standard model of Eq.~\ref{eq:vtanvradvf} into     
\begin{equation}
  v_{\rm obs} =   c_0 +  \sum^4_{\rm k=1} \left( c_{k}\cos k\theta + s_{k}\sin k\theta \right) \sin i  
  \label{eq:fouriervf}
,\end{equation}
where $c_k$ and $s_k$ are the Fourier velocity coefficients of harmonic order $k$ ($k$ is an integer). The $c_0$ coefficient is equivalent to the systemic velocity $v_{\rm sys}$ of 
Eq.~\ref{eq:vtanvradvf}.  
Unless constrained by direct methods,  the additional  vertical component \vz\ must not be used here   because its projection on the \los\ is degenerated with variations of the systemic velocity.

\citet{sch97} also showed that a perturbation of the potential of order $m$  induces kinematical components of order $k=m-1$ and $k=m+1$.
Therefore, as the $m=3$ order  is the weakest perturbing mode detected in the gravitational potentials of luminous matter, 
 kinematical harmonics up to $k=4$ can be derived. 
 We thus fitted Eq.~\ref{eq:fouriervf}  to the \ha\ data up to $k=4$ with 
obligatory uniform weightings, and fixed  $i$, $\Gamma$, and the coordinates of the kinematical center to the abovementioned values. 
The width of the radial bins is adaptive and chosen to  have more than 60 d.o.fs per ring for accurate fittings.
We derived the kinematical amplitudes $v_k$ of the harmonics   by $v_k=\sqrt{c_k^2+s_k^2}$ for $k>1$,  and 
$v_0=\sqrt{(c_0-v_{\rm sys})^2}$.  
No $k=1$ amplitude is derived because the  $s_1$ and $c_1$ coefficients are very similar to \vr\ and \vt\ from Eq.~\ref{eq:vtanvradvf} and Fig.~\ref{fig:harc}. 
This means that the decomposition is mostly efficient in modeling residual velocities  resulting from subtracting axisymmetric models to an observed velocity map, like the map shown in 
Figs.~\ref{fig:vfein2dasym} and ~\ref{fig:vfein2dasymshift}.

Figure~\ref{fig:fouriervfobs} shows the resulting profiles of the Fourier amplitudes. The phases are not shown as they do not present any particular interest for the 
analysis.
The amplitude becomes smaller for larger kinematical modes for $R< 7$ kpc. Beyond that radius, the
amplitude and the scatter increase and are larger for larger orders. 
Those variations at larger radii can be seen as direct responses to the increasing strength of perturbations of the stellar and neutral gas disk  potentials 
relative to the axisymmetric potentials (Fig.~\ref{fig:fracpotpert}) and,  likely ,  of dark matter as well. 
The number of d.o.fs  per radial bin decreases from $R \sim 4$ kpc, which explains the larger scatter of velocities at large radius. 
However, a smaller number of d.o.fs cannot be at the origin of the increase of the kinematical asymmetries. For instance, 
 the number of d.o.fs at $R \sim 2$ kpc is  similar to that at $R \sim 10.5$ kpc, while the asymmetries 
at $R \sim 2$ kpc are much smaller.
  The variation  of $v_0$,   principally in the center, does not mean that the systemic velocity varies with radius, but 
  the impact of the galaxy lopsidedness, and maybe of non-negligible  vertical  velocities. 
 The signature of the $m=1$ perturbation is confirmed by the detection of a $v_2 \sim 20$ \kms\ component out to $R = 7$ kpc.
  That $v_2$ component is very likely dominated by the effects of the $m=1$ pertubation, and not as 
  much  by the weakest $m=3$ perturbations. A confirmation of the small impact of the $m=3$ mode is the smaller component $v_4 \sim 5$ \kms\ 
  than $v_2$ for $R < 7$ kpc. 
  As for the $v_3$ component, it shows wiggles caused by  $m=2$ perturbing modes, among which are those identified 
  in the molecular gas and stellar disk  potentials.  Effects of the $m=2$ perturbations were already observed 
  in the $k=1$ order as departures of the 1D axisymmetric mass model from the rotation curve (Fig.~\ref{fig:harc}).    
 The kinematical asymmetries of M99 are  larger than those generally found in other dwarf and massive disks \citep{sch97,sch99,gen07,oh08}. This
discrepancy is explained by the lower angular resolution of their \hi\    velocity fields used to perform the  harmonic analysis. 
Large synthesized beams in radio interferometry indeed dilute the kinematics and  lower the  impact of asymmetries in the 
kinematical and dynamical modeling.  Similar amplitudes to those of M99 were found in other galaxies for $k=2$ components from  
\ha\ observations \citep[e.g.,][]{che06,spe07}, which 
confirms that high-resolution  data are more appropriate to estimate accurate amplitudes of kinematical asymmetries. 
 
 \section{Rotation curve, radial velocity,  and  velocity dispersion}
 \label{sec:vtvrsigmalos}

\begin{table*}
     \caption{M99 rotation curve, radial velocity,  and  velocity dispersion}
    \label{tab:vtvrsigmalos}
    \begin{tabular}{c|rr|rr|c||c|rr|rr|c}
\hline\hline
Radius & \vt & $\Delta_{v_\theta}$ & \vr & $\sigma_{v_R}$ & $\sigma_{\rm l.o.s}$  & Radius & \vt & $\Delta_{v_\theta}$ & \vr & $\sigma_{v_R}$ & $\sigma_{\rm l.o.s}$     \\ 
(kpc)  & \multicolumn{2}{c|}{(\kms)} &\multicolumn{2}{c|}{(\kms)} & (\kms)         & (kpc)  & \multicolumn{2}{c|}{(\kms)} &\multicolumn{2}{c|}{(\kms)} & (\kms)               \\
\hline                                                                                                                                    
 0.3 &  69.4 & 57.1 &   -8.4 &  8.6 &  24.5 & 6.0 & 262.1 &  5.3 &    5.9 &  4.5 &  20.7  \\
 0.5 & 114.5 & 35.6 &  -18.0 &  6.9 &  25.0 & 6.2 & 268.3 &  5.3 &    7.3 &  4.5 &  20.2  \\
 0.7 & 150.1 & 26.4 &  -15.1 &  5.9 &  24.5 & 6.4 & 267.5 &  6.9 &    8.3 &  4.7 &  19.7  \\
 0.9 & 166.0 & 21.1 &  -19.2 &  3.3 &  23.4 & 6.6 & 266.7 &  5.3 &    7.5 &  4.0 &  19.6  \\
 1.2 & 176.0 & 12.5 &  -18.5 &  2.9 &  23.7 & 6.8 & 267.2 &  4.9 &   -2.2 &  4.8 &  19.9  \\
 1.4 & 192.0 & 11.2 &  -23.6 &  3.1 &  23.5 & 7.0 & 272.3 &  3.3 &    1.0 &  4.6 &  20.8  \\
 1.6 & 203.5 & 10.2 &  -32.2 &  2.9 &  21.9 & 7.2 & 264.4 &  4.4 &   10.3 &  4.9 &  18.7  \\
 1.8 & 214.0 &  9.2 &  -24.2 &  4.3 &  21.2 & 7.4 & 269.5 &  5.0 &   15.6 &  5.4 &  19.4  \\
 2.0 & 233.0 &  5.7 &  -22.2 &  4.1 &  20.9 & 7.7 & 265.7 &  4.5 &   24.1 &  4.8 &  19.7  \\
 2.2 & 249.0 &  5.1 &  -19.6 &  2.9 &  20.8 & 7.9 & 258.6 &  4.2 &   16.1 &  5.2 &  18.4  \\
 2.4 & 258.3 &  4.8 &  -11.3 &  3.0 &  20.9 & 8.1 & 267.4 &  6.6 &   37.2 &  5.2 &  19.6  \\
 2.6 & 261.5 &  9.1 &   -6.3 &  3.0 &  20.5 & 8.3 & 263.7 &  5.0 &   25.8 &  5.3 &  19.7  \\
 2.8 & 267.7 &  9.0 &    6.2 &  3.0 &  21.0 & 8.5 & 265.3 &  6.7 &   24.2 &  5.0 &  20.2  \\
 3.0 & 274.6 & 12.8 &    7.6 &  2.9 &  21.7 & 8.7 & 278.9 &  6.3 &   30.6 &  5.3 &  20.0  \\
 3.3 & 273.5 & 13.4 &    7.3 &  3.1 &  21.7 & 8.9 & 268.8 &  6.4 &   29.0 &  6.2 &  19.6  \\
 3.5 & 269.3 & 13.0 &    2.9 &  3.2 &  21.2 & 9.1 & 273.0 &  7.0 &   22.0 &  6.8 &  19.7  \\
 3.7 & 265.1 & 13.3 &   -1.3 &  2.9 &  20.9 & 9.3 & 273.4 &  6.4 &    8.3 &  7.5 &  19.1  \\
 3.9 & 256.6 & 10.6 &    0.7 &  3.4 &  20.6 & 9.5 & 274.4 &  5.5 &   11.3 &  9.2 &  18.6  \\
 4.1 & 256.4 & 10.4 &    4.0 &  3.6 &  20.5 & 9.8 & 272.9 &  5.0 &   -2.1 &  8.3 &  17.9  \\
 4.3 & 252.1 &  7.4 &    7.4 &  4.0 &  20.6 &10.0 & 280.4 &  6.8 &    6.5 &  8.6 &  18.5  \\
 4.5 & 253.3 & 10.5 &    3.3 &  3.5 &  21.1 &10.2 & 271.3 &  5.4 &   -0.1 & 10.8 &  18.6  \\
 4.7 & 253.8 &  4.4 &    5.6 &  4.0 &  21.0 &10.4 & 271.2 &  8.5 &  -18.8 &  7.7 &  18.5  \\
 4.9 & 254.8 &  5.2 &   -6.5 &  4.3 &  20.6 &10.6 & 279.3 & 13.3 &   21.4 & 12.0 &  18.9  \\
 5.2 & 251.2 &  4.6 &   -2.9 &  3.7 &  20.2 &10.8 & 271.6 &  9.2 &   -4.2 & 11.1 &  18.1  \\
 5.4 & 252.1 &  3.5 &  -10.2 &  3.3 &  19.8 &11.0 & 263.4 &  9.2 &  -26.1 & 10.4 &  18.4  \\
 5.6 & 249.8 &  4.6 &   -8.8 &  4.3 &  20.2 &11.2 & 273.5 & 11.3 &  -15.1 & 16.1 &  16.9  \\
 5.8 & 258.1 &  4.7 &   -2.7 &  3.9 &  20.0 &11.4 & 278.3 & 10.3 &   29.6 & 20.6 &  18.4  \\
\hline                                                                                                                                    
\end{tabular}

Comments: $\sigma_{v_R}$ is the formal error from the fitting of \vr. 
\end{table*}
 
\section{Results of the mass distribution models}
\label{sec:tablesresults}

\begin{table*}
\centering
\begin{minipage}{\textwidth}
    \caption{Results of the mass distribution modeling of the rotation curve (1D) and velocity field (2D) of M99.}
    \label{tab:paramdm}
    \begin{tabular}{ll|ccc|cc}
\hline\hline
\multicolumn{2}{c}{}    &  \multicolumn{3}{c|}{Axisymmetric case}           &   \multicolumn{2}{c}{Asymmetric case} \\ 
\hline                                                                                                                                    
\multicolumn{2}{c|}{Centered-halo}  &        1D   & 2D  & 2D+\vr        &         2D &  2D+\vr  \\ 
\hline                                                                                                                                    
EIN     & $\rho_{-2}$   &  $0.9\pm3.7$    &  $1.0\pm1.6$  &      $1.3\pm1.8$    &   $3.1\pm2.8$  & $3.5\pm2.9$  \\
        &  $r_{-2}$     &  $48.30\pm126.13$    &  $43.11\pm40.88$  &      $38.43\pm32.56$    &      $22.31\pm12.01$   & $21.05\pm10.4$  \\
        &  $n$          &  $6.3\pm6.2$    & $6.1\pm2.3$   &      $5.8\pm2.0$    &         $4.3\pm1.3$    &   $4.2\pm1.2$  \\
        &  d.o.f       &     51          &     8999      &      8999           &    8999                &      8999         \\  
NFW     &$v_{200}$      &  $204.7\pm22.9$    &  $202.6\pm8.6$  &      $204.8\pm8.6$    &   $209.3\pm9.4$  &         $210.9\pm9.3$ \\
        & $c$           &  $14.5\pm1.9$    &  $14.7\pm0.7$  &      $14.6\pm0.7$    &       $14.2\pm0.7$   & $14.1\pm0.7$ \\
        &  d.o.f       &     52          &     9000      &      9000           &    9000                &      9000         \\  
PIS     & $\rho_0$      &  $274.9\pm45.0$    &  $275.3\pm17.2$  &      $270.6\pm16.3$    & $247.3\pm14.9$   & $245.1\pm14.3$      \\ 
        & $r_c$         &  $2.06\pm0.22$    &  $2.06\pm0.09$  &      $2.09\pm0.08$     & $2.22\pm0.09$   & $2.24\pm0.09$       \\                          
        &  d.o.f       &     52          &     9000      &      9000           &    9000                &      9000         \\  
\hline                                                                                                                                    
\hline                                                                                                                                    
\multicolumn{2}{c|}{Shifted-halo}        & --   & 2D  & 2D+\vr    &     2D &  2D+\vr  \\ 
\hline                                                                                                                                    
EIN     & $\rho_{-2}$   & --     &  $26.2\pm1.6$    &    $27.0\pm1.6$             &   $26.1\pm1.50$      &  $26.6\pm1.3$    \\
        &  $r_{-2}$     & --     &  $7.11\pm0.20$    &  $7.19\pm0.16$               &       $7.22\pm0.19$        & $7.33\pm0.16$ \\
        &  $n$          & --     &  $0.8\pm0.1$   &    $0.6\pm0.1$             &    $0.7\pm0.1$  & $0.6\pm0.1$    \\
        &  $\delta_x$   & --     &  $-0.65\pm0.06$   &     $-0.75\pm0.06$            &  $-0.46\pm0.06$  &  $-0.57\pm0.06$   \\
        &  $\delta_y$   & --     &  $-2.28\pm0.10$   &    $-2.51\pm0.10$             &     $-2.42\pm0.10$  &  $-2.64\pm0.10$   \\
        &  d.o.f        & --     &     8997      &      8997           &    8997                &      8997         \\  
NFW     &$v_{200}$      & --     &  $250.5\pm14.0$   &     $279.0\pm17.8$            &  $269.8\pm16.9$    & $302.6\pm22.0$  \\
        & $c$           & --     &  $11.5\pm0.7$   &    $10.3\pm0.7$             &         $10.6\pm0.7$ & $9.4\pm0.7$  \\
        &  $\delta_x$   & --     &  $-0.57\pm0.08$   &  $-0.65\pm0.07$               &     $-0.39\pm0.08$   & $-0.47\pm0.07$    \\
        &  $\delta_y$   & --     &  $-1.82\pm0.12$   &  $-2.15\pm0.12$               &      $-1.91\pm0.12$  &  $-2.23\pm0.12$   \\
        &  d.o.f       & --     &     8998      &      8998           &    8998                &      8998         \\  
PIS     & $\rho_0$      & --     &  $143.1\pm7.3$   &   $128.5\pm6.1$               &  $132.3\pm6.5$  & $120.1\pm5.6$    \\ 
        & $r_c$         & --     &  $3.25\pm0.12$   &      $3.54\pm0.13$           &        $3.44\pm0.13$  & $3.72\pm0.13$       \\                          
        &  $\delta_x$   & --     &  $-0.62\pm0.07$   &   $-0.69\pm0.07$              &     $-0.44\pm0.07$   &  $-0.53\pm0.06$   \\
        &  $\delta_y$   & --     &  $-2.13\pm0.10$   &  $-2.37\pm0.10$               &     $-2.23\pm0.10$   &  $-2.46\pm0.10$   \\
        &  d.o.f       & --     &     8998      &      8998           &    8998                &      8998         \\  
 \hline
\end{tabular}

Comments:  For the axisymmetric case, fitted data are the rotation curve (1D) and the velocity field (2D) of M99. 
For the asymmetric modeling, fitted data is only the velocity field.  
The 2D+\vr\ modeling differs from pure 2D by the addition of radial motions to tangential velocities.
The upper panel of the Table  (centered-halo) is for a dark matter halo whose dynamical center exactly coincides with that of 
the luminous matter ($x=y=0$), while the bottom panel (shifted-halo) is for a 
dark matter halo whose gravity center is offset from the  luminous center by $\delta_x$ and $\delta_y$ (kpc).
The parameter $v_{200}$ is in \kms, $r_c$ and $r_{-2}$ are in kpc, and $\rho_0$ and $\rho_{-2}$ in $10^{-3}$ \msol\ pc$^{-3}$. D.o.f  
 are the number of degrees of freedom.

\end{minipage}
\end{table*}

\begin{table*} 
\centering
\begin{minipage}{\textwidth}
    \caption{Comparisons between the various mass models: halo vs halo, axisymmetric vs asymmetric cases, with vs without the \vr\ component, and centered- vs shifted-halo.}
    \label{tab:diffaic1}
    \begin{tabular}{c|ccc|ccc|ccc}
\hline\hline
\multicolumn{1}{c|}{}  & \multicolumn{3}{c|}{$\rm AIC_{NFW}-AIC_{EIN}$}    &  \multicolumn{3}{c|}{$\rm AIC_{PIS}-AIC_{EIN}$}   &   \multicolumn{3}{c}{$\rm AIC_{NFW}-AIC_{PIS}$} \\ 
\hline                                                                                                                                    
Centered-halo & 1D & 2D  & 2D+\vr        &     1D &    2D &  2D+\vr  &    1D &     2D &  2D+\vr  \\ 
\hline                                                                                                                                    
Axisymmetric case  & 0.24 &   0.13 &   0.12 &  1.5 &    0.54 &    0.55 & -0.81 &   -0.42 &   -0.43  \\
Asymmetric  case & -- & 0.13 &   0.01  & -- & 0.33 &   0.24 &  -- & -0.21 & -0.22 \\
\hline                                                                                                                                    
\hline                                                                                                                                    
Shifted-halo & -- & 2D  & 2D+\vr        &     -- &    2D &  2D+\vr  &    -- &     2D &  2D+\vr  \\ 
\hline                                                                                                                                    
Axisymmetric case  & -- &   1.90 &   2.46  &  -- &  0.34 & 0.63  &  -- &   1.56 &  1.83  \\
Asymmetric  case & -- & 2.20 &  2.69  & -- & 0.42 & 0.67 &  -- &  1.79 &   2.02 \\
\hline
\end{tabular}

\begin{tabular}{c|cc|cc|c|cc}
 \hline
     \multicolumn{3}{c|}{$\rm AIC_{Axi.}-AIC_{Asym.}$}   &  \multicolumn{2}{c|}{${\rm AIC_{2D}}-{\rm AIC}_{\rm 2D+v_R}$} &    \multicolumn{3}{c}{$\rm AIC_{Centered-halo}-AIC_{Shifted-halo}$}  \\ 
\hline                                                                                                                                    
EIN  & 2D &  2D+\vr  &              Axi.  &  Asym.  &   &  Axi.  & Asym. \\ 
  Centered-halo   &1.62   & 0.72  & 6.19 &   5.30            & 2D   & 5.77  & 6.11 \\  
  Shifted-halo    &1.96   &   1.02 & 7.74 &   6.79           & 2D+\vr   & 7.31 & 7.60\\
\hline                                                                                                                                    
NFW  & 2D &  2D+\vr  &       Axi.  &  Asym.   &   &  Axi.  & Asym.\\ 
  Centered-halo   &1.62   & 0.71 &  6.20 &  5.30 & 2D  & 3.99  & 4.03 \\  
  Shifted-halo    &1.66   &   0.78 &  7.17 & 6.30       & 2D+\vr   & 4.96 & 5.03\\
\hline                                                                                                                                    
PIS  & 2D &  2D+\vr  &        Axi.  &  Asym.  &    &  Axi.  & Asym.\\ 
  Centered-halo   &1.83   & 0.92   & 6.19 & 5.28   & 2D  & 5.97  & 6.02 \\  
  Shifted-halo    &1.88   &   0.97  & 7.45 & 6.54 & 2D+\vr   & 7.23 & 7.28\\
\hline                                                                                                                                    
\end{tabular}

Comments: The numbers are the differences of the Akaike Information Criterion (AIC) for each configuration of mass model. A positive difference $\rm AIC_{Model1}-AIC_{Model2}$ means that 
Model2 is more likely  than the Model1. Differences of the AIC have been normalized to  $10^5$.
\end{minipage}
\end{table*}

\end{appendix}

\end{document}